\documentclass[12pt,preprint] {aastex}

\begin{document}
 
\title{THE VIRIAL BALANCE OF CLUMPS AND CORES IN MOLECULAR CLOUDS}

\altaffiltext {1} {Centro de Radioastronom\'{i}a y Astrof\'{i}sica, UNAM, Apdo. 72-3 (Xangari), 58089 Morelia, Michoac\'{a}n, Mexico; s.dib@astrosmo.unam.mx, e.vazquez@astrosmo.unam.mx}
\altaffiltext {2} {Korea Astronomy and Space Science Institute, 61-1, Hwaam-dong, Yuseong-gu, Daejon 305-764, Korea; jskim@kasi.re.kr}
\altaffiltext {3} {University Observatory Munich, Scheinerstrasse 1, D-81679 Munich, Germany; burkert@usm.uni-muenchen.de}
\altaffiltext {4} {School of Mathematical Sciences, Dublin City University, Glasnevin, Dublin 9, Ireland; mohsen.shadmehri@dcu.ie}
\altaffiltext {5} {Department of Physics, School of Science, Ferdowsi University, Mashhad, Iran}

\author{Sami Dib\altaffilmark{1,2}, Jongsoo Kim\altaffilmark{2}, Enrique V\'{a}zquez-Semadeni\altaffilmark{1}, Andreas Burkert\altaffilmark{3}, and Mohsen Shadmehri\altaffilmark{4,5}}

\begin{abstract} 

 We study the instantaneous virial balance of clumps and cores (CCs) in three-dimensional numerical simulations of driven, magnetohydrodynamical, isothermal molecular clouds (MCs). The models represent a range of magnetic field strengths in MCs from subcritical to non-magnetic regimes. We identify CCs at different density thresholds in the simulations, and for each object, we calculate all the terms that enter the Eulerian form of the virial theorem (EVT). A CC is considered gravitationally bound when the gravitational term in the EVT is larger than the amount for the system to be virialized, which is more stringent than the condition that it be large enough to make the total volume energy negative. We also calculate, quantities commonly used in the observations to indicate the state of gravitational boundedness of CCs such as the Jeans number $J_{c}$, the mass-to magnetic flux ratio $\mu_{c}$, and the virial parameter $\alpha_{vir}$. Our results suggest that: a) CCs are dynamical out-of-equilibrium structures. b) The surface energies are of the same order than their volume counterparts and thus are very important in determining the exact virial balance c) CCs are either in the process of being compressed or dispersed by the velocity field. Yet, not all CCs that have a compressive net kinetic energy are gravitationally bound. d) There is no one-to-one correspondence between the state of gravitational boundedness of a CC as described by the virial analysis or as implied by the classical indicators. In general, in the virial analysis, we observe that only the inner regions of the objects (the dense cores selected at high threshold levels) are gravitationally bound, whereas $J_{c}$, $\alpha_{vir}$, and $\mu_{c}$ estimates tend to show that they are more gravitationally bound at the lowest threshold levels and more magnetically supercritical. g) We observe, in the non-magnetic simulation, the existence of a bound core with structural and dynamical properties that resemble those of the Bok globule Barnard 68 (B68). This suggests that B68 like cores can form in a larger MC and then be confined by the warm gas of a newly formed, nearby \ion{H}{2} region, which can heat the gas around the core and confine it.      
                           
\end{abstract} 


\section{INTRODUCTION}

 Long standing questions in the process of star formation are the formation mechanisms, structure, and evolution of molecular clouds (MCs) and their substructure of clumps and cores (CCs), which are prone to star formation (e.g., Mouschovias 1995; McKee 1999; Williams et al. 2000; V\'{a}zquez-Semadeni et al. 2000; Mac low \& Klessen 2004; Elmegreen \& Scalo 2004; Burkert 2006; Ballesteros-Paredes et al. 2007). In MCs, which are observed to be turbulent (e.g., Zuckerman \& Evans 1974; Zuckerman \& Palmer 1974; Larson 1981; Blitz 1993; Williams et al. 2000; Engargiola et al. 2003; Rosolowsky \& Blitz 2005; Koda et al. 2006) and magnetized (Heiles et al. 1993; Crutcher 1999; Crutcher et al. 2004), it is essential to understand how the fragmentation process occurs. This is the necessary step on the way to understanding some of the most important results of the star formation process, a) the star formation efficiency, both locally in clouds and globally on galactic scales, b) stellar multiplicity and, c) the origin of the Initial Mass Function and its relationship with the prestellar cores mass distribution. In the last few decades, several theoretical models have been conceived in order to describe the evolving structure of MCs and there has been a debate on whether MCs and their substructure of CCs are dynamical out-of equilibrium structures or if they evolve in a state of quasi hydrostatic or magnetostatic equilibrium (e.g., Mac Low \& Klessen 2004; Mouschovias et al. 2006; Ballesteros-Paredes et al. 2007). 

There are two major classes of analytical models used to describe the evolving structure of MCs cores, concentrated around non-magnetic and magnetic configurations. In the non-magnetic case, Jeans (1902) derived the critical mass (i.e., the Jeans Mass) beyond which a system becomes gravitationally unstable against the support provided by the gas thermal motions. This analysis has been expended to include the support provided by isotropic, micro-turbulent motions (von Weizs\"{a}ker 1943,1951; Chandrasekhar 1951a,b; Sasao 1973). Since it was soon recognized that MCs are also turbulent on large scales of the order of their own size (e.g., Larson 1981; Basu \& Murali 2001; Ossenkopf \& Mac Low 2002; Brunt 2003; Heyer \& Brunt 2004), the assumption of micro-turbulence was replaced by turbulent motions that are scale dependent (e.g., Bonazzola et al. 1987; V\'{a}zquez-Semadeni \& Gazol 1995). In this class of non-magnetic equilibrium configurations, Bonnor (1956) and Ebert (1955,1957) derived analytical solutions for the equilibrium of radial density perturbations in a self-gravitating, isothermal ideal gas and which are solutions of the Lane-Emden equation (Lane 1870; Emden 1907). Column density observations show that the azimuthally averaged, radial density profiles of MCs cores have nearly flat or slow decreasing density in the inner parts and a surrounding envelope with a steeper density gradient (Ward-Thompson et al. 1994,1999; Andr\'{e} et al. 1996; Bacmann et al. 2000; Shirley et al. 2000,2002; Evans et al. 2001; Caselli et al. 2002; Young et al. 2003; Keto et al. 2004). Bonnor-Ebert (BE) sphere models have been applied to radial column density profiles observations of dense (average density of $\sim 10^{4}-10^{6}~$cm$^{-3}$) and cold ($T\sim 10$ K) cores (Bacmann et al. 2000; Johnstone et al. 2000; Alves et al. 2001; Evans et al. 2001; Harvey et al. 2001,2003a,b; Racca et al. 2002; Tafalla et al. 2004; Lada et al. 2004; Kirk et al. 2005; Kandori et al. 2005) and the dynamical evolution of BE spheres has been modeled numerically (McLaughlin \& Pudritz 1996; Ogino et al. 1999; Keto \& Field 2005; Aikawa et al. 2005). A study of isothermal cores whose structure is a solution to the Lane-Emden equation and which are embedded in a larger filamentary structure have been performed by Curry (2000).    

The good agreement in some cases between the models and the observational data has lead to the assumption that some cores are indeed BE equilibrium configurations, despite the fact that they have, in general, an obvious non-spherical morphology (e.g., Barnard 68; it should be noted however that Lombardi \& Bertin 2001, showed that the BE instability is more related to the core's density contrast than to it's shape). Ballesteros-Paredes et al. (2003) showed that azimuthally averaged column densities of dynamical, out-of-equilibrium cores formed in a turbulent, isothermal cloud, can be fitted by BE equilibrium profiles. Steinacker et al. (2004) argued that projection effects can be misleading about the true nature of the density distribution of MCs cores. Recently, Keto \& Field (2005) and Keto et al. (2006) argued that, not only the density profiles, but also the asymmetric molecular spectral lines observations in some cores (e.g., Ba 68, Lada et al. 2003; Redman et al. 2006) can be explained by non-radial oscillations around an equilibrium BE sphere (see also Galli 2005). An intriguing point about BE spheres models is that, as the volume of the core decreases and central density and density contrast increases, external pressure should increase (e.g., Figs 1, 2, and 3 in Keto \& Field 2005). Yet the derived values of external pressures, precisely from BE modeling of cores, by Kandori et al. (2005) do not show this effect (Fig. 7f in their paper). Rather, the external pressure they obtain for a sample of cores which are supposed to represent a evolutionary stage of equilibrium BE spheres, seems to be independent from the density contrast. 

A second class of models, introduces magnetic fields as an additional supporting mechanism against gravity (e.g., Mouschovias 1976a,b; Mouschovias 1987; Basu \& Mouschovias 1994,1995a,b; Shu et al. 1987; Hujeirat 1998). In the framework of this theory, magnetostatically supported cores, which can also account for thermal and micro-turbulent support, (e.g., Lizano \& Shu 1989; McLauglin \& Pudritz 1996), increase their masses by accreting neutrals that drift across the field lines from the surroundings onto the core's surface. This process is called ambipolar diffusion (AD, Mestel \& Spitzer 1956). AD will hence tend to increase the mass-to magnetic flux ratio in the core until it becomes larger than a certain critical value allowing for the denser inner parts of the core to collapse gravitationally. Whereas there is no doubt that AD can play an important role provided the ionization fraction inside MCs cores is small, the debate has concentrated on whether AD, which acts on it own timescale of $t_{AD} \sim 10~t_{ff}$ (McKee 1989; Fiedler \& Mouschovias 1992; Ciolek \& Basu 2001), where $t_{ff}$ is the free fall time, has time to play any role at all or whether dense cores which form in convergent flows in turbulent MCs proceed directly to collapse if their gravitational energy dominates all other forms of support (e.g., Balsara et al. 2001; Mac Low \& Klessen 2004; Mouschovias et al. 2006 and references therein). On the other hand, magnetic field strength measurements in cores, which can help constrain the AD timescale and the mass-to magnetic flux ratio (which is the main parameter in the theory of magnetically mediated star formation) of the cores, suffers from a major uncertainty which is due to the cores morphologies. Whether the cores are spherical or sheet-like structure can lead to an uncertainty by a factor of $1/3$ due to their unknown inclinations. Thus, cores that are classified as being magnetically supercritical, could be subcritical (Crutcher et al. 2004). Though observations tend to indicate that most cores are supercritical (Crutcher 1999; Sarma et al. 2000; Bourke et al. 2001; Crutcher \& Troland 2003; Crutcher et al. 2004; Mac Low \& Klessen 2004), Mouschovias et al. (2006) discuss the possibility that the latter estimates are biased towards the very innermost parts of the cores and that the supercritical cores are embedded in subcritical envelopes in agreement with the predictions of the AD theory (Cortes et al. 2005; Heiles \& Crutcher 2005).   
 
Whether magnetic or non-magnetic, analytical models suffer the fact that they apply only to idealized geometries, and neglect the complex interaction of the CCs with their environment. These facts, along with some observational and theoretical evidences that MCs may have lifetimes not longer than their turbulent crossing time and shorter than what would be implied by the AD-theory (e.g., Lee \& Myers 1999; Ballesteros-Paredes et al. 1999a; Palla \& Stahler 1999; Elmegreen 2000; Hartmann 2001; Hartmann et al. 2001) and that they must be magnetically supercritical (Nakano 1988), have prompted the development of mostly isothermal, magnetized and unmagnetized, turbulent or decaying MCs models in order to study the formation, statistics and properties of the dense structures formed within them (e.g., Padoan 1995,2001; Mac Low et al. 1998; Ostriker et al. 1999; Klessen et al. 2000,2001,2005; Smith et al. 2000; Heitsch et al. 2001; Gammie et al. 2003; Bate et al. 2003; Li et al. 2004; Tilley \& Pudritz 2004; Clark \& Bonnell 2005; Nakamura \& Li 2005). The emerging picture from these models is that dense structures (i.e., clumps) form  at the stagnation points of convergent turbulent flows with a tendency for the densest structures (i.e., cores) to form at the intersections of filaments in the clumpy distribution.  

In this picture of the dynamical evolution of MCs, the dynamical properties and detailed energy balance of their substructure remains poorly understood. Hunter \& Fleck (1982) showed that the gravitational boundedness of cores can be influenced by the presence of an external velocity field (i.e., a reduction of the Jeans mass). In the context of CCs forming and evolving in a turbulent parent MC, a valuable tool to quantify their instantaneous virial balance and state of gravitational boundedness is the virial theorem (Chandrasekhar \& Fermi 1953; McCrea 1957; Strittmatter 1963; Zweibel 1990). It has been widely assumed that MCs and their substructure are in virial equilibrium (e.g., de Jong et al. 1980; Shu et al. 1987; Henriksen 1991). This assumption is also widely spread in observational works (e.g., Larson 1981; Myers 1983; Solomon et al. 1987; Myers \& Goodman 1988a,b; Goodman et al. 1993), essentially because their observed kinetic, magnetic and gravitational energies are {\it of the same order of magnitude}. However, the assumption of the virial balance of CCs in MCs have been contested by several authors, particularly for the smallest clouds which are found to be, in general, non self-gravitating (e.g., Carr 1987; Blitz 1987,1993; Maloney 1988,1990; Herbertz et al. 1991; Bertoldi \& McKee 1992; Heyer et al. 2001). The observations by Swift et al. (2006) of a newly discovered core in L1551 (L1551 MC) suggest that this core is highly dynamical and its dynamics is not consistent with a slow gravitational contraction. However, it is not completely uncommon to see the detailed virial balance calculated for CCs in the observations (e.g., Crutcher 1999; Ward-Thompson et al. 2002;2006). In a study of a sample of 27 clouds and CCs in MCs, Crutcher (1999) found that the averaged values of the kinetic, magnetic, and gravitational volume energies in the virial theorem verify a static form of the virial theorem to $20~\%$. Individually, the different objects show a large scatter and do not satisfy the virial equation by $\gtrsim 30 \%$. Ward-Thompson et al. (2006) calculated the detailed virial balance for cores B33-SMM1 and B33-SMM2 in the Horsehead nebulae and found them to be out-of-equilibrium and in near virial equilibrium, respectively. Their estimates, however, took into account only the volume energies and neglected the surface terms in the virial theorem.

The aim of this work is to assess, for CCs which form in three-dimensional, isothermal, turbulent, magnetized MCs models, whether they are virialized and to quantify the importance of the different terms in the virial theorem equation. In \S~\ref{virial}, we present the virial theorem and discuss the different terms that enter it. The MCs simulations are presented in \S~\ref{data}, and in \S~\ref{anatheo}, the clump-finding algorithm used to identify the CCs in the simulations is presented. Results of the virial analysis for different MC models are described in \S~\ref{results}. In particular, we emphasize on the correspondence between the gravitational boundedness of CCs as diagnosed from their virial balance on one hand, on the other by classical gravitational boundedness indicators such as the Jeans number, the mass-to magnetic flux ratio, and the virial parameter, commonly used in theoretical and observational works. In \S~\ref{conc}, we summarize our results and conclude.

\section{THE EULERIAN VIRIAL THEOREM}\label{virial}

Cast in its Eulerian form (Parker 1979; McKee \& Zweibel 1992; McKee 1999), the virial theorem (EVT), which is nothing else but a re-writing of the momentum equation after it has been dotted by the position vector and integrated over a given volume of interest, has, for an object of arbitrary shape, the following expression

\begin{equation} 
\frac{1}{2} \ddot{I}_{E}=2 \left(E_{th}+E_{k}-\tau_{th}-\tau_{k}\right)+E_{mag}+\tau_{mag}+ W - \frac{1}{2}\frac{d \Phi}{dt},
\label{eq1}
\end{equation}

\noindent where $I_{E}=\int_{V}\rho r^{2} dV$ is the moment of inertia of the object, $r$ is the distance of each point in the object to its center of mass, and $V$ its volume. $E_{th}=\frac{3}{2}\int_{V} P dV$ is the volume thermal energy, with $P$ being the thermal pressure, $E_{k}= \frac{1}{2}\int_{V} m_{i} v_{i}^{2} dV$ the volume kinetic energy, $E_{mag}=\frac{1}{8 \pi}\int_{V} B^{2} dV$ the volume magnetic energy, $\tau_{th}=\frac{1}{2}\oint_{S} r_{i} P \hat{n}_{i} dS$ the surface thermal energy, $\tau_{k}=\frac{1}{2} \oint_{S} r_{i} \rho v_{i} v_{j} \hat{n}_{j} dS$ the surface kinetic energy, and $\tau_{mag}=\oint_{S} r_{i} T_{ij} \hat{n}_{j} dS$ the surface magnetic energy, where $T_{ij}$ is the Maxwell stress tensor which is given by 

\begin{equation} 
{\bf T} =  \frac{1}{4~\pi} \left({\bf B~B}-\frac{1}{2} B^{2} {\bf I}\right),
\label{eq2}
\end{equation}

\noindent where ${\bf I}$ is the diagonal identity tensor. The effects of the gravitational field are described by $W= - \int_{V} \rho r_{i} (\partial \phi/\partial r_{i}) dV$, where $\phi$ is the gravitational potential. This term is not exactly equal to the volume gravitational energy because the gravitational potential is a result of the distribution of matter inside (e.g., clump or core) and outside the object (e.g., the parent cloud), with the essential part of the gravitational acceleration being due to the mass contained inside the object. The last term on the right hand side in Eq.~\ref{eq1} is the first time derivative of the flux of moment of inertia through the boundary of the cloud, $\Phi=\oint_{S} \rho r^{2} v_{i} \hat{n}_{i} dS$. In all the previous quantities, ${\bf v}$ is the velocity of a point within the defined object relative to the velocity of the center of mass of the object. We chose to apply the Eulerian version of the virial theorem to the selection of CCs in our simulations simply because the numerical grid we are using is Eulerian. Using the Lagrangian form of the virial theorem would have forced us to trace fluid elements, which would have required to integrate a passive scalar equation (it would be a method of choice for Lagrangian methods such as smooth particle hydrodynamics). Additionally the Eulerian approach, using density-threshold defined CCs (i.e., see below), which are Eulerian because they trace a given volume rather than a given set of particles, is closer to what is regularly done in the observations. 

Few studies have attempted to evaluate the terms involved in Eq.~\ref{eq1} for CCs formed in a turbulent medium (i.e., for non-isolated clouds). Ballesteros-Paredes \& V\'{a}zquez-Semadeni (1997) and Ballesteros-Paredes et al. (1999b) evaluated some of the EVT terms in 2D numerical simulations and stressed the importance of the surface term $\tau_{k}$. Shadmehri et al. (2002) evaluated the different terms of the EVT in the case of dense structures formed in 3D, MHD simulations with grid resolutions of $100^{3}$. These studies made the assumption that an object is gravitationally bound if ($|W|>|2~(E_{th}+E_{k}-\tau_{k}-\tau_{th})+E_{mag}+\tau_{mag}|$). More recently, Tilley \& Pudritz (2004,2006), evaluated the non-time dependent terms in the EVT in 3D simulations, and have assumed that the sign of the sum of these terms determines whether an object is gravitationally bound or not (i.e., negative is bound).  

\begin{figure}
\plotone{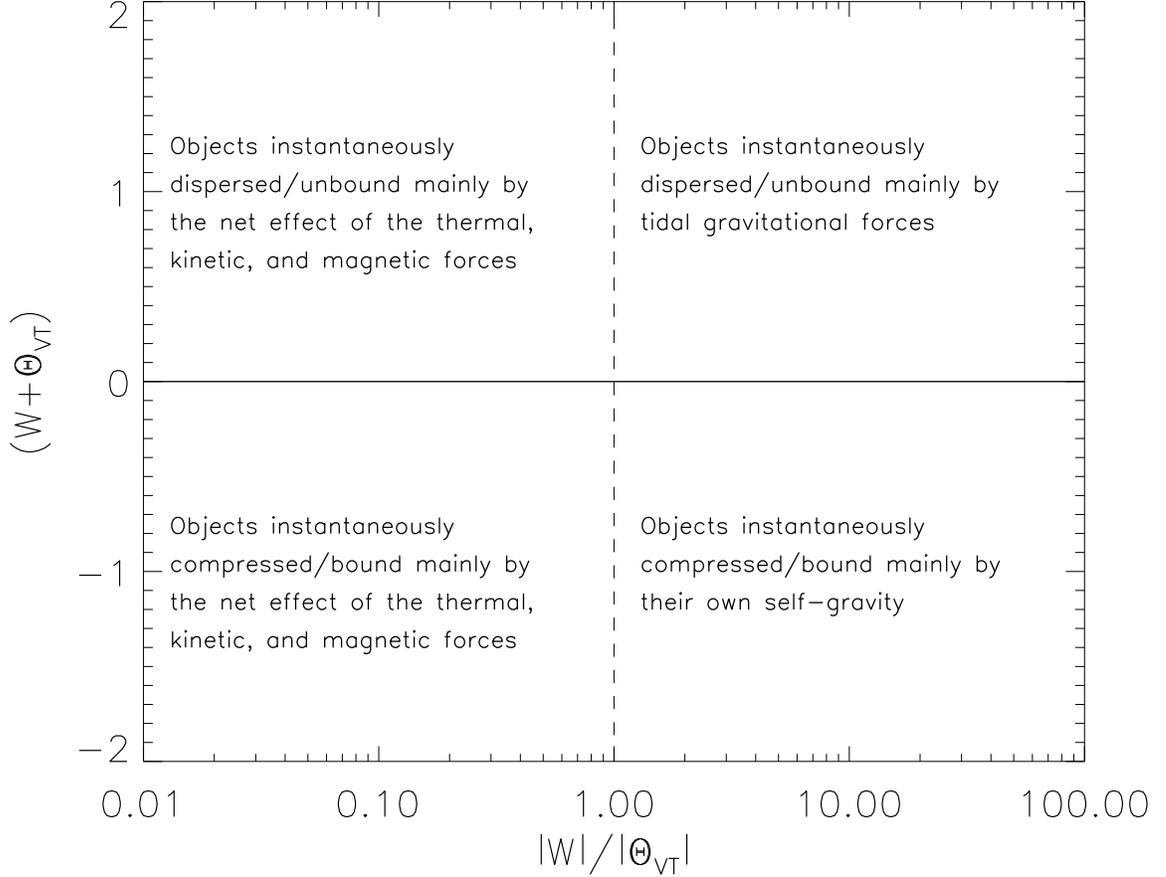}
\caption{A sketch showing the instantaneous gravitational boundedness state of an object according to its virial balance, shown here in arbitrary units. If the cloud in which the object forms is turbulent and magnetized, the general form of $\Theta_{VT}$ is $\Theta_{VT}= (2~(E_{th}+E_{k}-\tau_{k}-\tau_{th})+E_{mag}+\tau_{mag})$, where the $E_{i}$ and $\tau_{i}$ (with $i=th,k,mag$) terms are the volume and surface, thermal, kinetic and magnetic energies, respectively. The detailed expression of each term is given in \S~\ref{virial}. In the text, we use indiscriminately the terms 'bound' and 'compressed' for objects located in the lower quadrants, and the terms 'unbound' and 'dispersed' for objects located in the upper quadrants. Objects that fall on the horizontal $W+\Theta_{VT}=0$ are called indiscriminately 'instantaneously in equilibrium' or instantaneously virialized'. In isothermal systems such as the cores and clumps in our simulations, the notion of instantaneous (un)boundedness can be replaced by a notion of 'tendency to be permanently bound/unbound'. Thus, in isothermal systems an object located in the lower right quadrant will have a tendency of being permanently gravitationally bound by gravity and eventually proceed towards gravitational collapse (see \S 2).}  
\label{fig1}
\end{figure}

Due to the dual role of the surface+volume thermal, magnetic, and kinetic energies and also the gravitational field, which can, theoretically, both help confine an object or disperse it\footnote{In the case of an idealized axisymmetric filamentary structure, a purely poloidal magnetic field helps support the object against gravity while a purely toroidal field tends to aid gravity compress the object. The net effect of a helical magnetic field which has toroidal and poloidal components in such configurations will be of support/dispersion or compression depending on how much the field is poloidally or toroidally dominated (for more discussion on this point see Fiege \& Pudritz 2000).}, we argue that the criterion of instantaneous gravitational boundedness\footnote{In the rest of the text we will not repeat, for commodity, the term 'instantaneous' associated with the virial analysis, but it is always assumed to be the case. Also, in the literature it is common to find the term 'bound' associated with both the compressed/bound case (i.e,. $W+\Theta_{VT} < 0$, lower quadrants in Fig.~\ref{fig1}) and the virial equilibrium case ($W+\Theta_{VT} = 0$, horizontal line in Fig.~\ref{fig1}) case (both cases have a total volume energy $E_{tot} < 0$). In this paper, we make a clear distinction between the two cases by calling the first one, indiscriminately, bound or compressed (gravitationally bound, compressed, or captured if the CC is located in the lower right quadrant in Fig.~\ref{fig1}) and the second one 'equilibrium' or 'virialized'. Objects located in the upper quadrants in Fig.~\ref{fig1} are called indiscriminately 'unbound' or 'dispersed'.}, is not only $|W|>|\Theta_{VT}|$, but also $(W+\Theta_{VT}) < 0$, where $\Theta_{VT} =2~(E_{th}+E_{k}-\tau_{k}-\tau_{th})+E_{mag}+\tau_{mag}$. If for a given object $(W+\Theta_{VT}) < 0$, but $|W|<|\Theta_{VT}|$, the object is instantaneously being predominantly bound/compressed by forces other than gravity (i.e., magnetic stresses or dynamical compression). Fig.~\ref{fig1} shows, in arbitrary units, a diagram describing the state of instantaneous gravitational boundedness of an object according to the balance of the different virial terms and which we will apply in the next sections to the identified CCs in our simulations. Since the virial theorem ignores the spatial distribution of the different forces at work in a CC, its use as a criterion for CCs boundedness gives only correct qualitative behavior and scalings that deviate by some factor from the correct quantitative solution to the fluid equations (e.g., McCrea 1957; Zweibel 1990). For example, as reminded by Zweibel (1990), the use of a constant density model for an isothermal, self-gravitating, non-magnetized, and pressure truncated CC in the virial theorem gives results that deviate by factors 2-3 from the exact solutions (Spitzer 1978; see also McCrea 1957). Therefore the $W+\Theta_{VT}=0$ line which in our case draws the boundary between the bound (whether by gravity or other forces) and unbound regime may not be a very accurate description of the equilibrium of forces for objects which fall very close to the line (i.e. $W+\Theta_{VT} \sim 0$). However, to this date there is no analytical solution to describe the equilibrium of CCs with complex morphologies in the presence of magnetic fields and turbulence (i.e., like the Bonnor-Ebert theory for thermal gradient pressure force balancing the gravitational force), particularly when the velocity and magnetic fields have complex topologies in and around the objects.

 It is very important to stress that the virial theorem equation represents a snapshot of the transformed and spatially integrated balance of the different forces in a CC. The virial equation, without any additional physical information such as the local and time varying equation of state, rate of turbulent energy injection, and the strength and topology of the magnetic field, inside and at the boundaries of the object, can not generally be used to make a full prediction about the time evolution of CCs. Such predictions require the detailed knowledge on how the (magneto)hydrodynamical quantities evolve over time inside and at the boundaries of the objects and the full resolution of the time dependent (magneto)hydrodynamical equations. Thus, the diagram in Fig.~\ref{fig1} reflects the instantaneous statistics of CCs that are instantaneously bound mainly by gravity, bound mainly by other forces, or unbound, and mimics the observational situation where information at different epochs can not be obtained. Thus, in the most general case, the usefulness of the virial theorem for deriving quantities such as the star formation efficiency (for which a key element is the ratio of gravitationally bound objects to non-gravitationally bound objects) does not rely on its predictive power for individual objects (some predictions can be made in specific situations, e.g., if for example the gas is isothermal, see next paragraphs), but essentially on large statistical samples that can be found in each of the quadrants in Fig.~\ref{fig1}. 

Another way of defining the gravitational boundedness of a CC is by requiring that its total volume energy $E_{tot}$ be negative. In simplified systems with no effects of the environment (i.e., isolated systems), and for objects that are either in virial equilibrium or where the virial balance is tipped in favor of gravity (i.e., $W+2~E_{th} \gtrsim 0$), $E_{tot}$ is guaranteed to be negative. An example of such an isolated system is the case of a non-turbulent, non-magnetized gas sphere with no external compressive or dispersive agents. For an adiabatic index of the gas in the sphere of $\gamma > 4/3$ (e.g., a star) the sphere (i.e., star) is in equilibrium if $E_{g}+2~E_{th}=0$ (i.e., $E_{tot} < 0$). At equilibrium, the sphere will be located on the intersection of the $W+\Theta_{VT}=0$ and $|W|/\Theta_{VT}|=1$ lines in Fig.~\ref{fig1} (in this case $\Theta_{VT}$ is reduced to the volume thermal energy contribution $2~E_{th}$ and $W$ to the gravitational energy $E_{g}$). The gas may condense in a first stage in reaction to a compression (and will have temporarily $E_{g}+2~E_{th} < 0$, and $E_{tot} < 0$), but as in this case $2~E_{th}$ grows faster than the gravitational energy (i.e., thermal gradient pressure force grows faster then the gravitational force), the compression will come to a halt and the gas will expand ($E_{g}+2~E_{th} > 0$, and still with $E_{tot} < 0$) and return to equilibrium (e.g., Hayashi 1966; Larson 1969; Low \& Lynden-Bell 1976; Masunaga \& Inutsuka 1999). Conversely if the sphere (i.e., star) is perturbed and starts to expand ($W+2~E_{th} > 0$ and $E_{tot} < 0$), $|E_{g}|$ decreases slower than $2~E_{th}$ and the expansion will be halted and the system will contract again and return to equilibrium. In all cases, such a system always has $E_{tot} < 0$. 

However, if the adiabatic index of the equation of state of the gas is $\gamma < 4/3$ (such as for an isothermal, $\gamma=1$, gas; i.e., like our simulations, and eventually MCs), as soon as an object is captured/compressed/bound by the gravitational force (i.e., the lower right quadrant in Fig.~\ref{fig1}), the pressure gradient force will not be able to balance the gravitational force. In the absence of any other support agent, the CC will be compressed further as time proceeds, and if it becomes dense enough, will eventually proceed towards gravitational collapse. The same object  (i.e., with $\gamma < 4/3$) will expand indefinitely if perturbed by an expansion off the equilibrium (even if in an initial stage it still has $E_{tot} < 0$), as in this case $|E_{g}|$ decreases faster than $2~E_{th}$. If a CC is found in the lower right quadrant in our simulations (i.e., instantaneously gravitationally bound), and eventually in the observation, and if the gas is known to be isothermal or nearly-isothermal and if the time varying surface terms are assumed not to vary substantially as time evolves, and in the absence of stellar feedback, then this CC will very likely remain gravitationally bound until it becomes dense enough to undergo  gravitational collapse. Thus, the definition of boundedness based on the total energy applies only to isolated systems for which the future behavior of the forces can be predicted. In non-isolated system, the fact that $E_{tot} < 0$ and the system be defined as 'bound' at one instant according to this criterion does not imply that it will remain bound at a subsequent epoch. Therefore, in these limiting cases, there is no discrepancy between the virial analysis and the $E_{tot}$ argument for objects which are truly instantaneously captured by the gravitational force or that are in equilibrium. Both cases can be merged to define the 'term' bound in the $E_{tot}$ argument. However, calling an object that is instantaneously being dispersed by the gradient pressure force (i.e., $E_{g}+2~E_{th} > 0$) 'bound' if it has $E_{tot} < 0$ is only possible if the adiabatic index and eventually its time dependence at each location of the object allows such conclusion. More generally, predictions based on the total energy, similarly to the virial analysis, require the knowledge of how external pressure and also the time varying turbulence and magnetic fields at the boundaries of the object will affect the total volume energy. The advantage of using the virial analysis over the total energy is that even if it cannot predict the time evolution of the magnetohydrodynamical quantities in general systems inside and at the surface of the object, it incorporates an evaluation of the instantaneous effects of the surface terms while the total energy criterion does not.

\section{SIMULATIONS}\label{data}

Aspects of the three-dimensional numerical simulations analyzed in this paper are described in V\'{a}zquez-Semadeni et al. (2005a). We use the first 4 simulations summarized in Table 1 of that paper, keeping the same nomenclature for the different runs. Here, we recall their basic features. The ideal MHD equations are solved using a total variation diminishing scheme (TVD), which is a second-order-accurate upwind scheme. Its implementation for isothermal flows is described in detail in Kim et al. (1999). Periodic boundary conditions are used in the three directions. The Poisson equation is solved to account for the self-gravity of the gas using a standard Fourier algorithm. In order to achieve second-order accuracy in time, an update step of the momentum density due to the gravitational force is implemented, as in Truelove et al. (1997). Following the method described in Stone et al. (1998), turbulence is constantly driven in the simulation box and the kinetic energy input rate is adjusted as to maintain a constant specified rms sonic Mach number $M_{s}=10$. Kinetic energy is injected at large scales, in the wave number range $k=1-2$. In physical units, all 4 simulations have a box size of 4 pc, an average number density of $500$ cm$^{-3}$ (i.e., thus the column density is $\sim 10^{21.7}$ cm$^{-2}$ which is similar to that of many MCs except for those associated with massive star formation (e.g., Crutcher 1999), a temperature of 11.4 K, a sound speed of $0.2$ km s$^{-1}$, and an initial {\it rms} velocity of 2 km s$^{-1}$ (i.e., therefore the initial sonic Mach number is $M_{s}=10$). The Jeans number of the box is $J_{box}=4$ (i.e., number of Jeans masses in the box is $M_{box}/M_{Jeans,box}=J_{box}^{3}=64$). The 4 simulations vary by the strength of the initial magnetic field ranging from a magnetically subcritical cloud model to a nonmagnetic cloud. The initial magnetic field strength in the box for the subcritical, moderately supercritical, and strongly supercritical magnetized models are $B_{0}=45.8, 14.5$, and $4.6$ $\mu$G, respectively. Correspondingly, the $\beta$ plasma and mass-to-magnetic flux values of the box for these runs are, $\beta=0.01, 0.1$ and $1$, and $\mu_{box}=0.9, 2.8$, and $8.8$, respectively.

Several issues pertaining to the set of simulations presented in this paper such as the assumption of isothermality, the neglect of ambipolar diffusion, resolution considerations, and the choice of a driven turbulence regime have been discussed in some detail in V\'{a}zquez-Semadeni et al. (2005a). We complement this discussion by justifying our selection of simulations where turbulence is driven on the largest scales. Our motivation for analyzing such simulations is due to the increased evidence from recent theoretical work and observations that turbulence is most likely driven on large scales in the interstellar medium of different galaxies (e.g., Stanimirovi\'{c} \& Lazarian 2001 for the LMC; Dib \& Burkert 2004,2005 for Holmberg II; Koda et al. 2006 in the Galaxy). An analysis of the turbulent velocity structure by Ossenkopf \& Mac Low (2002) and the filamentary density structure in Taurus by Hartmann (2002) also suggest that MCs are driven on scales which are of the order of their own size, or larger. 

The nature of the large scale turbulence driver(s) in the ISM remains an unresolved issue. A minimum level of turbulence might be due to supernova explosions (e.g., de Avillez 2000; Kim et al. 2001; Kim 2004; Dib et al. 2006), but several large scale instabilities are suspected to play an additional role in that respect, such as instabilities that occur in spiral shocks (Wada \& Koda 2004; Bonnell et al. 2005; Kim \& Ostriker 2006; Kim, Kim \& Ostriker 2006; Dobbs et al. 2006), large scale gravitational and thermal instabilities (e.g., Wada et al. 2002; Dib \& Burkert 2005), generic convergent flows (Heitsch et al. 2005,2006; Audit \& Hennebelle 2005; Folini \& Walder 2006; V\'{a}zquez-Semadeni et al. 2006a,b) and the magneto-rotational instability (e.g., Brandenburg et al. 1995; Selwood \& Balbus 1999; Dziourkevitch et al. 2004; Dziourkevitch 2005; Piontek \& Ostriker 2005). Nevertheless, recent studies also show that, in star forming clouds, the feedback from protostellar outflows (Li \& Nakamura 2006; Matzner 2007) and \ion{H}{2} regions (Krumholz et al. 2006) can, not only enhance the level of turbulence inside the clouds, but eventually lead to a self-sustained level of turbulent motions.  
  
The simulations start with a uniform density field, but are evolved for half a sound crossing time (the sound crossing time is $t_{s}=20$ Myr), equivalent to 5 turbulent crossing times (the turbulent crossing time is $t_{c}=2$ Myr), before self-gravity is turned on. This is a necessary step in order to allow for the full development of the turbulent cascade. In some simulations (i.e., those with a weak and a zero magnetic field), CCs form and reach the stage of gravitational collapse at epochs which are shorter than a turbulent crossing time. In the absence of a prior established turbulent cascade, those CCs could not be considered as forming and evolving in a driven medium, but rather in a medium with decaying turbulence, which would be contrary to the formation of CCs in a driven turbulence regime we want to investigate here. For example, in the subcritical simulation M10J4$\beta$.01, we focus on timestep $223$ which corresponds to the epoch $223 \times 0.002 \times t_{s}=0.446~t_{s}=4.46~t_{c}$ after self-gravity has been turned on (i.e., time dumps of the physical quantities are performed at intervals of $0.002~t_{s}$). For this run the turbulent cascade would have been already established at this timestep, without any initial driving. However, for the moderately supercritical (i.e., run M10J4$\beta$.1), and strongly supercritical (i.e., M10J4$\beta$1) and hydrodynamical runs (i.e., M10J4$\beta \inf$), we focus the analysis on timesteps 40, 30, and, 20 respectively. This timesteps would correspond to epochs of $0.8, 0.6$, and $0.4~t_{c}$, respectively. It is clear that at these epochs, if no turbulent cascade was already established, the medium in which CCs would be forming and evolving in these simulations could not be considered a driven one. 
  
When applying a virial balance analysis to CCs in simulations such as ours, a crucial question is: until which time in the evolution of the simulation does the virial analysis remain valid. Due to the limitations of the numerical resolution, the internal dynamics of a collapsing object can not be accurately followed if the mass of the object exceeds a Jeans mass and if it is not resolved by at least 4 cells in one direction (Truelove 1997). This criterion translates into a maximum density ($n_{res} \sim 256~\bar{n}$), which when reached locally in one cell, indicates that the advection of the flow around this cell is numerically biased. Because we are precisely interested in this work in the internal dynamics of the dense structures forming in MCs, we choose to present the detailed analysis only for time-steps which precede the appearance of any collapsed structure in the simulation box. Fig. 2 in V\'{a}zquez-Semadeni et al. (2005a) shows the evolution of the maximum density in the sample of simulations we analyze in this paper. This figure shows that with a decreasing initial magnetic field strength, CCs tend to proceed towards collapse at earlier epochs. It should be noted, however, that the simulations are evolved beyond the epoch of appearance of the first collapsed object in the box. The global dynamics in the cloud, outside the collapsing CC, remains accurately described by the code. Any selection of CCs after the appearance of the first collapsing objects, remains valid in order to investigate the statistics of CCs in the box with the caveat that it does not properly describe the internal dynamics of the population of cores which are gravitationally collapsing.   
  
\section{SIMULATIONS ANALYSIS}\label{anatheo}

\subsection{Clump-finding algorithm}\label{method}

We developed a clump-finding algorithm that is based on a density threshold criterion and a friend-of-friend approach. The clump finding proceeds as follows: cells that have a density higher than the density threshold are selected. Among the selected cells, the maximum is sorted out and all cells spatially connected to it are saved in a separate list that defines the first clump. The density in the first clump cells are then assigned a value close to zero and the second maximum is searched for in the data cube and the procedure is repeated until all cells that define the second clump are found. This is repeated until all cells that have densities higher than the selected density threshold are assigned to a clump. Since our simulations use periodic boundary conditions, the algorithm also checks for clumps that extend across the boundaries of the box. Once the cells of each clump are defined, it becomes easy to calculate, for each clump, all the physical quantities that enter the EVT, in addition to other physical quantities that characterize its structure and dynamics such as its mass $M_{c}$, volume $V_{c}$, average density $\bar{n_{c}}$, turbulent velocity dispersion $\sigma_{c}$\footnote{The velocity dispersion, $\sigma_{c}$, is density weighted, and calculated as being $\sigma_{c}^{2}=\int_{V_{c}} n_{l} (v_{l}-\bar{v})^{2} dx dy dz/\int_{V_{c}} n_{l} dx dy dz$, where $n_{l}$ and $v_{l}$ are the local density and local velocity module of each cell, respectively, and $\bar{v}$ is the average velocity module inside the CC.}, and angular momentum $J_{c}$. The density thresholds $n_{thr}$ used to define the CCs are, in units of the average density (i.e., $\bar{n}=500$ cm$^{-3}$), 7.5, 15, 30, 60, and 100 $\bar{n}$\footnote{In the figures shown later, different symbols are used to discriminate between the data points at different thresholds. Namely, the $7.5~\bar{n}$ threshold level is shown with a (+), the $15~\bar{n}$ with a ($\ast$), the $30~\bar{n}$ with a ($\diamond$), the $60~\bar{n}$ level with a ($\triangle$), and the $100~\bar{n}$ level with a ($\square$).}. A few time-steps in some runs have additional clumps determined with thresholds of 500 $\bar{n}$. The latter range of density thresholds (i.e., from $3.75 \times 10^{3}$ to $5\times 10^{4}$ cm$^{-3}$) mimics a large range of critical densities that are necessary to excite various molecular lines in real MCs such as the $^{13}$CO (2-1) and CS (1-0) lines which have critical densities of $\sim 6.2$ and $18 \times 10^{3}$ cm$^{-3}$, respectively (Rohlfs \& Wilson 1996). The cores in our simulations do reach peak densities of $\sim 256~\bar{n}~\sim 1.25~\times 10^{5}$ cm$^{-3}$ at the limit of the Truelove resolution criterion. This value is close to the critical densities of the N$_{2}$H$^{+}$ (1-0, F$_{1}$=2-1,F=3-2) and HCO$^{+}$ (1-0) lines. However, in pre-collapse situations, very few cells that have a density of $\sim 250~\bar{n}$ are present in the simulation box, and CCs made of cells selected at this threshold are entirely dominated by numerical noise.     
  
In order to lower numerical noise, we transform the surface terms in the EVT into volume integrals using the Gauss theorem (i.e., $\oint_{S} \vec{X} \vec{dS}=\int_{V} \vec{\nabla} \vec{X} dV$)\footnote{We calculate the x-direction component of the gradient of a quantity $A$ at position (i,j,k) as being $(A(i+1,j,k)-A(i-1,j,k))/2$. The components in the y and z directions are calculated in a similar way.}. The volume integrals are replaced by summations running over all cells belonging to each object. The time derivatives appearing in Eq.~\ref{eq1} are calculated by assuming that the CC, within a time lapse $dt$, has moved over small distances $d{\bf r} = {\bf v_{CM}}\times dt$ following a uniform linear translation (i.e, with no acceleration), where $\bf v_{CM}$ is the velocity of the center of mass of the clump at instant $t$. Since we can not control $\bf v_{CM}$ of each object, we use the smallest time separation of time-dumps that is available to us and which amounts to $dt=0.002~t_{s}$, where $t_{s}=20$ Myr is the simulation box sound crossing timescale. Thus, the second time derivative of the moment of inertia is calculated using estimates of $I_{E}$ at three different time-steps $t-dt$, $t$, and $t+dt$. The flux of moment of inertia $d\Phi/dt$ is an average of the two values calculated from the three consecutive time-steps. The typical shifts in one direction are of the order of 1-2 cells, never exceeding 4 cells. Since the shifts in position associated to shifts from time-step $t$ to time-steps $t-dt$ and $t+dt$ generally  correspond to non-integer values, the physical quantities at the shifted time-steps are obtained by using a trilinear interpolation at the new position using all neighboring cells to this new position in the three directions. Hence, the approximation we make is that the volume of the cloud $V_{c}$, follows a linear uniform\footnote{This approximation amounts to assuming that the real motion of the object's center of mass: $d{\bf r}=1/2~{\bf a_{CM}}~dt^{2}+{\bf v_{CM}}~dt$ (where ${\bf a_{CM}}$ is the center of mass acceleration), over small time intervals $dt$, is dominated by the second term on the right hand side. The acceleration of small, law density CCs is expected to be larger than for massive, dense CCs. However, the former ones are those whose quantities are more affected by numerical noise.} motion on distances equal to $d{\bf r} = {\bf v_{CM}}\times dt$ around the central time-step $t$. This amounts to applying a Galilean transformation to each core, thus preserving the Eulerian nature of the virial theorem.       
  
\begin{figure}
\plotone{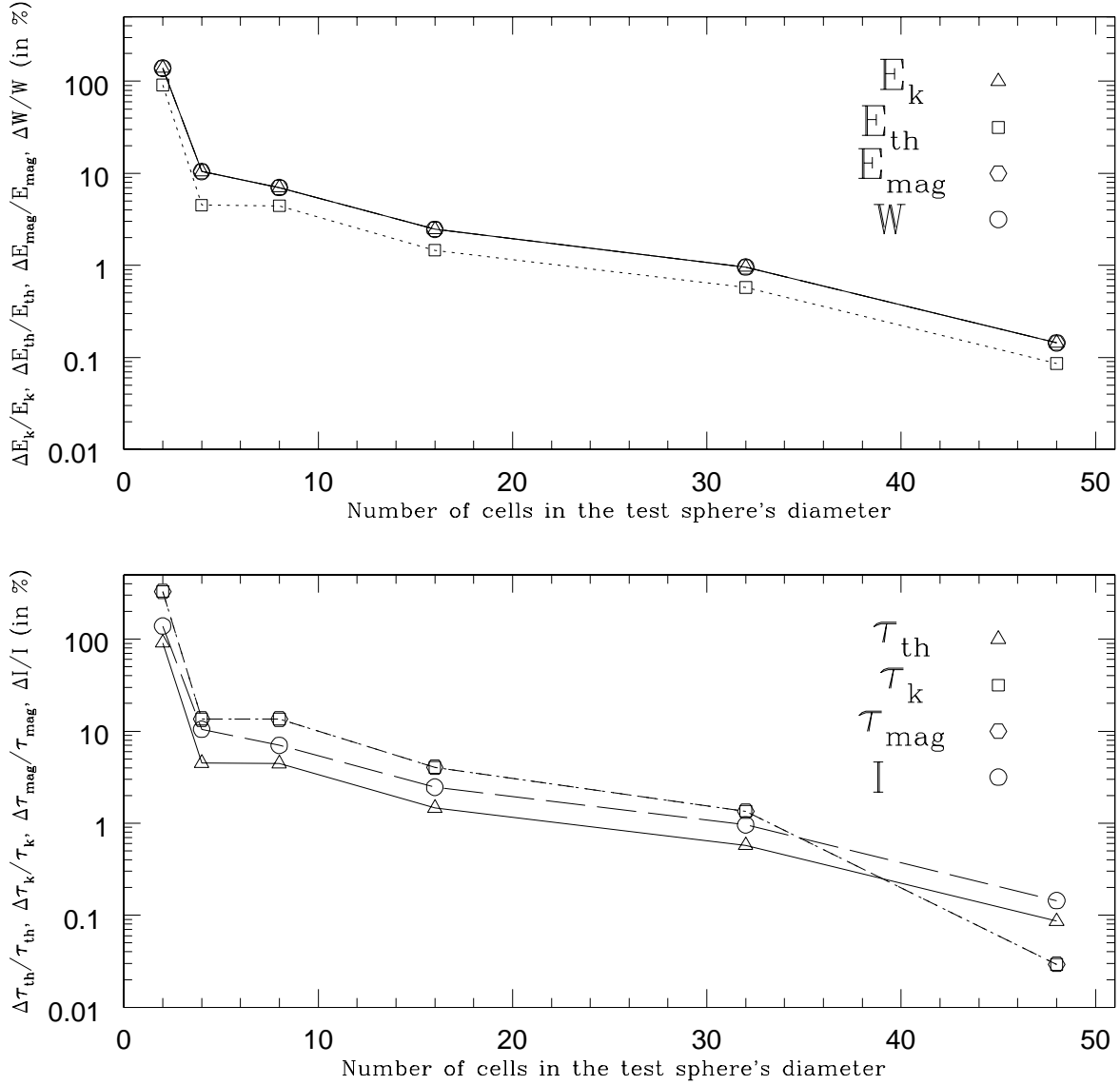}
\caption{Error estimates for the terms involved in the Eulerian Virial Theorem for a test setup. The test setup is a sphere with a uniform density distribution. The velocity and magnetic fields possess, both, a pure (unphysical) radial dependence with arbitrary amplitudes which facilitates the comparison to the analytical solution. $E_{k}$, $E_{th}$, $E_{mag}$, and $\tau_{k}$, $\tau_{th}$, $\tau_{mag}$ are the volume and surface kinetic, thermal, and magnetic energies, respectively.$I$ is the moment of Inertia and $W$ the gravitational term.}
\label{fig2}
\end{figure}

In order to check the accuracy of our clump-finding algorithm for the computation of the different physical quantities, the program was tested against a simplified, albeit unphysical, test case problem, consisting of a uniform-density sphere with a pure radial dependence for both the velocity and magnetic fields. The values of the density and the amplitudes of the velocity and magnetic field are arbitrary. The sphere is placed on a uniform, lower density grid with a specified resolution. Fig.~\ref{fig2} shows, in percent, the discrepancy between the analytical solution for each of the physical quantities involved in the EVT and the numerical solution yielded by our algorithm as a function of the number of grid cells present in the diameter of the test sphere. As seen in Fig.~\ref{fig2}, 32 cells in each direction are needed in order to bring the uncertainties below the one percent level. Nevertheless, with more than 4 cells per direction, the errors on the EVT quantities are not larger than $\sim 15$ percent. It should be stressed though, that the relative errors shown in Fig.~\ref{fig2} are only order of magnitude error estimates on the quantities implied in the EVT, because such uncertainties are model dependent. In the case of clumps that have a complex, non-spherical morphology, which is the case in our simulations, the uncertainty on the physical quantity will be dominated by the uncertainty in the direction with the lowest number of cells. Nevertheless, in order to stay consistent with the results of Fig.~\ref{fig2}, we must keep in mind that cores with small number of cells (typically with less than 3 cells in one direction) can be dominated by numerical noise.     

In our evaluation of the EVT terms, we have neglected the role played by the driving force which should also appear on the right-hand side of Eq.~\ref{eq1}. As the three random components of the driving force for the whole computational domain have not been stored, it is impossible to recover this information since there is no analytical description of this force. However, we do not expect this omission to be a problem since in the simulations, the driving is performed at the largest scales possible, while the CCs are structures with a much smaller size, in each direction, than the box size. Therefore, the driving mainly advects the CCs, without greatly impacting their internal dynamics. Nevertheless, we have tested this conclusion by running anew a simulation of our sample (the strongly supercritical cloud model M10J4$\beta$1, but with different turbulent random seed numbers) and dumped the forcing components of the velocity field. From this data, we have calculated the ratio of the kinetic energy due to instantaneous forcing ($\delta E^{'}_{k}$) to the bulk total kinetic energy of the cloud $E^{'}_{k}$ (in $E^{'}_{k}$ and  $\delta E^{'}_{k}$, the velocity components are bulk velocities: i.e., unsubstracted from the center of mass velocity components). For the largest and most massive cloud found in this model at $t=30=1.2$ Myr, this ratio is (9.08,7.07,3.71,3.29,3.35) $\%$ at the threshold levels of (7.5,15,30,60,100) $\bar{n}$, respectively. In general, we find that the percentage of the turbulent kinetic energy associated with the forcing in CCs is between 3-9 $\%$ of the total kinetic energy with an average value around $4 \%$. 
 
\subsection{Calculation of the classical indicators}\label{classical}

For each identified CC, we calculate other quantities commonly used in theoretical and observational work to assess the state of gravitational boundedness of the objects, namely, the Jeans number $J_{c}$, the mass-to-magnetic flux ratio $\mu_{c}$, and the virial parameter $\alpha_{vir}$. It should be mentioned at this stage that these indicators, when used in observational studies, intrinsically suggest that the thermal, kinetic, and magnetic energies act in support against gravity, making two basic simplification to the virial equation: a) The neglect of the surface terms\footnote{The concept of magnetic critical mass (similar to the Bonnor-Ebert critical mass in the non-magnetic case, includes the surface magnetic term; e.g., McKee et al. (1993). However, the surface magnetic term is not commonly estimated in observational studies.}, and b) the assumption that turbulent motions inside the clouds are isotropic. The Jeans number is defined as $J_{c}=R_{c}/L_{J,c}$, where $R_{c}$ is the characteristic size of the CC, $L_{J,c}=(\pi c_{s}^{2}/G \bar{\rho_{c}})^{1/2}$ is the mean Jeans length of the CC and $c_{s}$ the sound speed. The mass-to-magnetic flux ratio $(M_{c}/\phi_{c})$ is generally expressed in units of the critical value for collapse, calculated, in the linear regime, for a sheet-like structure, $(M/\phi)_{cr} \approx (4 \pi^{2} G)^{-1/2}$ (Nakano \& Nakamura 1978). Krasnopolsky \& Gammie (2005) showed that this criterion for magnetic criticality holds for turbulent magnetized clouds in the non-linear regime. Note that Mouschovias \& Spitzer (1976) derived a value of  $(M/\phi)_{cr} \approx 0.126~ G^{-1/2}$ for disks with only thermal support along field lines, and Tomisaka et al. (1988) found, from extensive numerical simulations, that $(M/\phi)_{cr} \approx 0.12~G^{-1/2}$. In principle, the magnetic flux of a CC should be computed as $\phi_{c}=\int_{S_{b}} {\bf B~n}~dS$, where ${\bf B}$ is the mean magnetic field in the CC and ${\bf n}$ is the unit vector normal to a surface $S_{b}$ that bisects the CC. In complex geometries such as those of the CCs found in our simulations, $S_{b}$ is difficult to evaluate, particularly for small, non-spherical cores which have a limited number of pixels along one direction. Instead, as in V\'{a}zquez-Semadeni et al. (2005a), we use a simpler approach by defining $\phi_{c}=\pi R_{c}^{2} {\bf B}$, where $R_{c}$ is the characteristic CC size. We make the common observational assumption that CCs are spherical and compute $R_{c}$ as being $(3~V_{c}/4~\pi)^{1/3}$. Alternatively, one could consider that $R_{c}= V^{1/3}$, or by making the assumption that CCs are flattened, $R_{c}$ would be the maximum separation between the position of the center of mass and any point (i.e., cell) in the CC. Fig.~\ref{fig5} shows that not all condensations are flattened and even if some are more flattened than others (e.g., Fig.~\ref{fig12}), they are not completely cylindrical objects. The interpretation of $J_{c}$ and $\mu_{c}$ is that $J_{c}$ measures if a core is gravitationally unstable with respect to the thermal support (i.e., $J_{c} > 1$), whereas $\mu_{c}$ measures the importance of the magnetic support against gravity. A core is collapsing when $J_{c} > 1$ and $\mu_{c} > 1$, while cores with $J_{c} > 1$ and $\mu_{c} < 1$ are gravitationally bound but remain in a stable magnetostatic state (under ideal MHD conditions such as the simulations analyzed in this work). Cores with $J_{c} < 1$ are Jeans stable, regardless of the value of $\mu_{c}$ and are likely to re-disperse due to their internal dynamics or loose their identity in subsequent local compressions and dispersions of the local medium by large scale flows. The virial parameter (also called gravitational parameter) is calculated as  

\begin{equation}
\alpha_{vir}= \frac{5~\sigma^{2}~R_{c}}{G~M_{c}},
\label{eq3}
\end{equation}

\noindent where $M_{c}$ is the mass of the CC and $\sigma$ is the one-dimensional velocity dispersion inside the object, calculated as 

\begin{equation}
\sigma^{2}=\frac{\sigma_{c}^{2}}{3}+c_{s}^{2}.
\label{eq4}
\end{equation}

\noindent where $\sigma_{c}$ is the density weighted turbulent three-dimensional velocity dispersion. The virial parameter is often used in observational and theoretical studies in order to measure the balance between the CCs self-gravity and their kinetic+thermal energies (e.g., Leung et al. 1982; Magnani et al. 1985; Keto \& Myers 1986; Herbertz et al. 1991; Bertoldi \& McKee 1992; Falgarone et al. 1992; Dobashi et al. 1996; Yonekura et al. 1997; Kawamura et al. 1998; Tachihara et al. 2000,2002; Heyer et al. 2001; Krumholz \& McKee 2005). A CC is assumed to be gravitationally bound if $\alpha_{vir} \lesssim 1$, unbound otherwise. We make the common observational assumption that turbulent motions are isotropic in the CCs and calculate, in Eq.~\ref{eq4}, the one-dimensional turbulent velocity dispersion as being $\sigma_{c}/3^{1/2}$. An $\alpha_{vir}$-Mass relation is commonly used in observational studies in order to assess the gravitational boundedness of CCs in MCs (e.g., Bertoldi \& Mckee 1992; Williams et al. 2000). The $\alpha_{vir}-M_{c}$ relation is observed to follow $\alpha_{vir} \propto M_{c}^{\epsilon}$. Tab.~\ref{tab3} displays the values of the exponent $\epsilon$ found by Bertoldi \& McKee (1992) and Williams et al. (2000) for CCs in different MCs. A value of $\epsilon = 2/3$ is expected for pressure bounded objects (Bertoldi \& McKee 1992). Associated to the virial parameter is the concept of virial mass, also commonly used in observational studies in order to obtain an estimate of the real mass (e.g., Kramer \& Winnewisser 1991; Jijina \& Myers 1999; Caselli et al. 2002). The virial mass is defined as $M_{vir}=\alpha_{vir}~ M_{c}$. CCs whose measured mass is comparable to the virial mass are usually assumed to be in virial equilibrium (e.g., Caselli et al. 2002).

\begin{figure}
\plotone{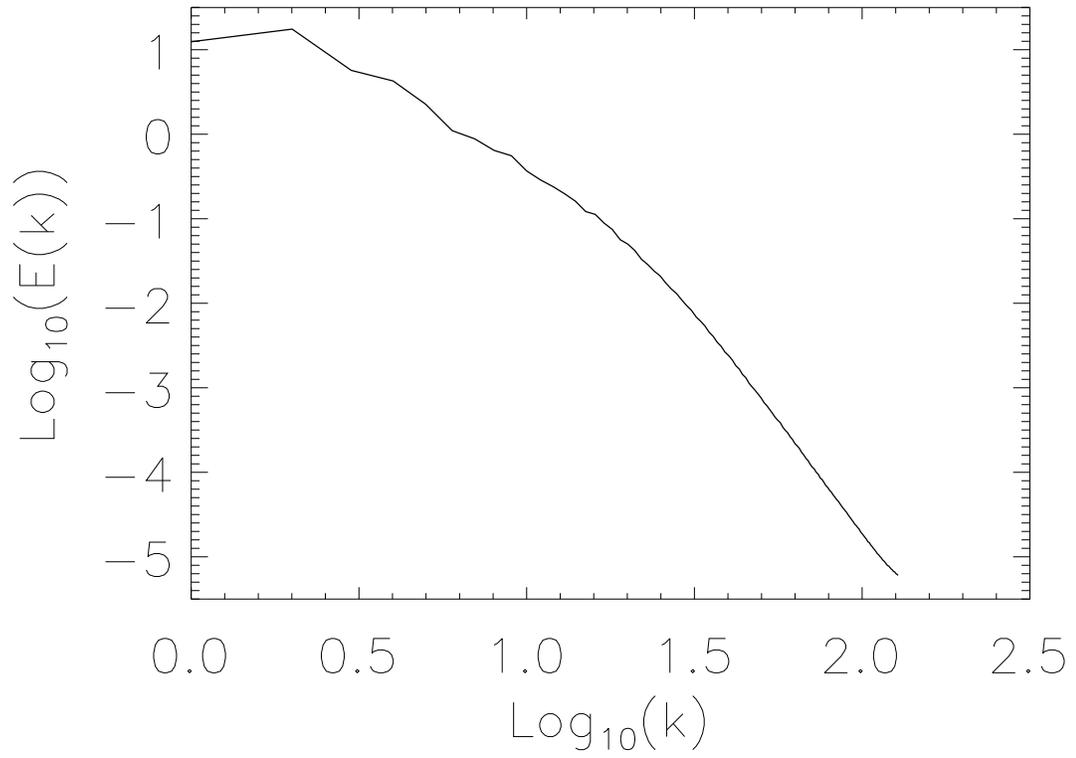}
\caption{Kinetic energy power spectrum for the supercritical run M10J4$\beta$1 at $t=30=1.2$ Myr. The turnover to the diffusive regime occurs at wave number $k \sim 30$ equivalent to $\sim 8$ grid cells.} 
\label{fig3}
\end{figure}

\begin{figure}
\plotone{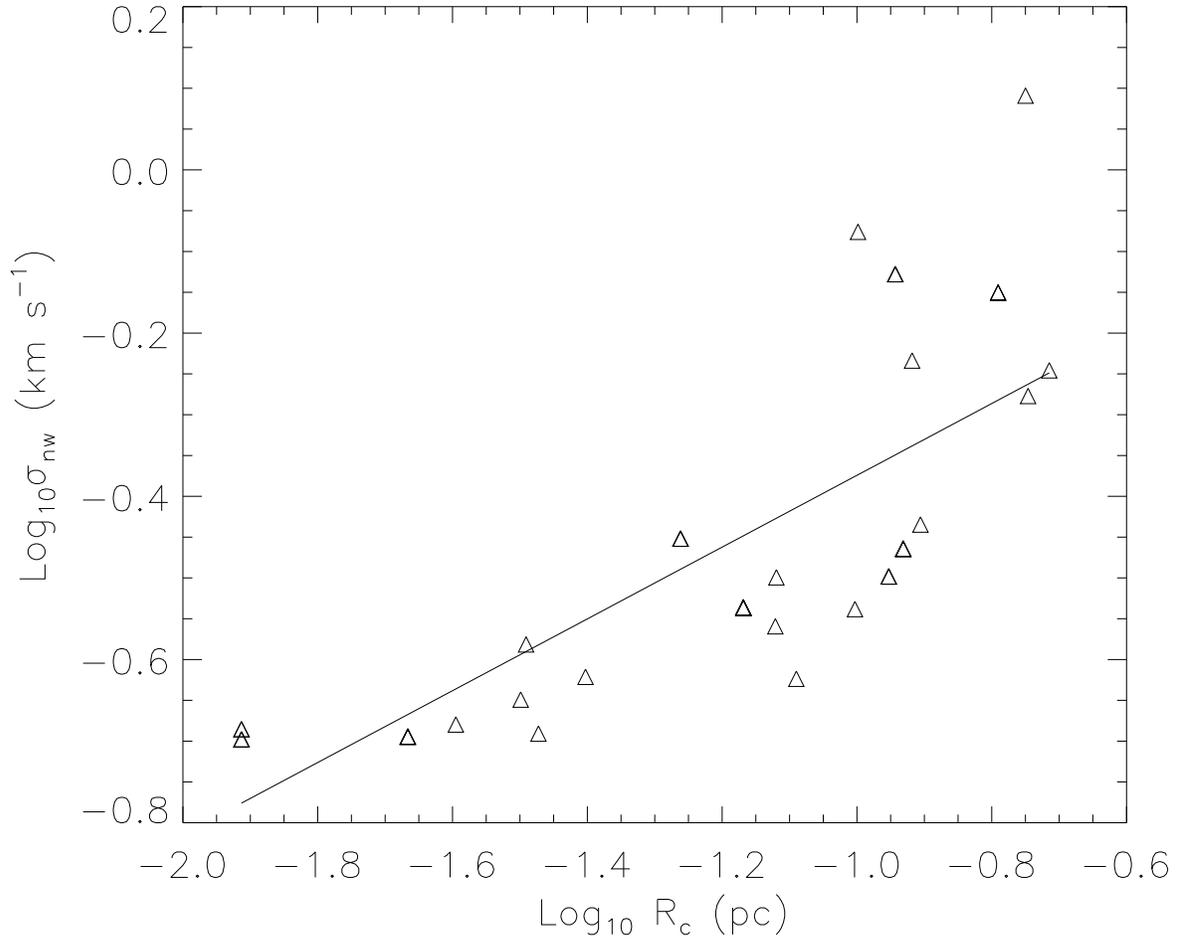}
\caption{Velocity dispersion-Size relation for the ensemble of CCs in run M10J4$\beta$1 at $t=30=1.2$ Myr. The quantity $\sigma_{nw}$ includes a thermal component and a three-dimensional, non-density weighted, turbulent component. This relation is shown in order to evaluate at  which scale the effects of numerical diffusion become important for the CCs.} 
\label{fig4}
\end{figure}

An unavoidable aspect of our simulations is that turbulence inside CCs will be affected by the effects of numerical viscosity, particularly for the smallest objects (in addition to the discretization effects), which for our numerical code, start to be important at around 8 grid cells. As an example, Fig.~\ref{fig3} shows the kinetic energy power spectrum for the strongly supercritical run at timestep $t=30=1.2$ Myr. As can be seen from this figure, the turnover from the linear regime to the diffusion dominated regime occurs at around wave number $k \sim 30$ which corresponds to $\sim 8$ grid cells (i.e., in physical units to sizes of $0.125-0.140$ pc). However, the kinetic energy power spectrum does not single out dense structures on a given scale. Therefore, it is also important to verify how does the three-dimensional non density-weighted velocity dispersion, $\sigma_{nw}$, (which also includes a thermal component and is only a fraction of percent to a few percents smaller than $\sigma_{c}$ in all cases) depends on the CCs sizes. Fig.~\ref{fig4} shows this relation\footnote{The exponent $\beta$ in the relation $\sigma_{nw}= R_{c}^{\beta}$ is found to be $\beta=0.43 \pm 0.06$, in very good agreement with Larson's (1981) observational result.} for the supercritical run at timestep $t=30=1.2$ Myr. Despite the discreteness of the data, it is possible to observe a change of regime at around $Log_{10}(R_{c}) \sim -1$ which corresponds to $\sim 6.4$ grid cells. It is difficult to evaluate precisely how dissipation will affect the velocity dispersion inside individual CCs because of their non-idealized structures. However, from Fig.~\ref{fig3}, for scales of $\sim$ 5 cells ($k \sim 50$), the deviation at this location to the inertial range regime is about a factor $\sim 3$ which is roughly the factor by which estimates of the virial parameter will be affected. Note that this argument applies essentially to small and dense cores (the dense inner parts of larger clumps). For small but low density objects, we find that their virial parameter are much larger than unity and thus they are unbound. Applying any correction to take into account for the effects of the numerical dissipation of turbulence inside the latter cores will make them even more unbound.      
 
\section{VIRIAL BALANCE OF CLUMPS AND CORES}\label{results}

In this section, we present the results concerning the application of the EVT to our set of simulations using the algorithm described in \S~\ref{method}. Tab.~\ref{tab1} summarizes the ensemble of CCs found in each simulation, for different density thresholds and at different selected time-steps\footnote{It is important to note that, for a given set of parameters of the simulations (i.e., same scale of turbulence driving, initial magnetic field strength, number of Jeans masses and Mach number), the numbers of CCs and the mass they contain may slightly change from simulation to simulation due to the different seed of the random turbulence driver. An example of this effect is given in V\'{a}zquez-Semadeni et al. (2005b). Nevertheless, the general properties of the simulations are conserved: That is, no collapsed objects form in magnetically subcritical clouds, near critical or moderately supercritical runs will always form a moderate number of CCs with a few collapsing ones, and the supercritical and non-magnetic runs will always form a large number of CCs with many of them proceeding to gravitational collapse.}. At a given time-step, each CC is assigned a number corresponding to its place in the cumulative sum of CCs at all thresholds (i.e., for example in run M10J4$\beta$.01 at $t=223$, 23 cores have been found, counting all thresholds. Core number 20 corresponds to the first core found at threshold $n_{thr}=30~\bar{n}$). Tab.~\ref{tab2} summarizes the numbering of CCs for time-steps which are analyzed in detail in the next sections.   

\subsection{subcritical cloud}\label{subcloud}

\begin{figure}
\vspace*{-20mm}
\plottwo{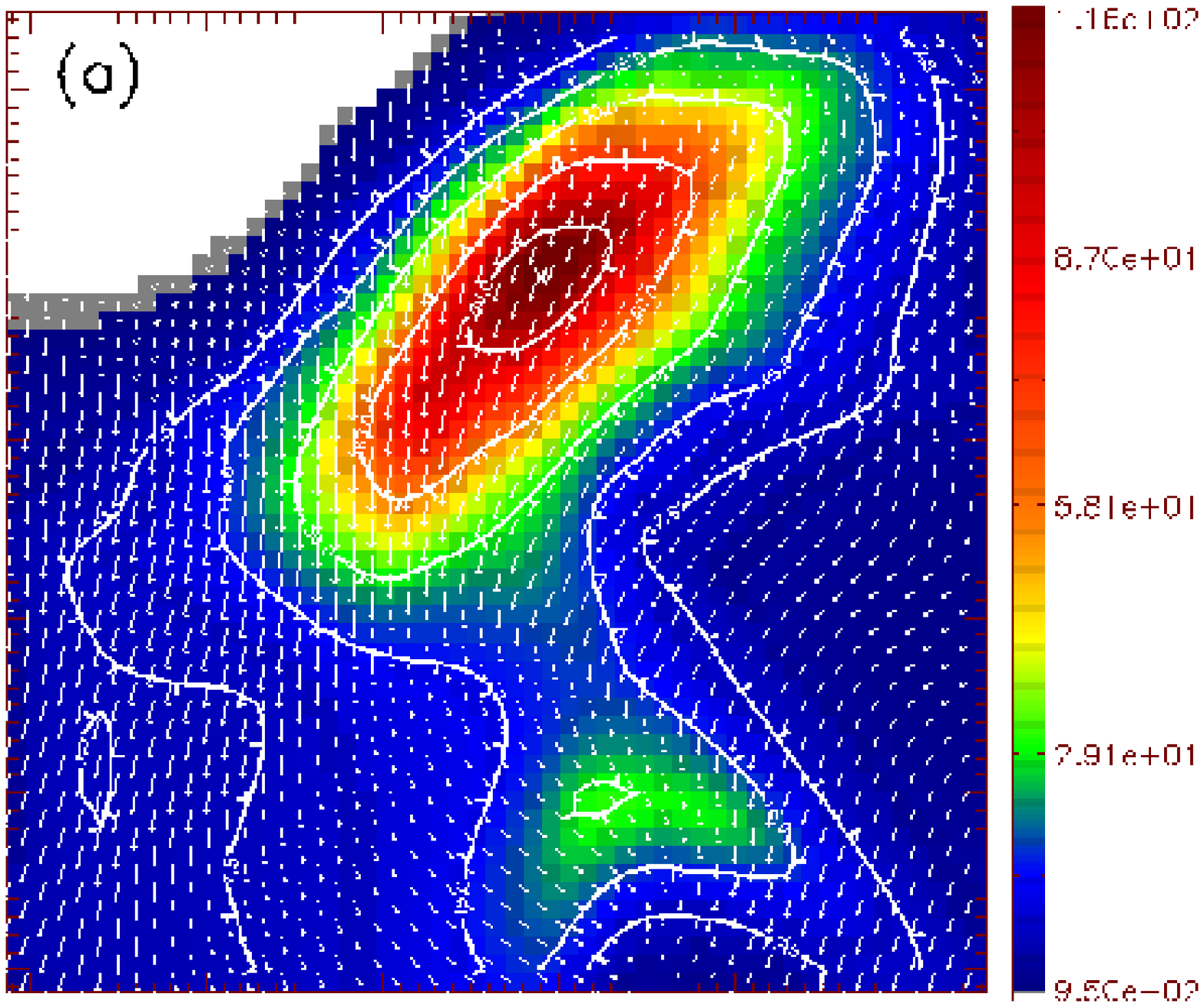} {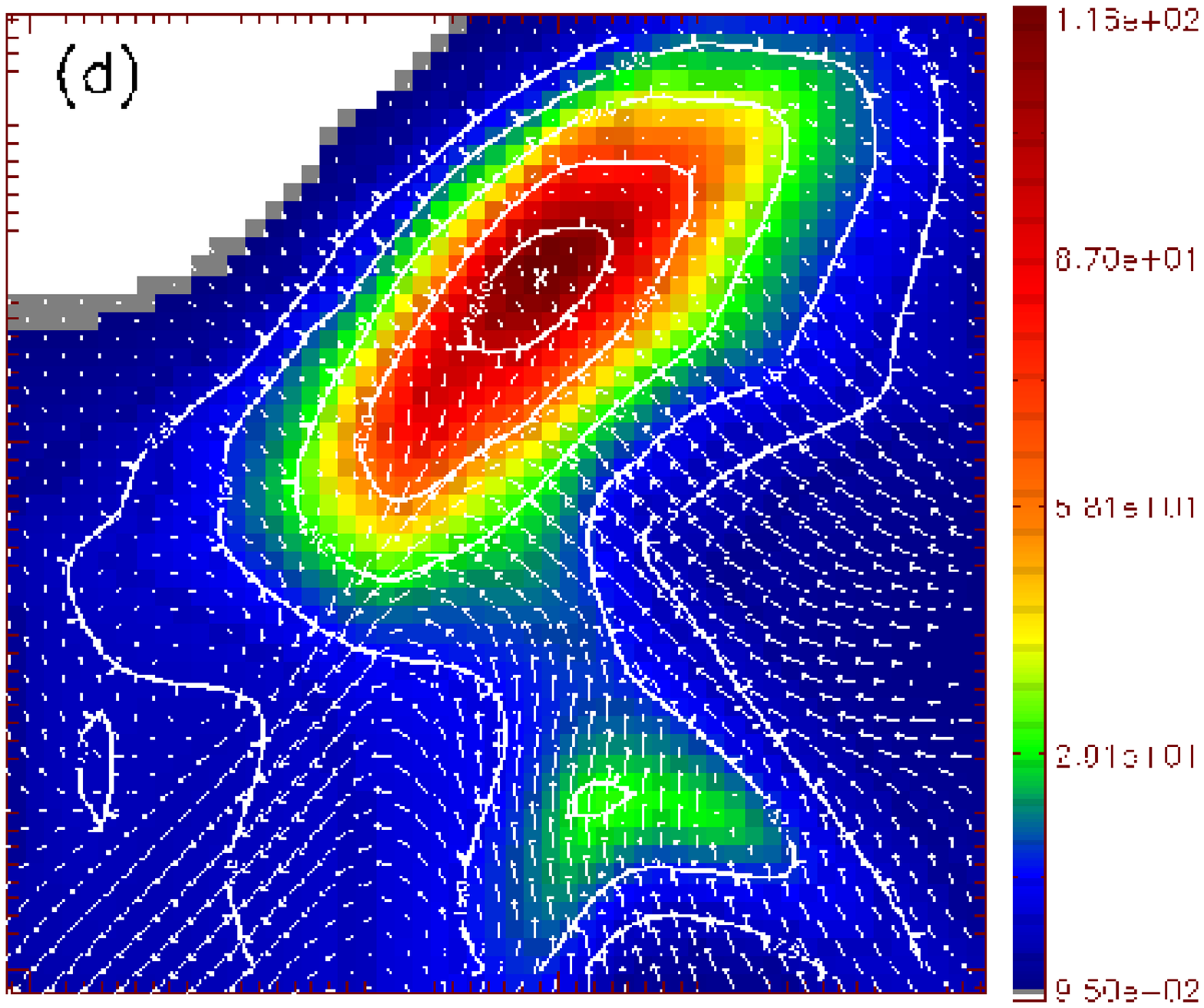}
\plottwo{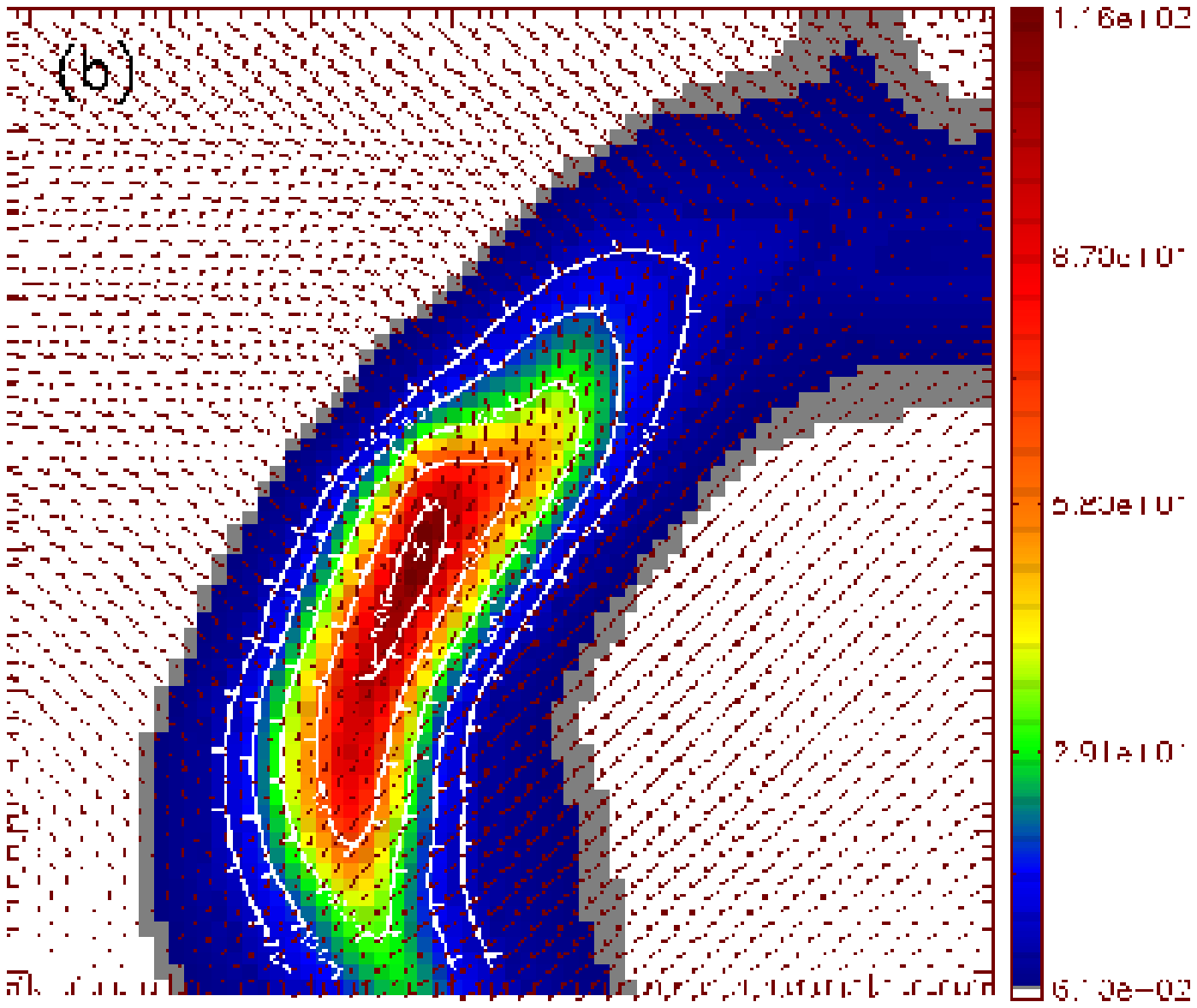} {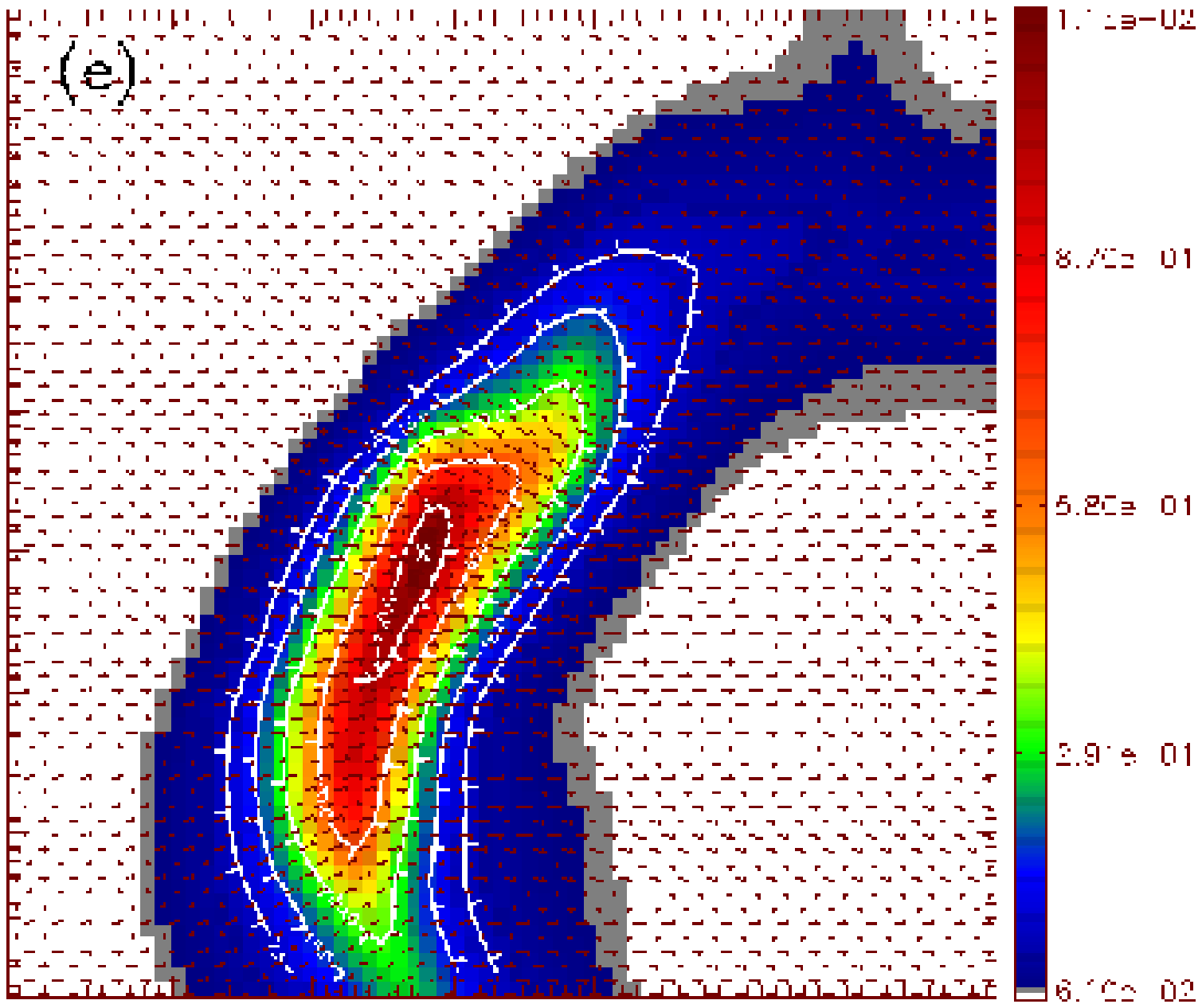}
\plottwo{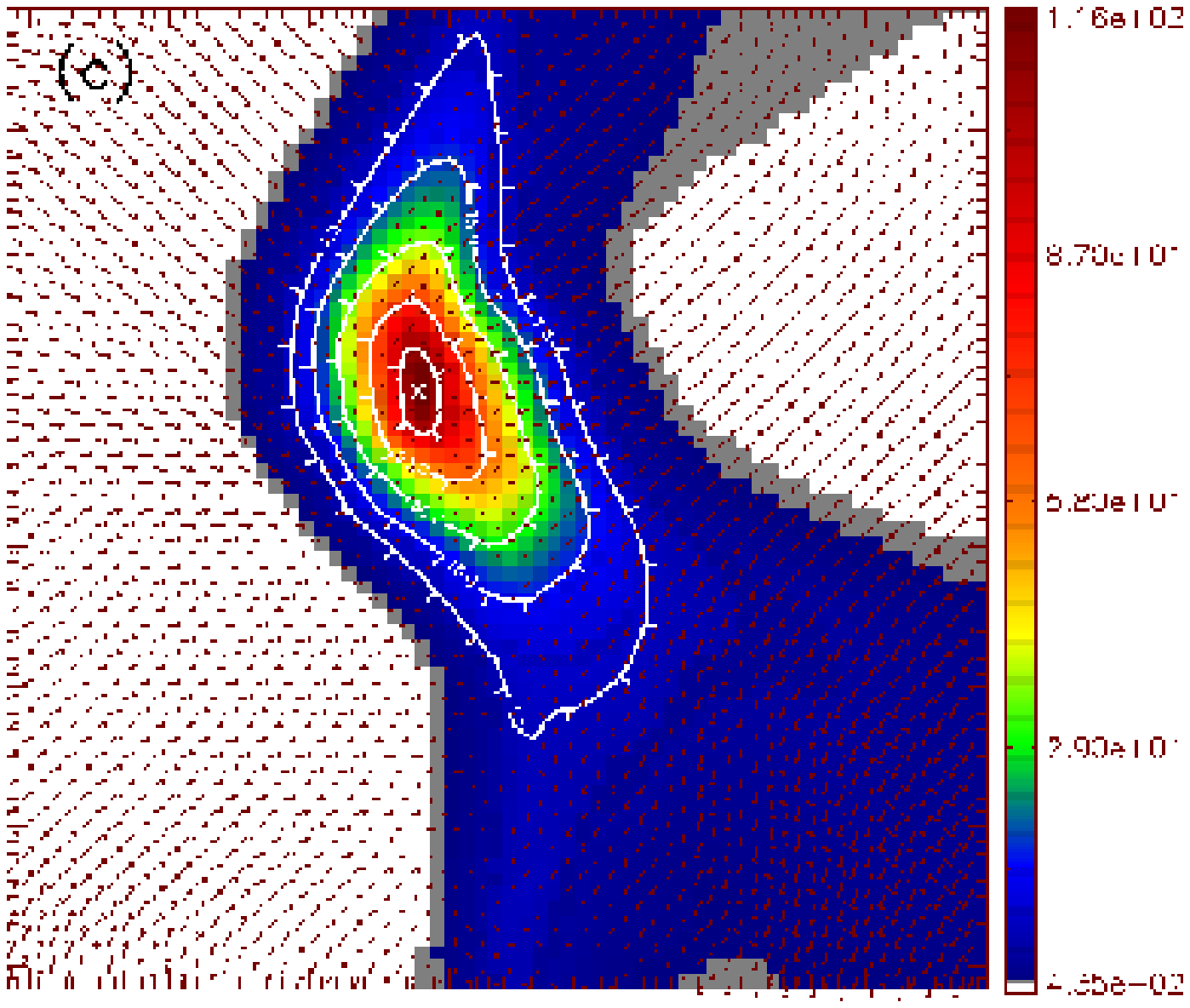} {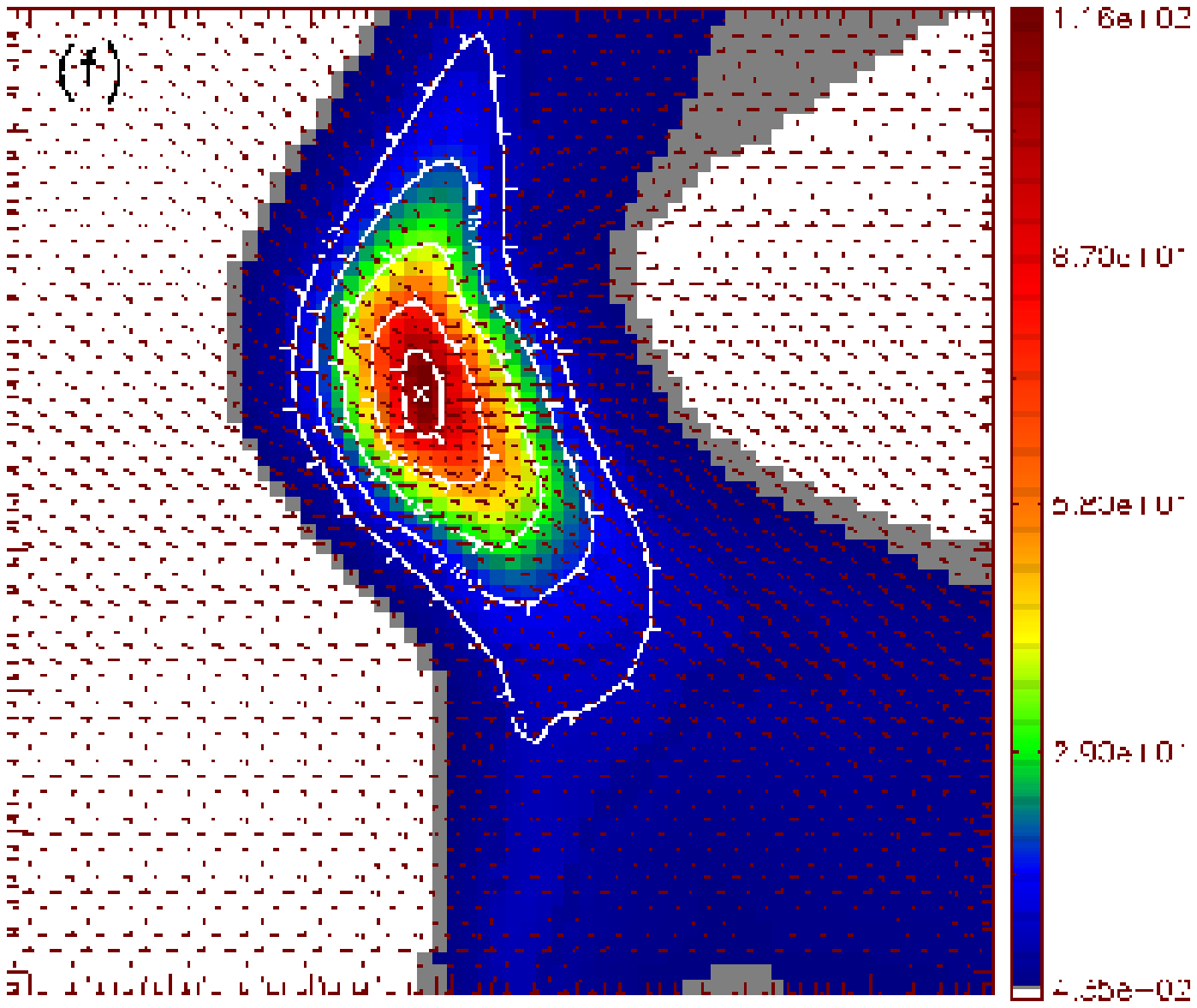}  
\caption{Density cut in the x (top), y (middle), and z (bottom) directions at the position of the density maximum for the densest clump in run M10J4$\beta$.01 at $t=223$. This object corresponds to clump 1, 12, 20, 22 and 23 in Tab.~\ref{tab1}. The contours show the boundary of the object at the density thresholds of 7.5, 15, 30, 60 and 100 $\bar{n}$. The cross marks the position of the density maximum. Arrows show the projected velocity (left) and magnetic (right) fields at the position of the two-dimensional cut, scaled to the maximum value on the grid. In order to better highlight the structure of the cloud, the resolution of the map has been artificially multiplied by a factor 2 and the map smoothed. }
\label{fig5}
\end{figure}

In the subcritical run M10J4$\beta$.01 ($\mu_{box}=0.9$ and initial magnetic field strength of 45.8 $\mu$G), CCs are observed to be very transient (see the animation in V\'{a}zquez-Semadeni et al. 2005a), surviving only for time-spans of $0.01-0.075~t_{s} \sim 0.2-1.5$ Myr before being dispersed by their internal dynamics, by large scale flows, or by merging with other clumps, and thus, lose their identity. We choose to evaluate the EVT terms in the subcritical run at time-step 223 corresponding to $t=8.85$ Myr after self-gravity has been turned on, which is the epoch at which the most massive clump appears in this run. Fig.~\ref{fig5} displays two-dimensional cuts at the position of the peak density of the most massive object found in the simulation, viewed along the three directions of the computational box. Overlayed on the density map are the projected velocity (right column) and magnetic (left column) field components and density contours which show the extent of the clump when selected at the different density thresholds.

\begin{figure}
\plotone{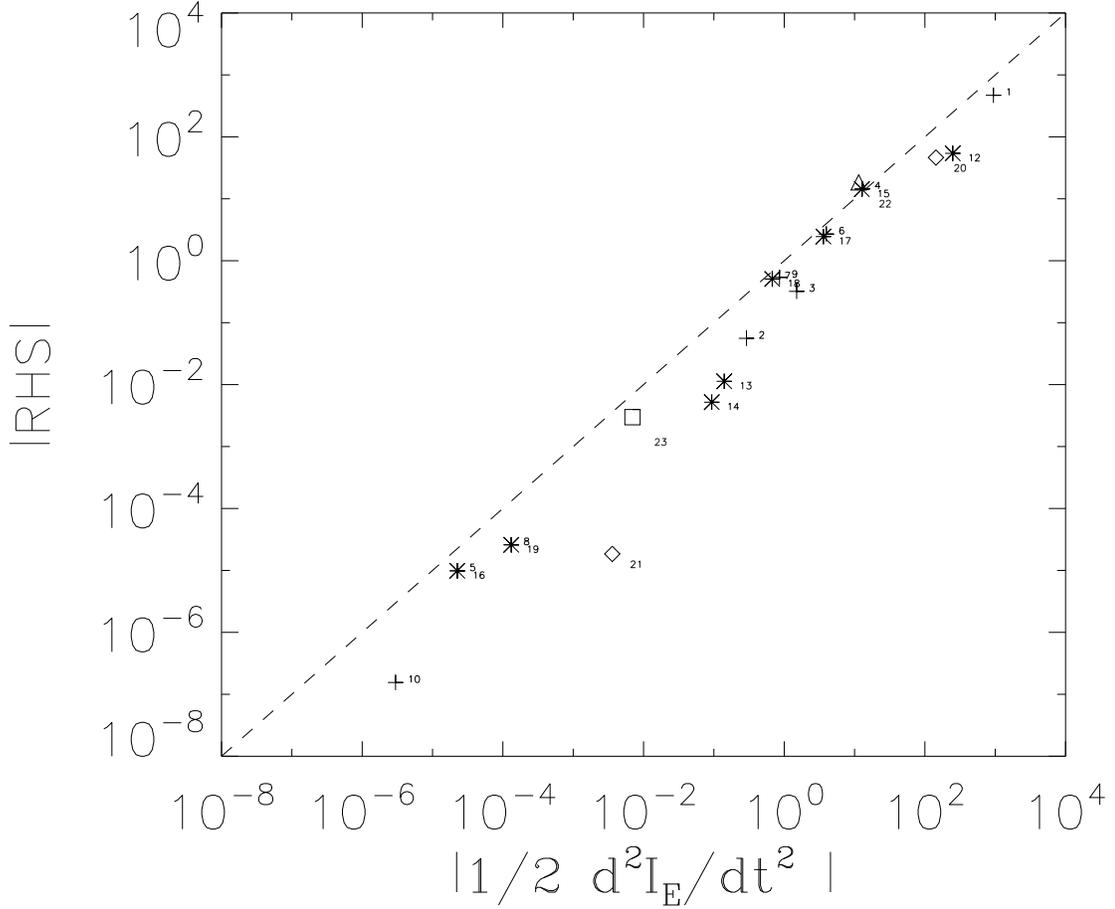}
\caption{Right Hand Side terms in Eq.~\ref{eq1} (RHS) vs. the Eulerian second time derivative of the moment of Inertia $d^{2}I_{E}/dt^{2}$ in run M10J4$\beta$.01 at $t=223=8.92$ Myr. The virial theorem is verified to within $\sim 0.5$ dex for the largest clouds. For the smaller clumps (i.e., with small numbers of cells), the larger scatter is due on the one hand to the effects of the turbulent driving, unaccounted for in Eq.~\ref{eq1} and to numerical noise (see Fig.~\ref{fig1}). The $7.5~\bar{n}$ threshold level is shown with a (+), the $15~\bar{n}$ with a ($\ast$), the $30~\bar{n}$ with a ($\diamond$), the $60~\bar{n}$ level with a ($\triangle$), and the $100~\bar{n}$ level with a ($\square$).} 
\label{fig6}
\end{figure}

\begin{figure}
\plotone{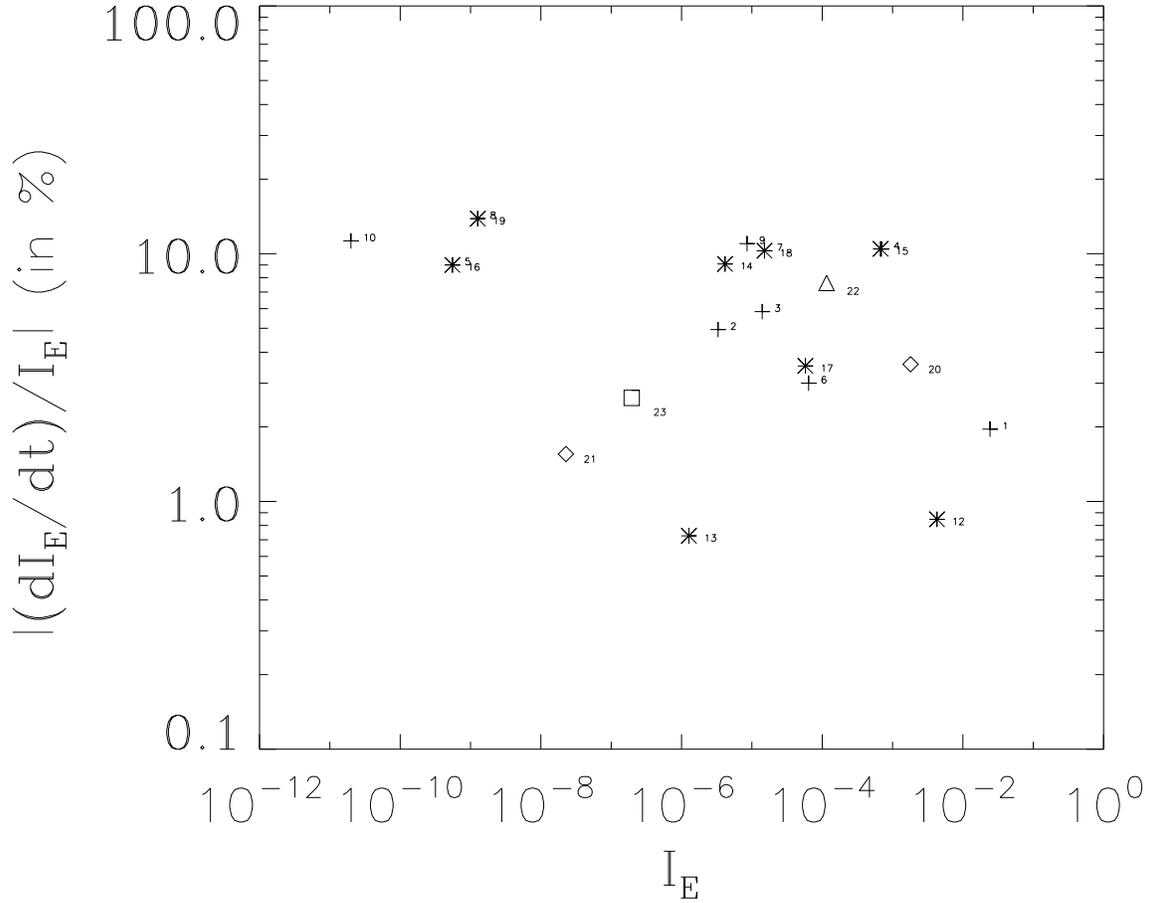}
\caption{This plot shows the temporal rate of change, in percentage, of the moment of inertia for the ensemble of clumps and cores in run M10J4$\beta$0.01 at $t=223=8.92$ Myr. The smallest clumps exhibit larger temporal variations of their moment of inertia, which is indicative of their more dynamical nature. }
\label{fig7}
\end{figure}

\begin{figure}
\plotone{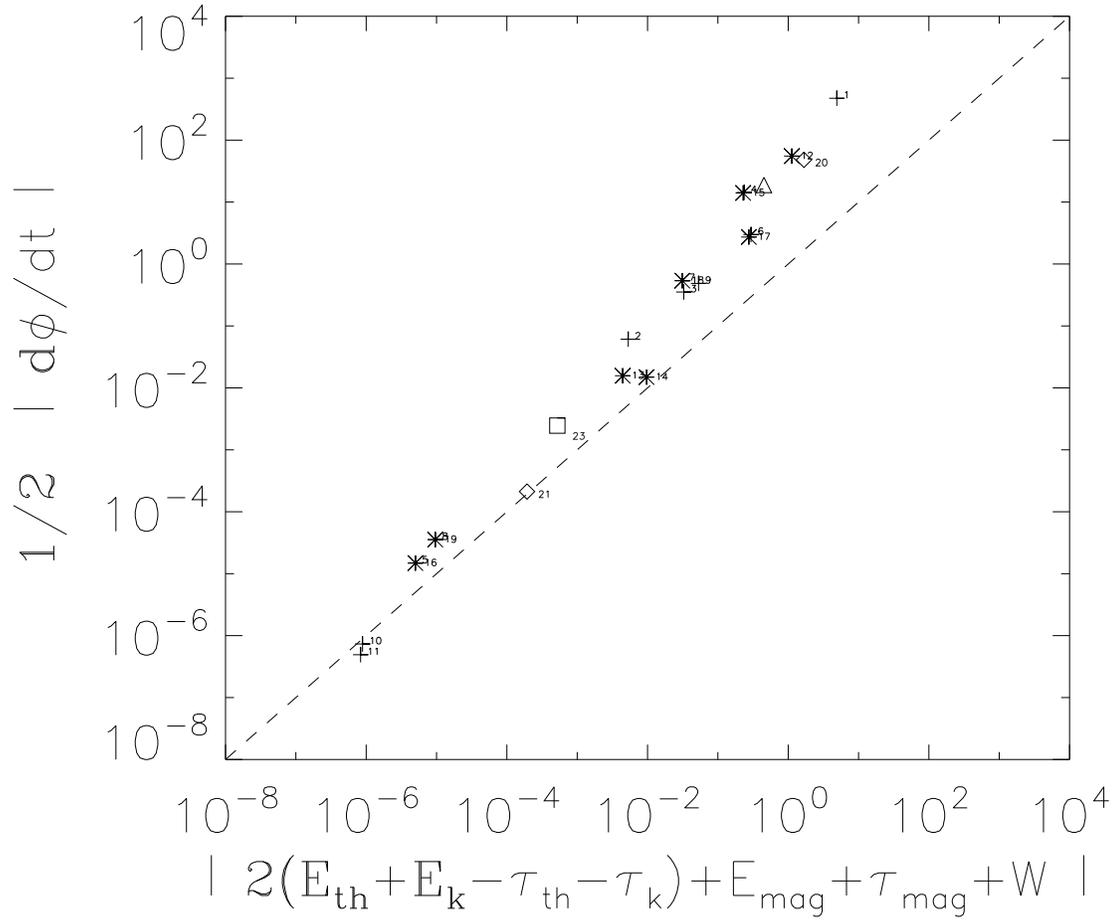}
\caption{Comparison of the temporal term $d\Phi /dt$ to the other terms in the RHS of the EVT. The figure is for run M10J4$\beta$0.01 at $t=223=8.92$ Myr.}
\label{fig8}
\end{figure}

\begin{figure}
\plotone{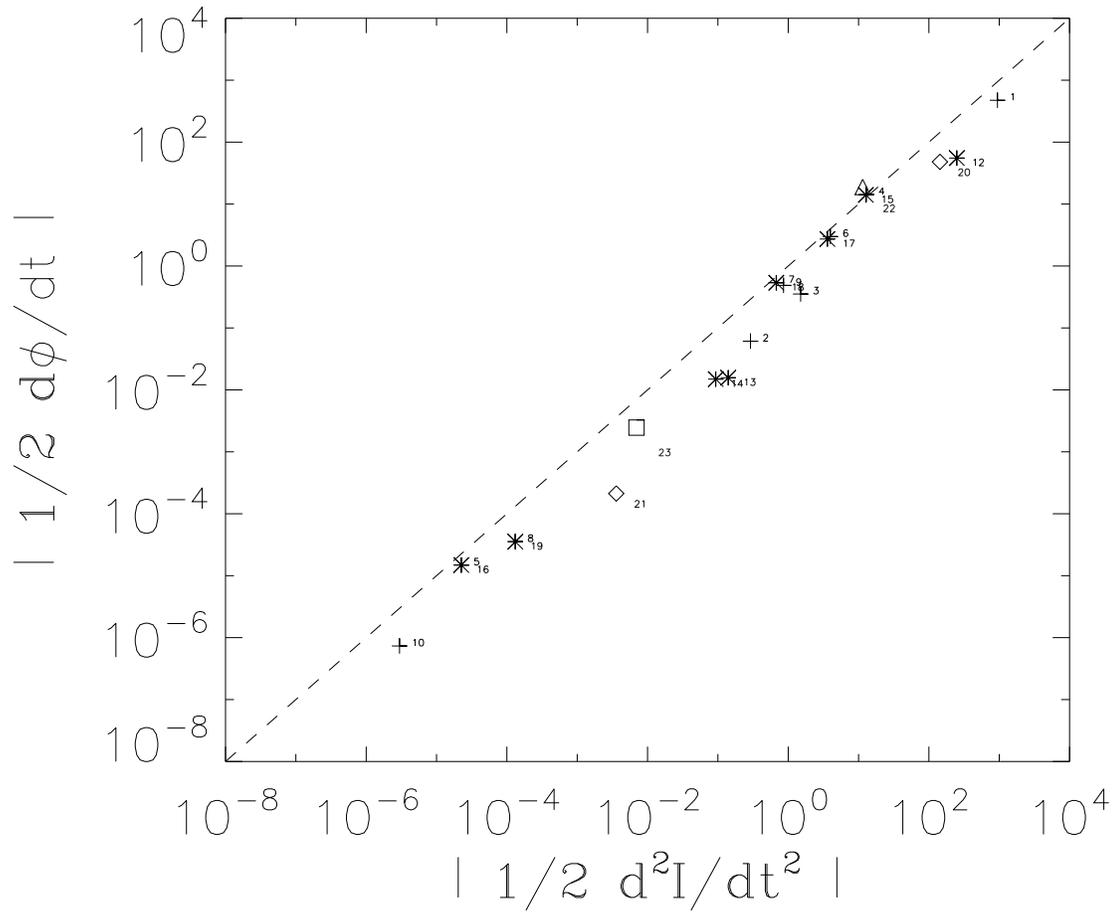}
\caption{Contribution of the $d\Phi /dt$ term to the total second time derivative of the moment of inertia. The figure is for run M10J4$\beta$0.01 at $t=223$.}
\label{fig9}
\end{figure}

Fig.~\ref{fig6} displays the virial balance for all CCs selected at this time-step at all thresholds, i.e., $d^{2}I_{E}/dt^{2}$ vs. the right hand side (RHS) in Eq.~\ref{eq1}. Since Eq.~\ref{eq1} is nothing else but a re-writing of the momentum equation, all points should, theoretically, fall on the equality line. Deviations, which are at most of the order of a factor $\sim 5$ are due to two effects: 1) Discretization errors, due to the limited numbers of cells in the CCs, and 2) the neglect of the contribution from the driving force. A legitimate question is: can the small values of $d^{2}I_{E}/dt^{2}$ for small CCs be considered an indication of dynamical equilibrium? An alternative view to the fact that $d^{2}I_{E}/dt^{2}$ and the RHS are of the same order and which answers the latter question is given in Fig.~\ref{fig7}. This figure shows that, despite the fact that the moment of inertia of the different CCs spans over $\sim 10$ orders of magnitude, the temporal rate of change, normalized to its own value, in percent, i.e., $|(dI_{E}/dt)/I_{E}|$, changes by less than one order of magnitude, irrespective of their masses or sizes. In fact, {\it the normalized rate of change of the moment of inertia of the smallest clumps is observed to be larger, thus indicating their more transient nature}. In Fig.~\ref{fig8}, we evaluate the relative importance of the time derivative of the flux of moment of inertia, i.e. $1/2~(d\Phi/dt)$ of all CCs versus the other terms in the RHS in Eq.~\ref{eq1}. This term is observed to be the dominant one for almost all CCs, particularly for the largest ones, and is of the same order of magnitude as $d^{2}I_{E}/dt^{2}$ (i.e., Fig.~\ref{fig9}). This fact implies that $(dI_{E}/dt) \sim -\Phi$, which in turn implies that the time variation of the moment of inertia of a clump is essentially determined by the flux through its boundary, which is another indication of the dynamical and morphing nature of CCs formed in the simulations.

\begin{figure}
\plottwo{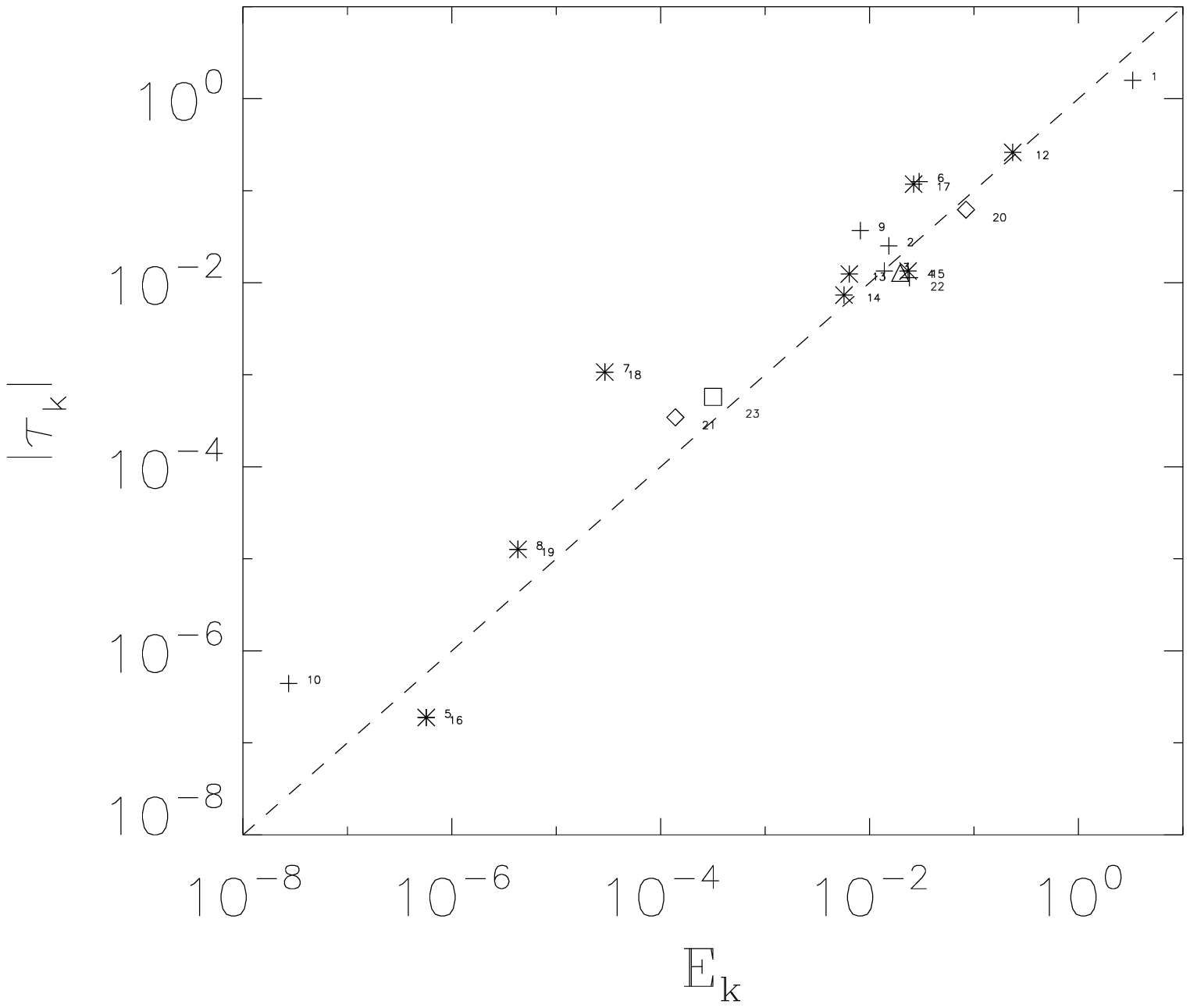} {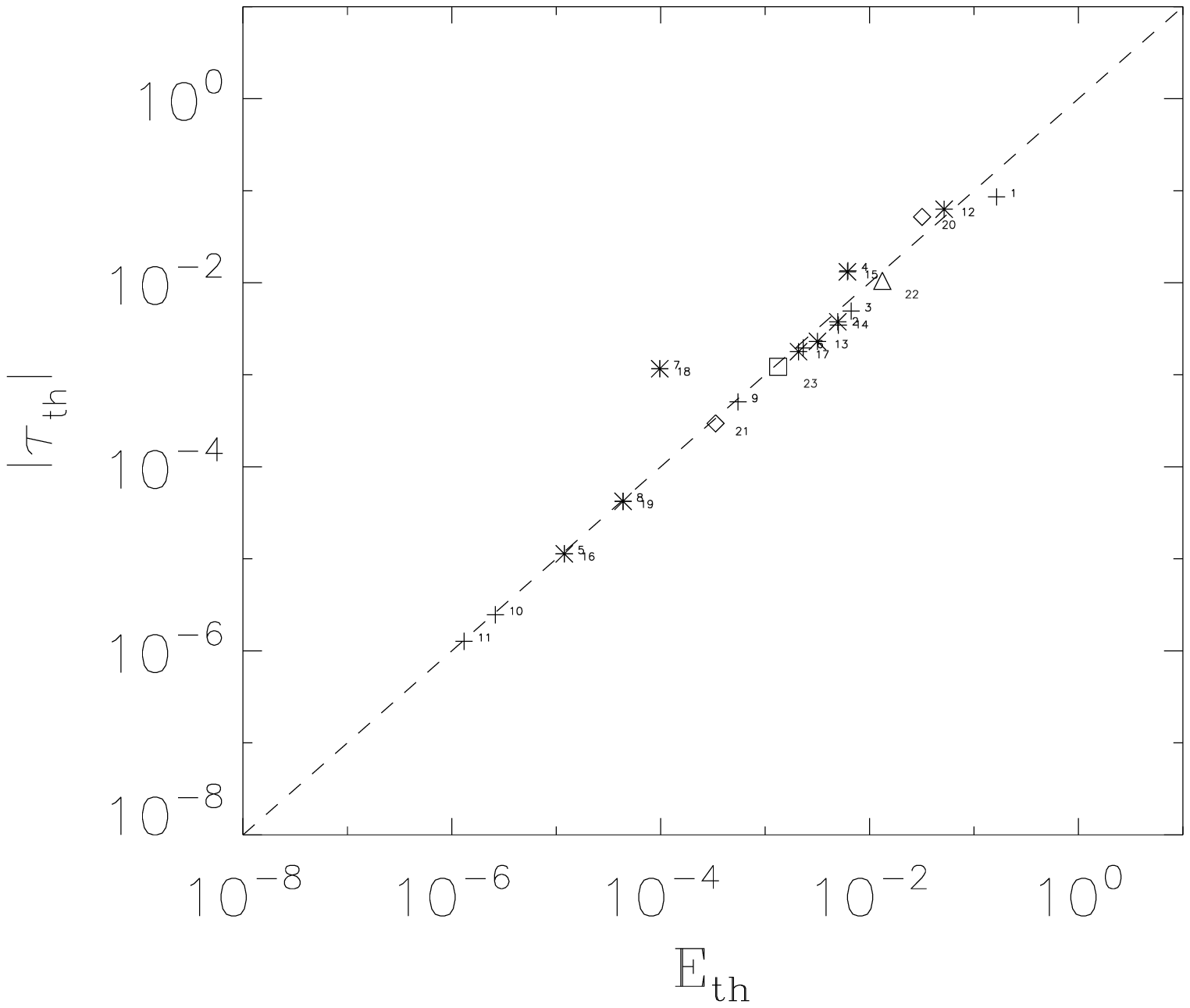}  
\plottwo{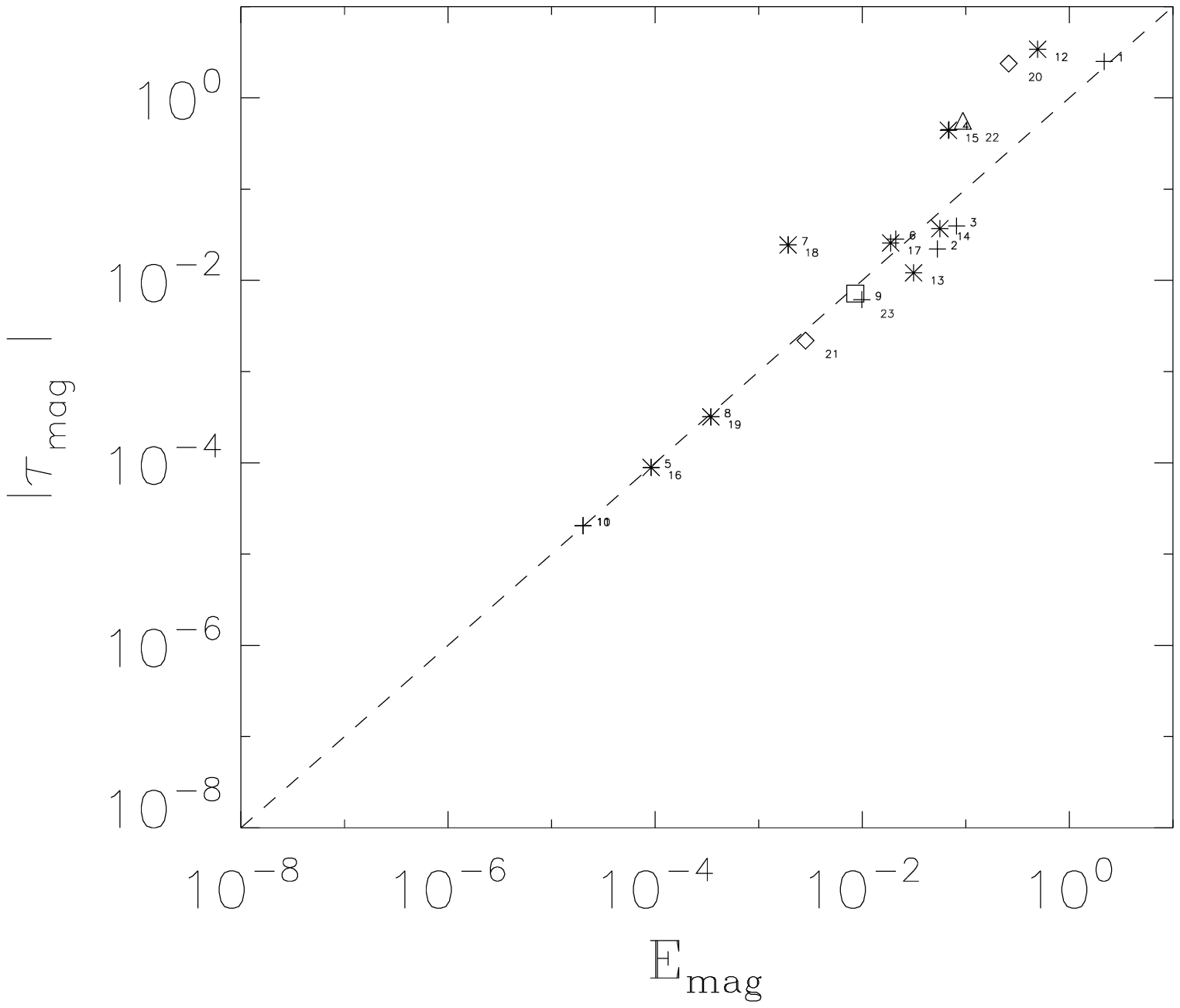} {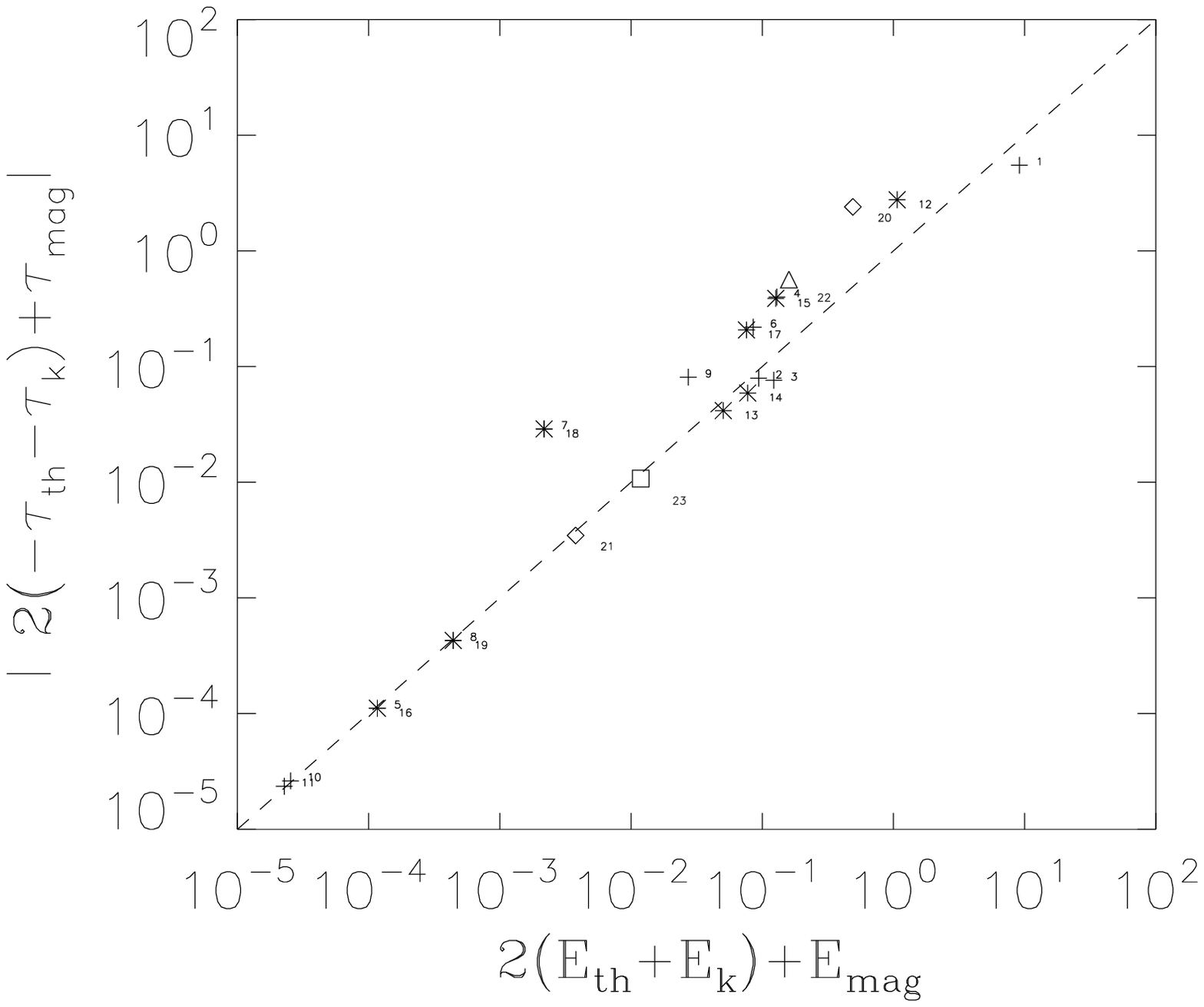}
\caption{Surface energy terms versus volume energy terms a) kinetic, b) thermal, c) magnetic and d), total for run M10J4$\beta$.01 at $t=223=8.92$ Myr.}
\label{fig10}
\end{figure}

Since the time dependent term in the RHS is a purely geometrical term which takes into account the morphing nature of the CC and its changing boundary and which adapts such that the EVT is always verified, the evolution of a CC will be determined by the non-time dependent terms in the RHS of the EVT. Namely, by the competition between confining forces (negative energies) on the one hand and dispersive ones on the other hand (positive energies). Figs.~\ref{fig10} a,b,c, and d compare the importance of the individual surface energy terms to their volume counterparts, and the total surface and volume energy terms, respectively. In general, the surface and volume energy terms are of the same order of magnitude, yet with scatter around the equality line, particularly for the case of the larger clumps. The difference is more important for the kinetic and magnetic energy pairs than for the thermal one. The origin of the differences in the $E-\tau$ relations is twofold. On the one hand, the scatter among different objects is due to the difference of the thermal pressure, velocity and magnetic field between the surface of the clumps and their interior. Second, for the same object, the difference in the scatter at the different density thresholds is simply the result of the energy profile of that object and the position of the surface with respect to the density peak imposed by the density threshold. The larger differences observed in the kinetic and magnetic energy pairs are due to the anisotropic nature of the kinetic and magnetic surface energy terms and are stronger wherever the velocity or magnetic field do not cross smoothly across the surface of the CC. For example, In Fig.~\ref{fig5}, it is possible to observe how the velocity field penetrates the clump's boundaries at the thresholds of 7.5~$\bar{n}$ and 15~$\bar{n}$ becoming more parallel to the clump surface at the higher thresholds of 30~$\bar{n}$, 60~$\bar{n}$ and 100~$\bar{n}$. 

\begin{figure}
\plotone{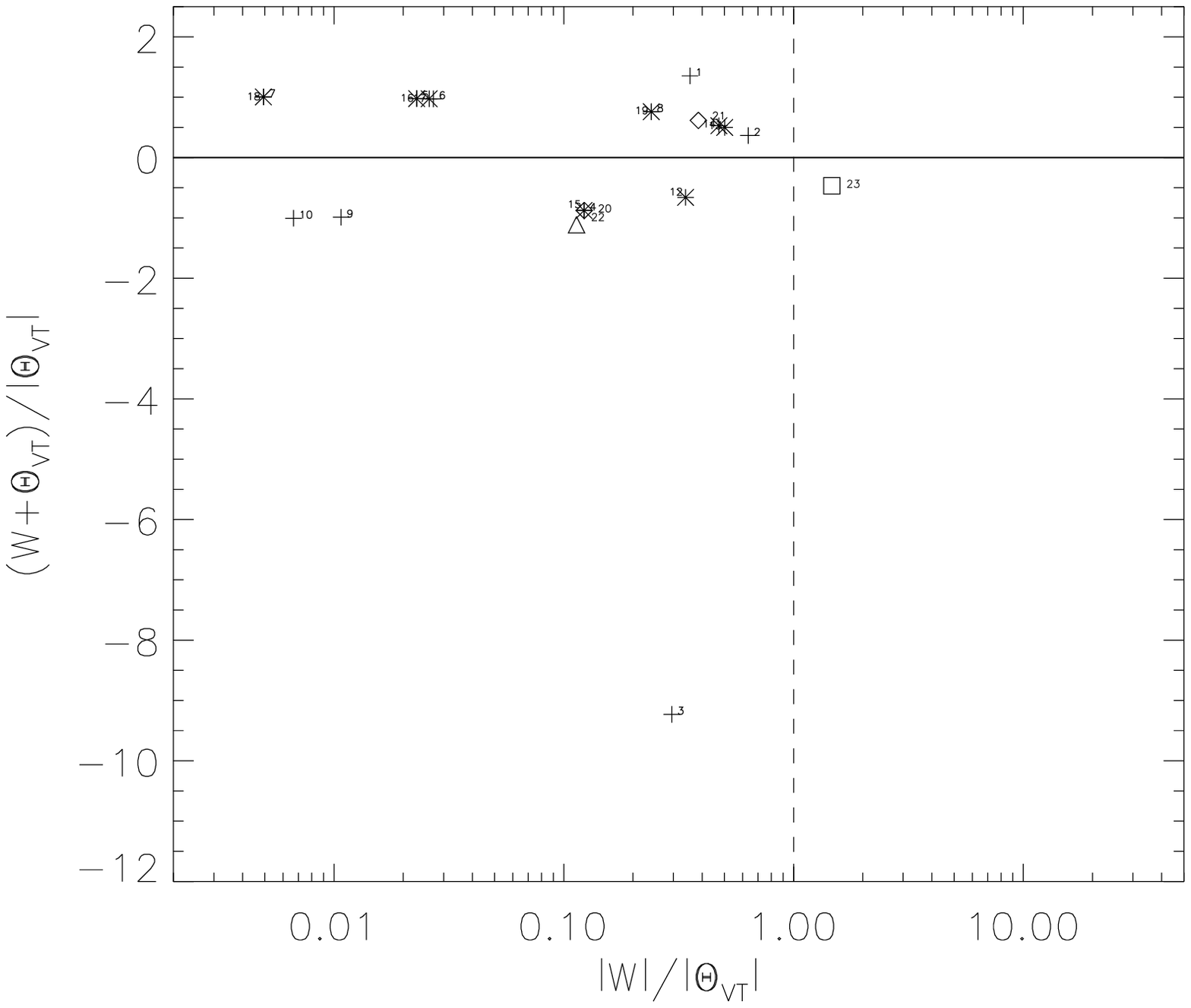}
\caption{This figure compares the importance of the gravitational term $|W|$ (y axis) to the net amount of thermal, kinetic, and magnetic energies that appear in the EVT (x axis) in run M10J4$\beta$0.01 at $t=223=8.92$ Myr.}
\label{fig11}
\end{figure}

\begin{figure}
\vspace*{-22mm}
\plottwo{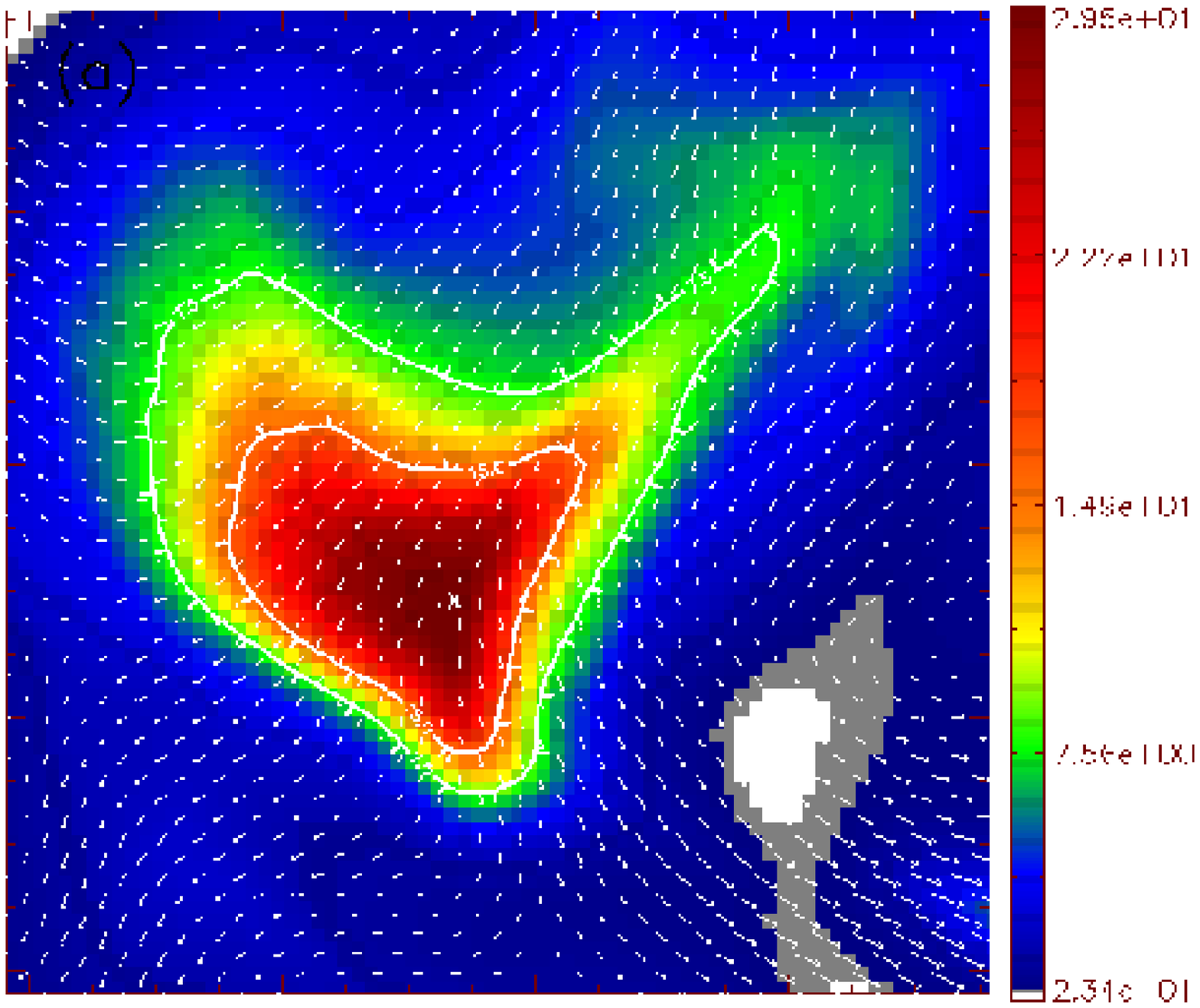}{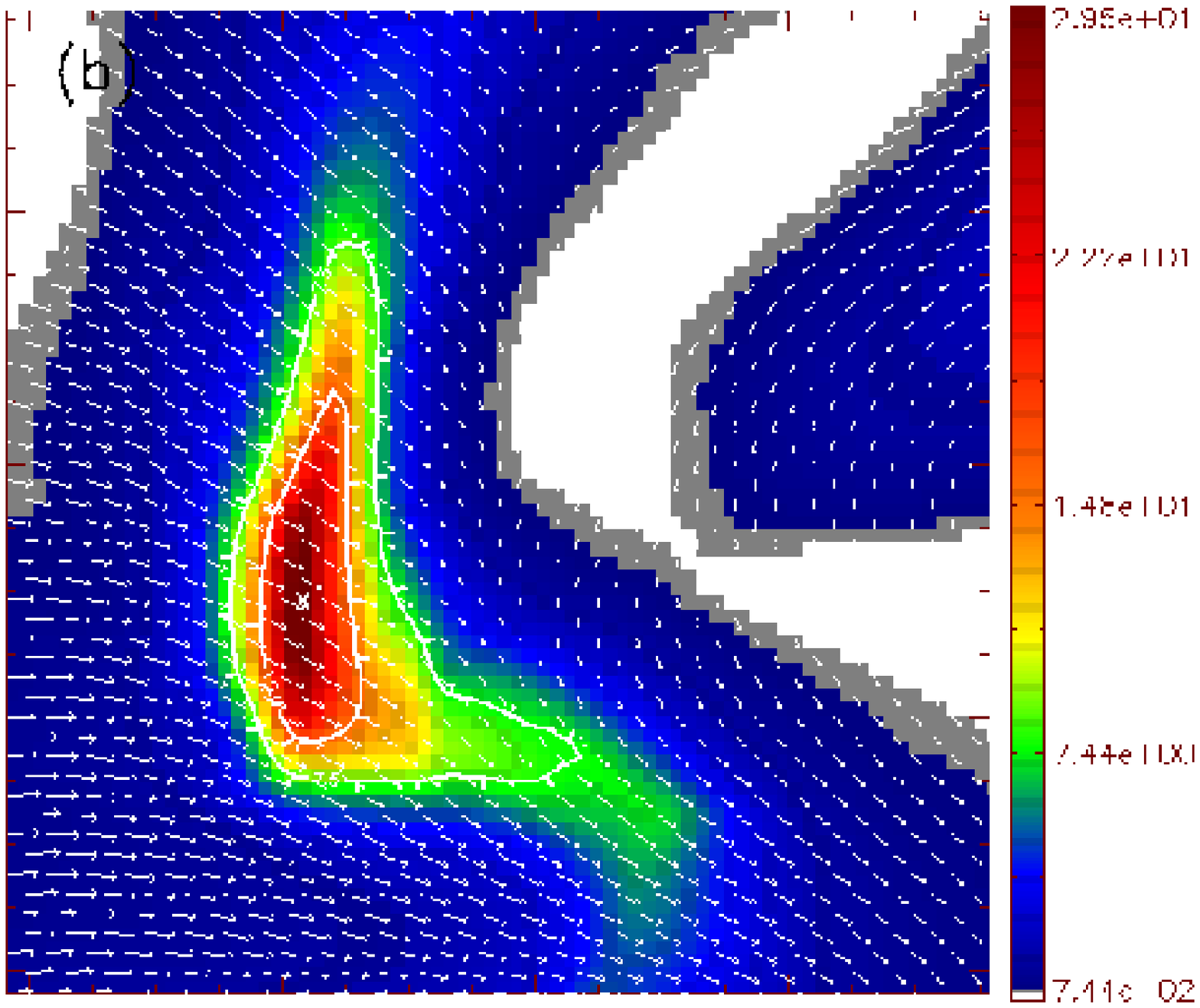} 
\plotone{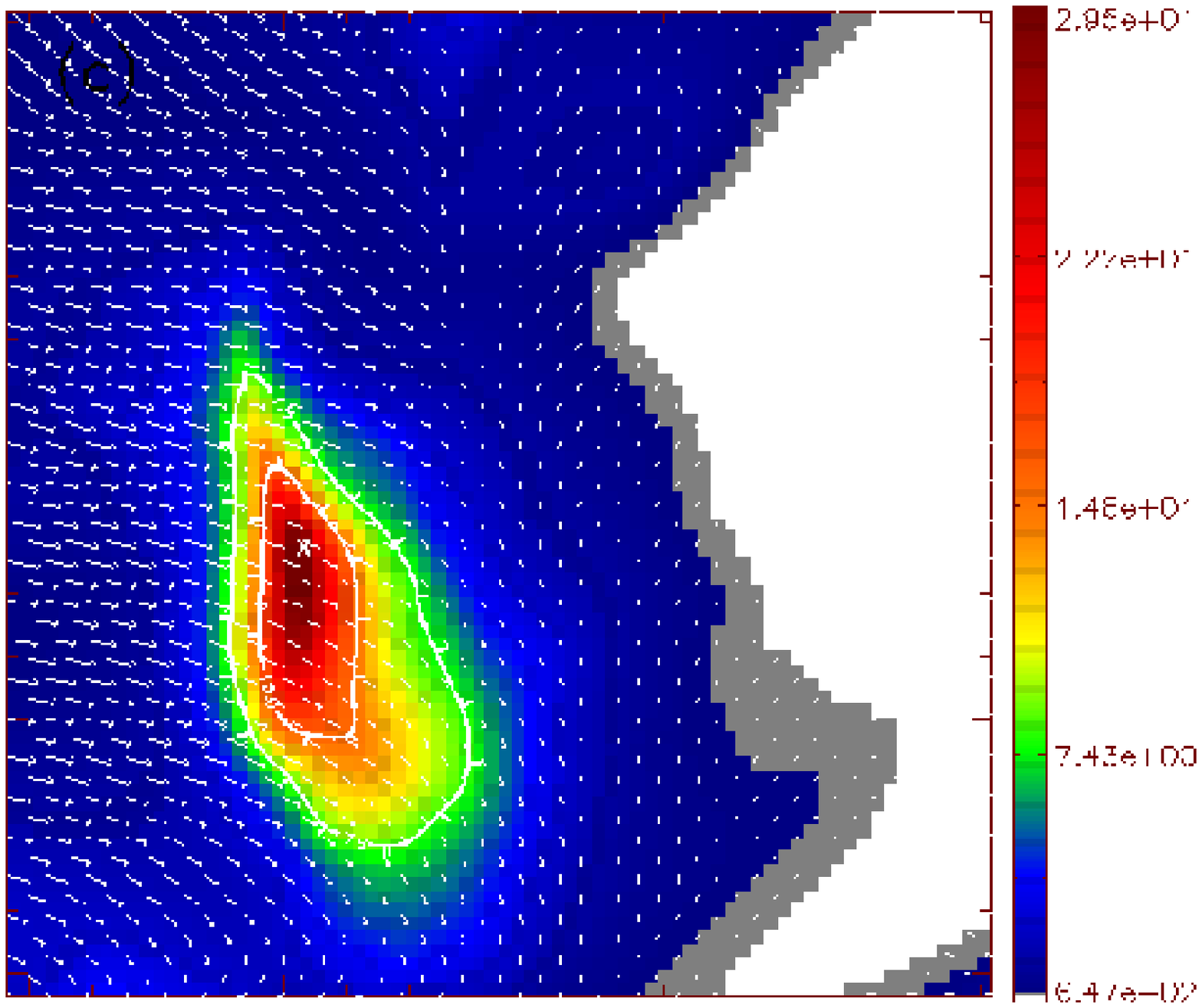}
\caption{Density cuts in the x (top), y (middle), and z (bottom) directions at the position of the peak density for the condensation corresponding to clumps number 3 and 14 in run M10J4$\beta$.01 at $t=223=8.85$ Myr, defined at the $n_{th}=7.5~\bar{n}$ and $15~\bar{n}$ threshold levels, respectively. Arrows represent the projected velocity field. In order to better highlight the structure of the cloud, the resolution of the map has been artificially multiplied by a factor 3 and the map smoothed.}
\label{fig12}
\end{figure}

 Fig.~\ref{fig11} describes the state of gravitational boundedness of the CCs following the approach given in \S~\ref{virial}. Objects 1,12,20,22, and 23 correspond to the same condensation seen at the thresholds of 7.5,15,30,60 and 100 $\bar{n}$, respectively (see Tab.\ref{tab2}). At high density thresholds (30,60 and 100 $\bar{n}$), the clump has an ellipsoidal structure, with axis ratios of $\sim 1:2$. At the lower thresholds of 15 and 7.5 $\bar{n}$ , the clump is linked by a small bridge to a protuberance to the south which causes its morphology to strongly deviate from an ellipsoidal configuration (the bridge is only on one side and not fully seen in the cut at the position of the peak density). The outer envelope is being dispersed by forces other than gravity ($W_{1}+\Theta_{VT,1} > 0$ and $|W|_{1} < |\Theta_{VT,1}|$) causing the clump to be 'peeled' from the outside. At the intermediate threshold levels of 15, 30, and 60 $\bar{n}$, the clump is being confined by forces other than gravity ($W+\Theta_{VT} < 0$ and $|W| < |\Theta_{VT}$). Material is being redistributed from the main clump towards the southern condensation; see, in Fig.~\ref{fig5}a, the projected velocity field pointing southward in the bridge. The inner part of the condensation observed at the highest threshold level of $100~\bar{n}$ (i.e., core 23) is seen to be slightly gravitationally bound. However, core 23 has only 141 cells and due to its elongation, is dominated by numerical noise along its minor axis. Therefore, the condensation is essentially not bound by gravity. This is why this object, the most massive that forms in this simulation, does not proceed towards gravitational collapse. The condensation corresponding to clumps (3,14), defined at the thresholds of 7.5 and 15 ~$\bar{n}$, respectively, is found to be bound by forces other than gravity at the lowest threshold level. This object is a low density clump (i.e., its peak density is $29~\bar{n}$) and is very flattened. Fig.~\ref{fig12} shows density cuts at the position of the clump peak density in the three directions of the box along with the projected velocity field. The velocity field is compressing the object along its smallest dimensions and would it not be for the magnetic support (discussed below), it would end up collapsing into a thin sheet, eventually following the description given in Burkert \& Hartmann (2004).
 
\begin{figure}
\plotone{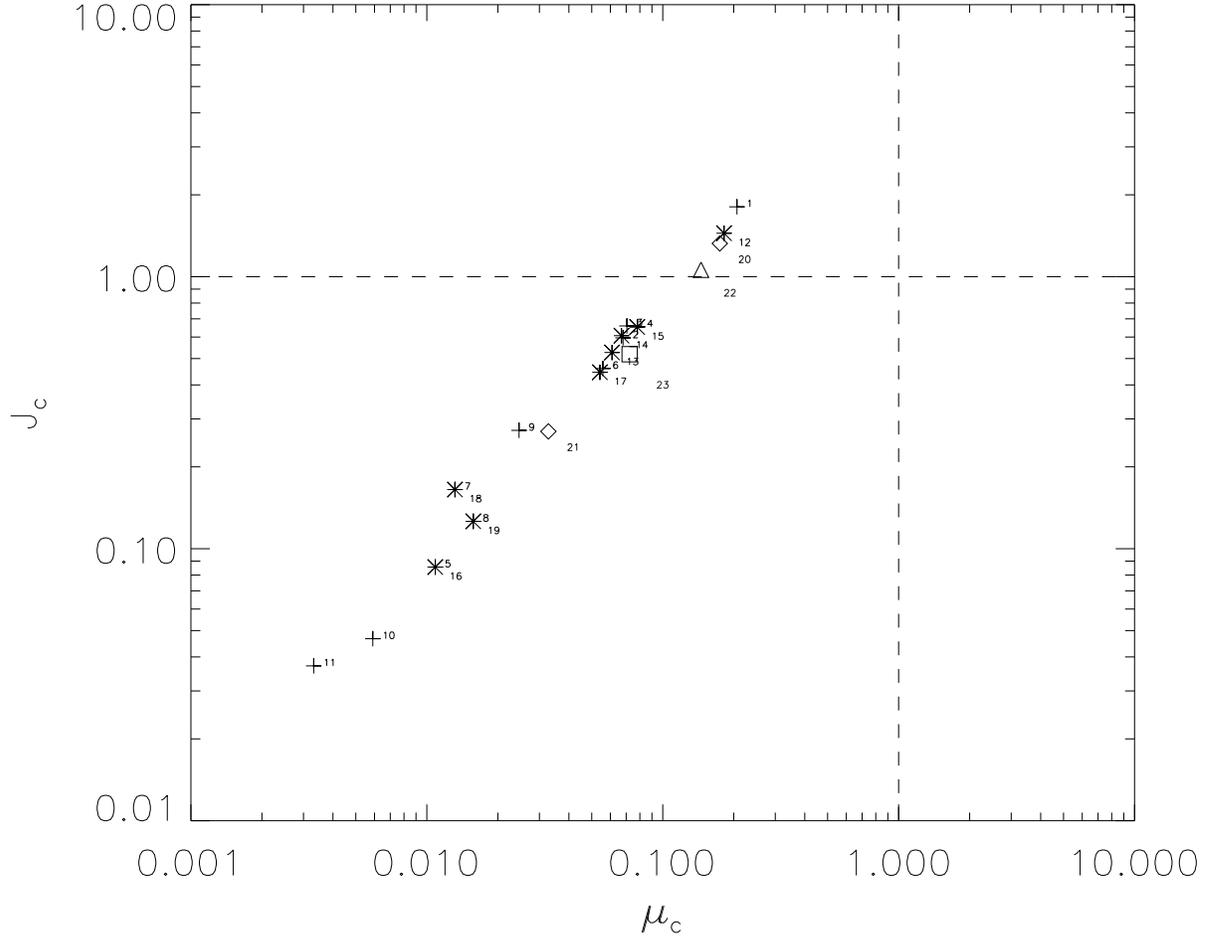}
\caption{Jeans numbers $J_{c}$ and mass-to-magnetic flux ratio (normalized to the critical value for collapse) $\mu_{c}$ of the ensemble of clumps and cores found in run M10J4$\beta$.01 at $t=223=8.92$ Myr.}
\label{fig13}
\end{figure}

\begin{figure}
\plotone{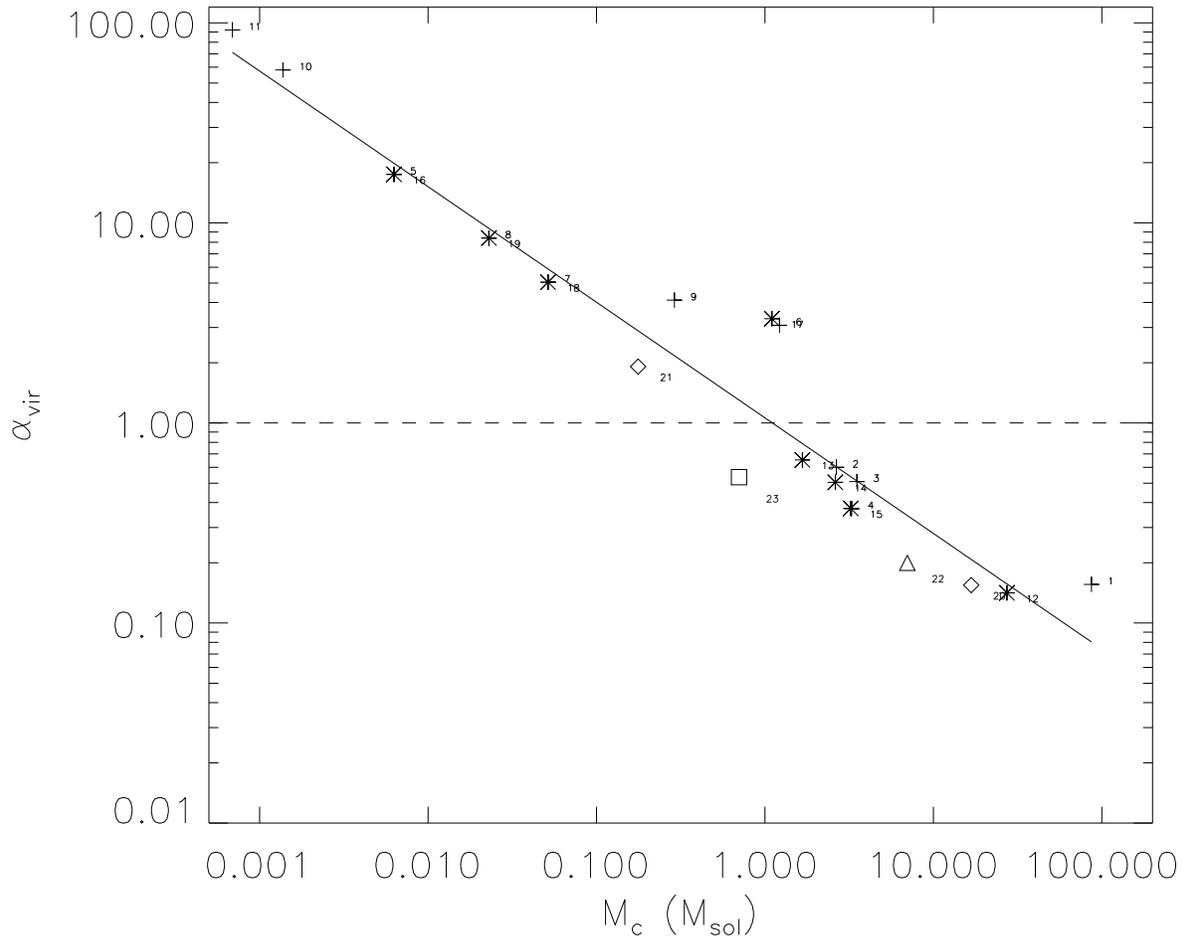} 
\caption{The virial parameter as a function of the mass of the clumps and cores in run M10J4$\beta$.01 at $t=223=8.92$ Myr.}
\label{fig14}
\end{figure}

 Figs.~\ref{fig13} and \ref{fig14} display the values of the Jeans number $J_{c}$, the mass-to magnetic flux ratio (normalized to the critical value for collapse) $\mu_{c}$, and the virial parameter $\alpha_{vir}$ for the ensemble of CCs in this simulation (at $t=223=8.85$ Myr). The $J_{c}$ and $\alpha_{vir}$ values tend to indicate that the most massive condensation (i.e., clumps 1,12,20,22,23) is gravitationally bound, practically at all threshold levels. The object corresponding to clumps (3,14) is marginally bound ($J_{c} \lesssim 1$ and $\alpha_{vir} \gtrsim 1$). However, both objects are observed to be magnetically supported ($\mu_{c} < 1$) at all threshold levels, which is in agreement with the virial analysis of not being gravitationally bound. It is important to note that the $\alpha_{vir}$ estimates (Fig.~\ref{fig14}) tend to indicate a larger number of gravitationally bound objects with respect to their true virial balance or $J_{c}$ estimates. This is the case for the second most massive object in the simulation box (corresponding to clumps 2,13,21 at threshold levels of 7.5,15, and 30 $\bar{n}$) as well as the object corresponding to clumps (4,15 at threshold levels of 7.5 and 15 $\bar{n}$, respectively). Fig.~\ref{fig11} indicates that the first object is unbound, whereas the second one is confined by forces other than gravity. These results tend to suggest that {\it the assessment of the dominance of self-gravity in a clump or core using virial parameter estimates is at the least dubious and should be used in observational studies with caution}. A fit to the $\alpha_{vir}-M_{c}$ data in Fig.~\ref{fig14} yields a relation $\alpha_{vir} \propto M_{c}^{-0.57 \pm 0.03}$, which is, within error bars, in good agreement with the $\alpha_{vir}-M_{c}$ relations found by Bertoldi \& McKee (1992) and Williams et al. (2000) for CCs in several Giant Molecular Clouds and especially with CCs in Cepheus OB3.  

\begin{figure}
\plotone{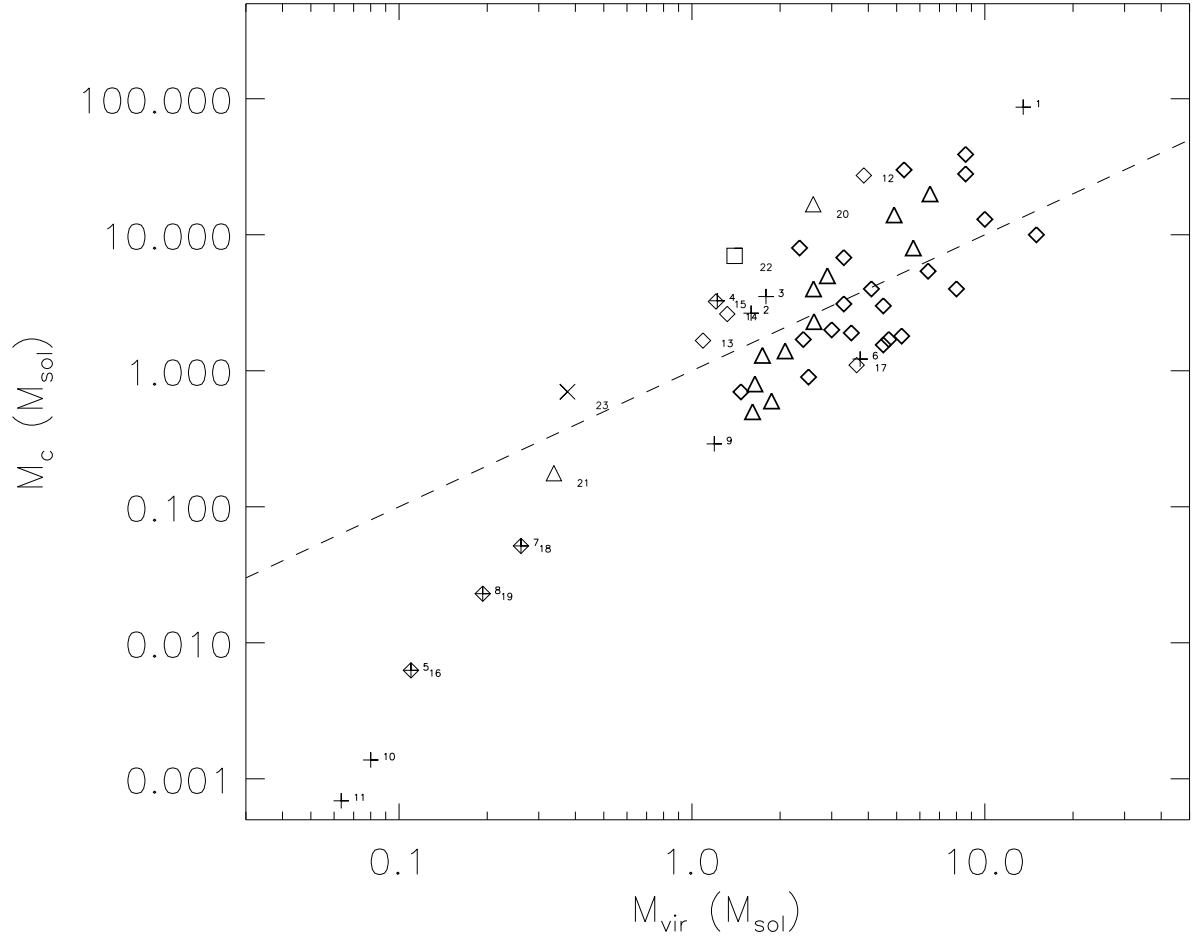}
\caption{The real clumps and cores mass $M_{c}$ is compared to the virial mass $M_{vir}$ which is calculated by comparing the clumps gravitational and kinetic+thermal energies in run M10J4$\beta$.01 at $t=223=8.92$ Myr. Thick triangles and thick diamonds correspond to the starless CCs and CCs which contains stars in the sample of Caselli et al. (2002), respectively.}
\label{fig15}
\end{figure}

Fig.~\ref{fig15} compares the virial mass $M_{vir}$ to the true mass $M_{c}$. The range of masses over which $M_{vir} \sim M_{c}$ is $\sim$ [$0.3-4$] M$_{\odot}$ and overlaps the one observed in Caselli et al. (2002) with a similar scatter around the equality line. We also note a tendency, both in the observations and in our results, for the most massive objects ($M_{vir} \gtrsim 4$ M$_{\odot}$) to have $M_{vir} < M_{c}$ (i.e., $\alpha_{vir} < 1$). Finally we observe a population of small, low density objects for which $M_{vir} >> M_{c}$ which indicates their highly dynamical nature. Unfortunately, neither in the N$_{2}$H$^{+}$ cores sample of Caselli (2002) nor in the $^{13}$CO and C$^{18}$O of Bertoldi \& McKee (1992) and  Tachihara et al. (2000,2002) does the CCs masses extend to this low values that could enable a comparison with our results. The absence of smaller mass clumps in low density tracers observations is very likely due to projection effects. This has been demonstrated by Gammie et al. (2003) who compared the mass spectra of three-dimensional clumps to the mass spectra of two-dimensional clumps obtained from the projected column density map. 

A similarity between the real mass and the virial mass does not imply that the objects found in this mass range are in virial equilibrium, but they merely indicate, on the one hand that our simulations correctly represent real MCs, and that there is an approximate equipartition between the volume kinetic+thermal and gravitational energies in that mass range though the objects are dynamically evolving and, finally, that if the assumptions (i.e., see \S~\ref{classical}) made in the observations are applied to the CCs found in the simulations, it is possible to recover similar results. 

\subsection{Moderate supercritical cloud}\label{msupercloud}

\begin{figure}
\plotone{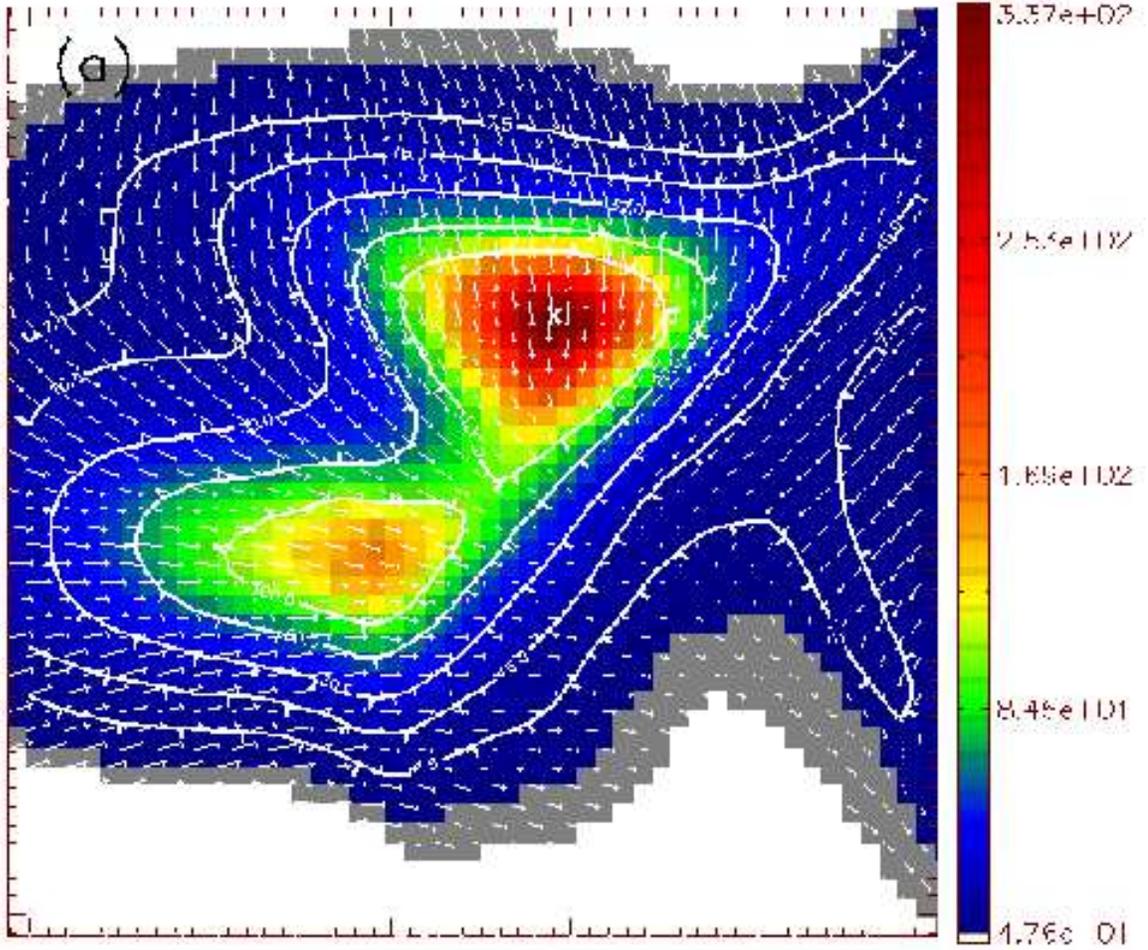}
\caption{Density cut in the x-direction showing two merging core cores in run M10J4$\beta$.1 at $t=40=1.6$ Myr, respectively. Arrows represent the projected velocity field scaled to its maximum value on the map.}
\label{fig16}
\end{figure}

We now turn to the moderately supercritical cloud (i.e., run M10J4$\beta$.1, $\mu_{box}=2.8$ and intial magnetic field strength of 14.5 $\mu$G). In this simulation, we observe the formation of collapsed objects with high peak densities ($n_{peak} \gtrsim 5000~\bar{n}$) at three distinctive time-steps, i.e., frames 44, 130 and 210, corresponding to the epochs of 1.76, 5.2 and 8.4 Myr after the simulation has been started (see animation in V\'{azquez-Semadeni et al. 2005a}). We have analyzed the virial balance of CCs in this simulations at frames 30, 40, 50 and 210 (see Tab.~\ref{tab1}). Only at time-steps 30 and 40, which precede gravitational collapse in any core, is the internal dynamics accurately described for all CCs according to the Truelove criterion. The diminishing  number of clumps detected at the lower thresholds of 7.5 and 15 $\bar{n}$ in Tab.~\ref{tab1} between frames 30 and 210 clearly shows that clumps are merging to form denser structures. An example is shown in Fig.~\ref{fig16} which displays, at $t=40=1.6$ Myr, the ongoing merger of the two most massive cores in the simulation box at this epoch. The merger of these two condensations leads to the formation of a massive core which proceeds towards gravitational collapse at $\sim t=44=1.76$ Myr. 

\begin{figure}
\plotone{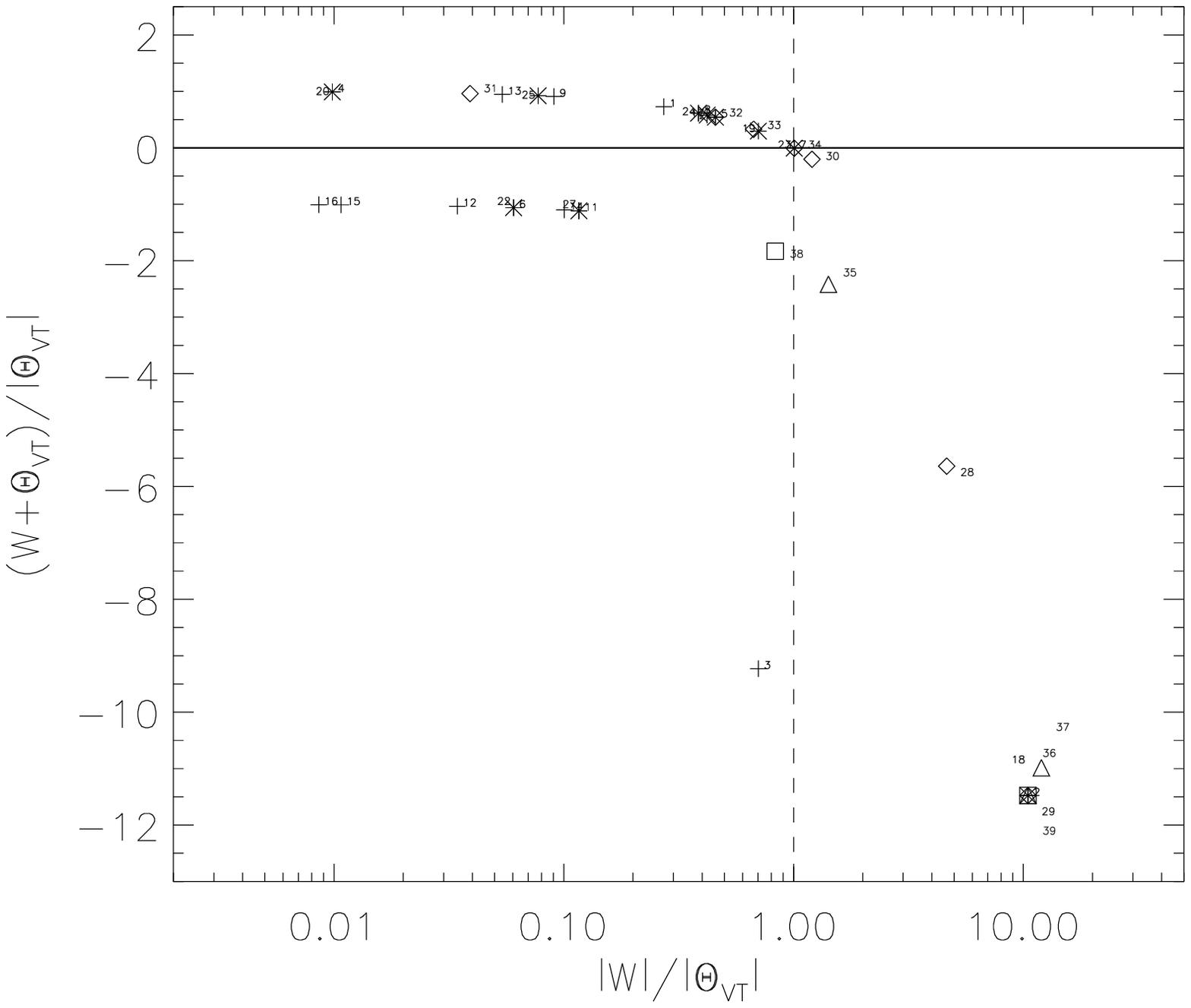} 
\caption{Same as Fig.~\ref{fig11} but for the moderately supercritical run M10J4$\beta$.1 at $t=40=1.6$ Myr.}
\label{fig17}
\end{figure}

\begin{figure}
\plotone{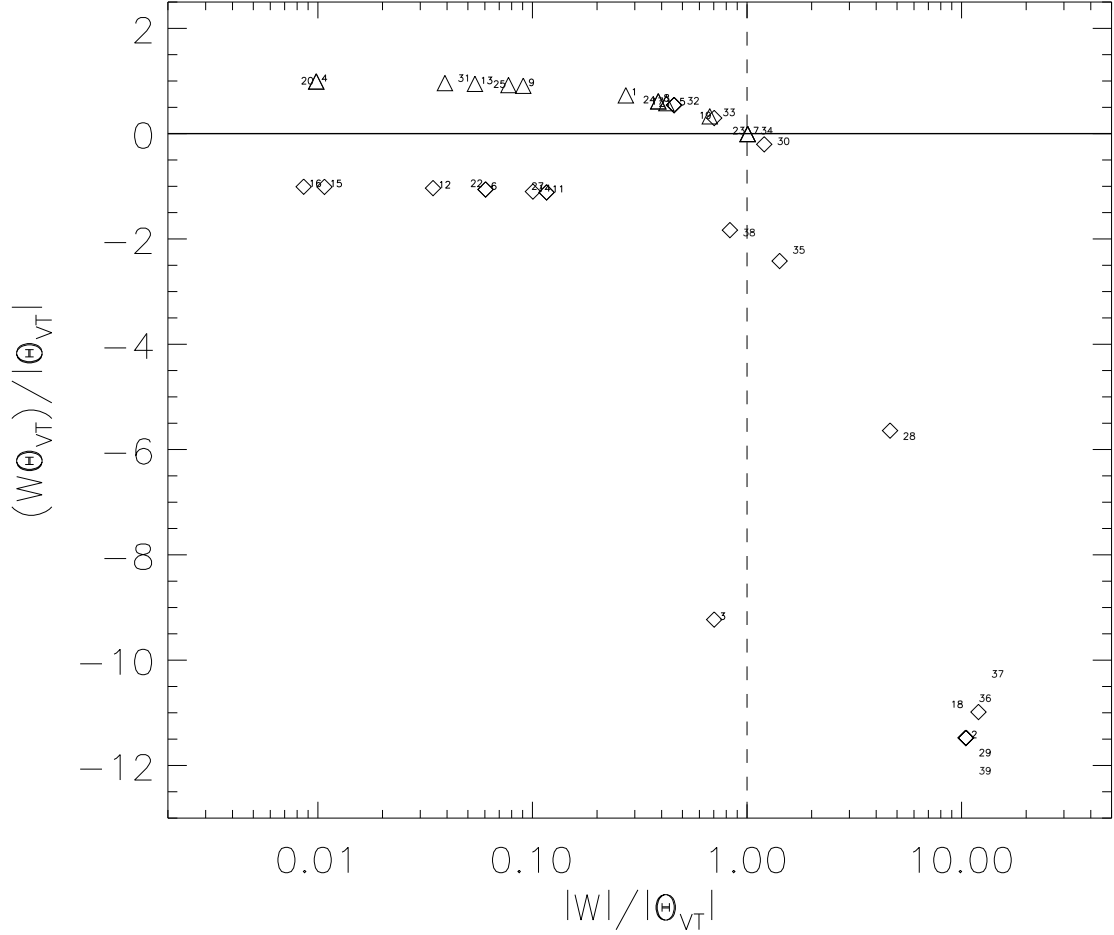}
\caption{Same as Fig.~\ref{fig17} but where clumps are cataloged by whether they have ($E_{k-}\tau_{k}) > 0$ (triangles) or ($E_{k}-\tau_{k}) < 0$ (diamonds). A positive/negative ($E_{k}-\tau_{k}$) is indicative that the clump or core is experiencing a net dispersive/compressive effect of the velocity field.}
\label{fig18}
\end{figure}

\begin{figure}
\plotone{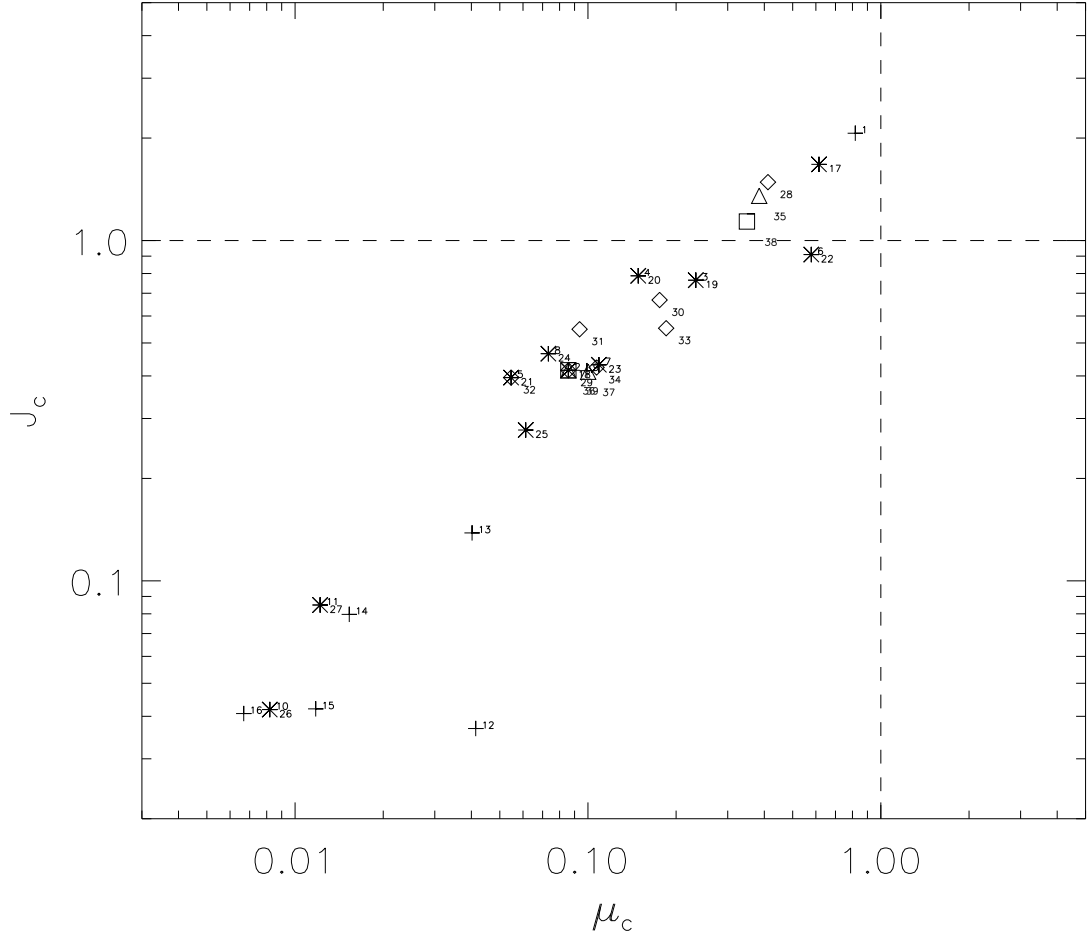}
\caption{Jeans numbers $J_{c}$ and mass-to-magnetic flux ratio (normalized to the critical value for collapse) $\mu_{c}$ of the ensemble of clumps and cores found in the moderately supercritical run M10J4$\beta$.1 at $t=40=1.6$ Myr.}
\label{fig19}
\end{figure}

\begin{figure}
\plotone{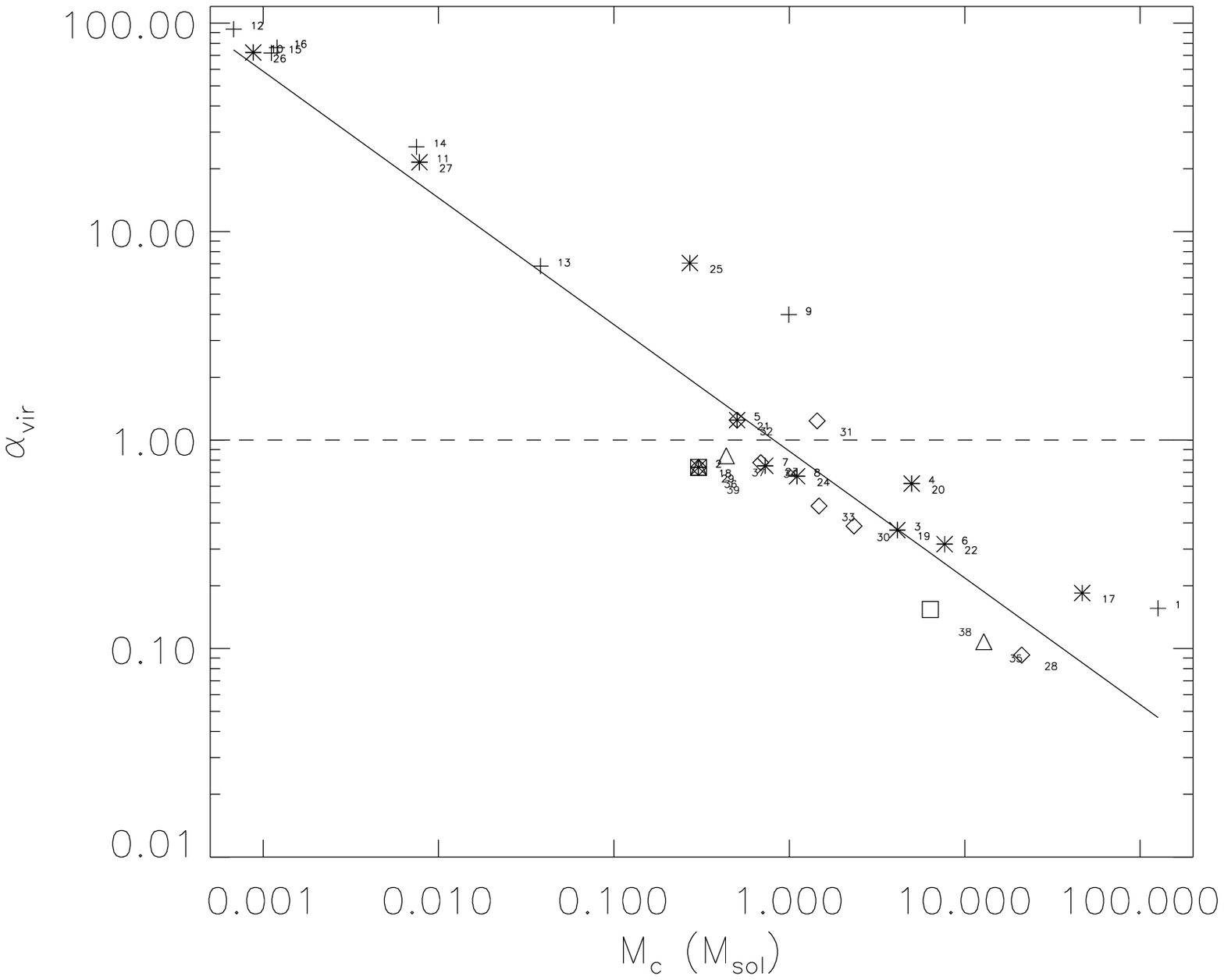}
\caption{Same as Fig.~\ref{fig14} but for run M10J4$\beta$.1 at $t=40=1.6$ Myr.}
\label{fig20}
\end{figure}

The detailed virial analysis is similar to the one presented for the subcritical cloud. The general conclusions drawn from the earlier case such as the verification of Eq.~\ref{eq1}, the importance of the time dependent and surface energy terms in the virial equation, and the more transient nature of the smaller and less dense clumps remain valid for the CCs found in this run. However, it is interesting to investigate the energy balance of the CCs at the onset of the cloud merger observed in Fig.~\ref{fig16}. The energy balance of the ensemble of CCs at this epoch is displayed in Fig.~\ref{fig17}. The most massive condensation corresponds to clumps 1,17,28,35 at the threshold levels of 7.5,15,30,60 $\bar{n}$, respectively, and 38 and 39 at the threshold of 100 $\bar{n}$ which are the two merging cores observed in Fig.~\ref{fig16}. The condensation as a whole is observed to be unbound at the lowest thresholds (i.e., clumps 1 and 17) but is in the process of being compressed by the velocity field (i.e., Fig.~\ref{fig18}). At higher threshold levels, the condensation is gravitationally bound (clumps 28 and 39) or marginally bound (clumps 35 and 38). On the other hand, the calculated $J_{c}$, $\mu_{c}$, and $\alpha_{vir}$ values for this condensation (i.e., Figs.~\ref{fig19} and \ref{fig20}) suggest that it is gravitationally bound at almost all thresholds (except for core 39 which has $J_{c,39} \lesssim 1$), and close to magnetic criticality ($\mu_{c} \lesssim 1$)\footnote{If the characteristic radius of the clumps were calculated as the cubic root of the volume and not as the radius of a sphere of equivalent volume, the magnetic flux would be larger by a factor of $\sim 3$ and as a result the clumps would be slightly supercritical $\mu_{c,1,17,28,35,38} \gtrsim 1$.}. 

In contrast to the subcritical run, a larger number of condensations are observed to be gravitationally bound (i.e., in the lower right quadrant). The condensation corresponding to clumps (2,18,29 and 36) is observed to be gravitationally bound at all threshold levels and the inner parts of the condensation corresponding to clumps (3,19,30, and 37) are gravitationally bound (cores 30 and 37). Condensation (2,18,29,36), however, is numerically under-resolved with a total number of cells of 43 (i.e., at the lowest threshold) with a size of two cells in one direction and thus its energy balance is unreliable according to the rough uncertainty estimates in Fig.~\ref{fig2}. Condensation (3,19,30, and 37) is a well resolved object (i.e., 2794 cells at the lowest threshold level of 7.5 $\bar{n}$). the virial analysis tends to indicate that the inner parts of this condensation are more strongly gravitationally bound (i.e., $|W|/|\Theta_{VT,37}| > |W|/|\Theta_{VT,30}| >|W|/|\Theta_{VT,3,19}|$). On the other hand, the classical indicators tend to show that the condensation is more gravitationally bound or marginally bound as a whole than its inner parts as indicated by the $\alpha_{vir}$ and $J_{c}$ estimates ($\alpha_{vir,3,19} < \alpha_{vir,30} < \alpha_{vir,37}; J_{c,3,19} > J_{c,30} > J_{c,37}$) and the fact that the inner parts are less magnetically supported than the condensation as a whole according to the $\mu_{c}$ estimates ($\mu_{c,37} < \mu_{c,30} < \mu_{c,3,19}$).  

 In Fig.~\ref{fig18}, we reproduce the data of Fig.~\ref{fig17} but where clumps are cataloged by whether they have $E_{k}-\tau_{k} > 0$ (triangles), which is an indication of a net dispersive effect by the velocity field or $E_{k}-\tau_{k} < 0$ (diamonds), which is indicative of a net compressive effect by the velocity field. From Fig.~\ref{fig18}, it is interesting to note that unbound CCs, not very surprisingly, have $E_{k}-\tau_{k} > 0$, which is an indication that the velocity field is playing a role in the dispersion of these objects. It is also important to note that not all of the CCs that are being globally compressed by the velocity field $(E_{k} - \tau_{k} < 0)$ are necessarily dominated by gravity as is assumed in some studies (e.g., Field et al. 2006).  

\begin{figure}
\plotone{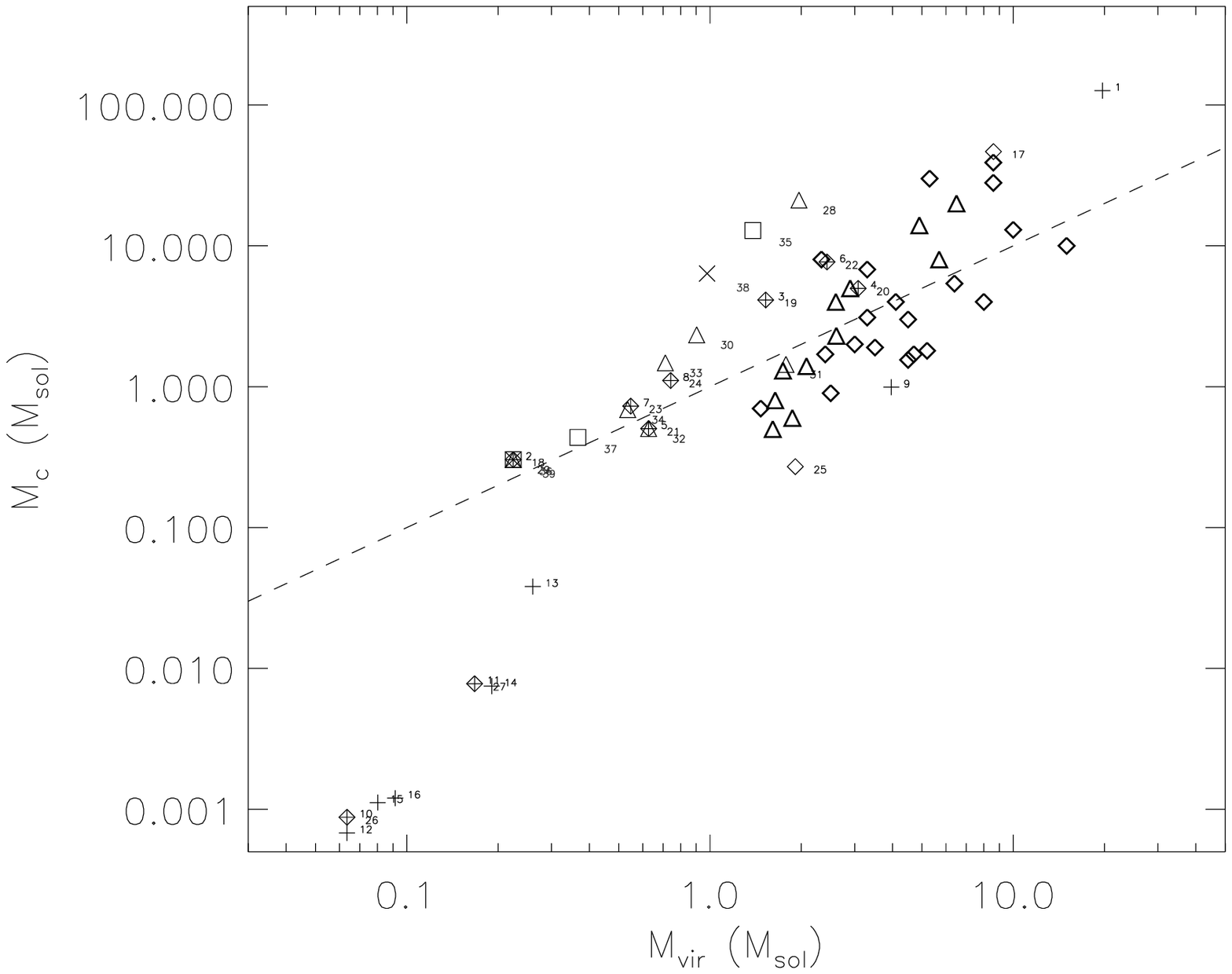}
\caption{Same as Fig.~\ref{fig15} but for run M10J4$\beta$.1 at $t=40=1.6$ Myr.}
\label{fig21}
\end{figure}

 As in the subcritical cloud case, the $\alpha_{vir}$ estimates tend to indicate a larger number of gravitationally bound CCs than that implied by the virial and Jeans number estimates. An example is the case of the condensation corresponding to clumps 6,12, and 33 (defined at the 7.5,15 and 30 $\bar{n}$ threshold levels, respectively) which is seen to be gravitationally bound according to its $\alpha_{vir}$ values (i.e., Fig.~\ref{fig20}) whereas Fig.~\ref{fig17} indicates that this condensation is bound by forces other than gravity (outer parts, clumps 6, 22) or unbound (inner parts, core 33). The relationship between the $\alpha_{vir}$ gravitational boundedness estimator and the real mass $M_{c}$ in Fig.~\ref{fig20} is well fitted by $\alpha_{vir} \propto M_{c}^{-0.60 \pm 0.03}$ which is also in very good agreement with the $\alpha_{vir}-M_{c}$ relations of Bertoldi \& McKee (1992) and Williams et al. (2000) (see Tab.~\ref{tab3}). An additional interesting point, both in this model and in the subcritical cloud model, is that the $^{13}$CO like cores (i.e., at the lowest threshold level of 7.5 $\bar{n}$) can be described as gravitationally bound according to their $\alpha_{vir}$ estimates starting from masses of a few solar masses (objects 2,3, and 4 in Fig.~\ref{fig14} and objects 3,4, and 6 in Fig.~\ref{fig20}; objects 2 and 5 are poorly resolved objects), typically [3-6] M$_{\odot}$. Fig.~\ref{fig21} displays the same general trends in the $M_{c}-M_{vir}$ relation as in the subcritical cloud model with the exception that the mass range over which $M_{c} \sim M_{vir}$ extends downs to $M_{vir} \sim M_{c} \sim 0.25$ M$_{\odot}$ and that the data points of Caselli et al. (2002) are better encompassed by our sample of CCs found in this model. 
 
\subsection{Strongly supercritical cloud}\label{ssupercloud}

 In this section, we discuss the gravitational boundedness of CCs that form in the strongly supercritical cloud model M10J4$\beta$1 (i.e., weak magnetic field strength of 4.6 $\mu$G; $\mu_{box}=8.8$). A visualization of the time evolution of the dense structures formed in this cloud model (see animation in V\'azquez-Semadeni et al. 2005a) shows the existence of both collapsed, and other, non-collapsing, long-lived condensations. The first of these long-lived condensations appears at time-step $t=16=0.64$ Myr and disperses around frame $t=84=3.36$ Myr. At the epoch at which we perform our analysis (i.e., at time-step $t=30=1.2$ Myr), no collapsing object is present in the simulation box. The clump-finding algorithm finds three condensations at all thresholds levels (i.e., 7.5,15,30,60 and 100 $\bar{n}$), namely the condensations corresponding to clumps (1,12,23,32,37), (2,13,24,33,38), and (3,14,25,34,39). Two condensations are observed at the first four threshold levels (i.e., clumps 4,15,26,35 and 5,16,27,36, respectively), and six condensations are observed at lower threshold levels (i.e., Tab.~\ref{tab1}).

\begin{figure}
\plotone{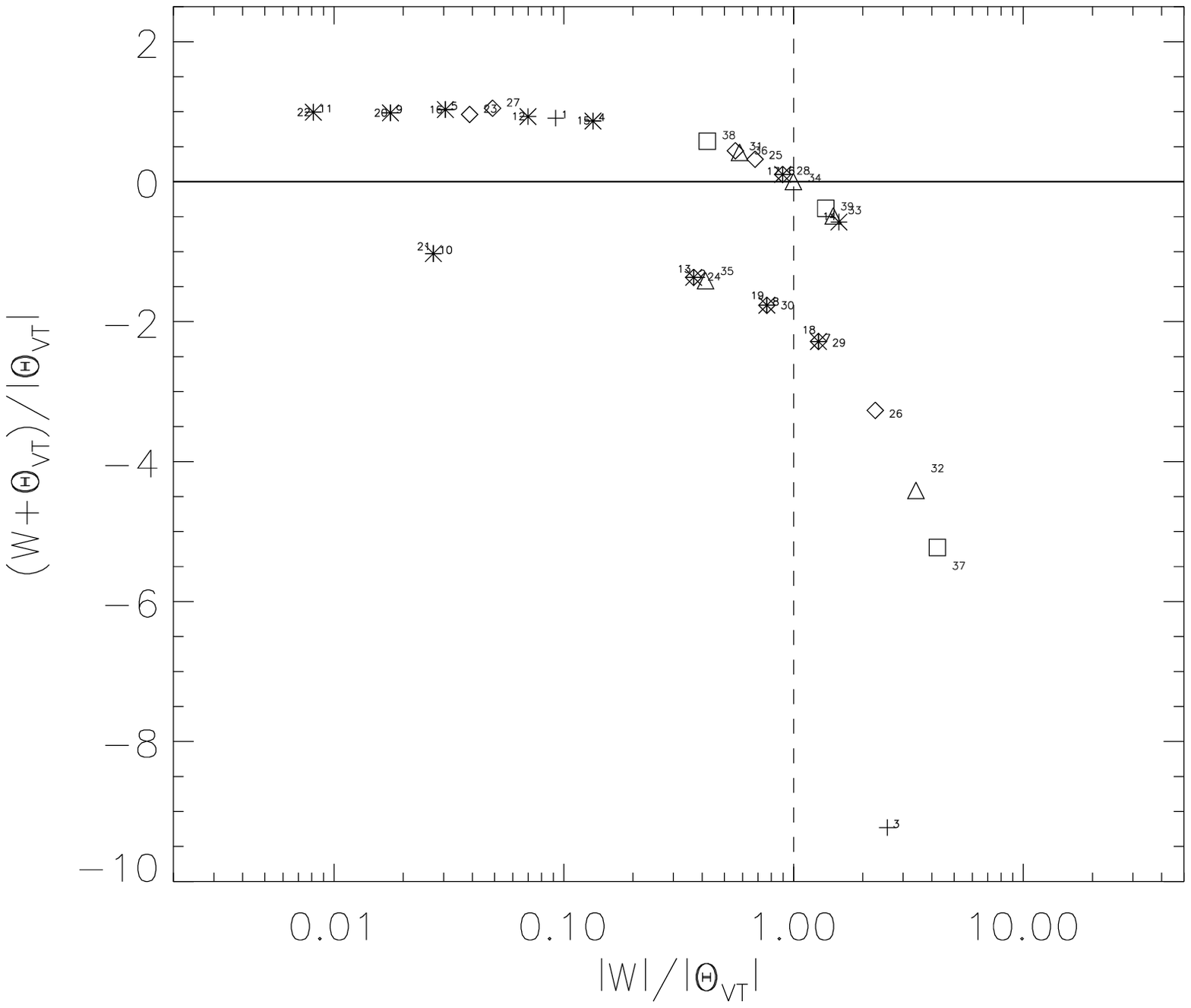}
\caption{Same as Fig.~\ref{fig11} but for the strongly supercritical run M10J4$\beta$1 at $t=30=1.2$ Myr.}
\label{fig22}
\end{figure}

\begin{figure}
\plotone{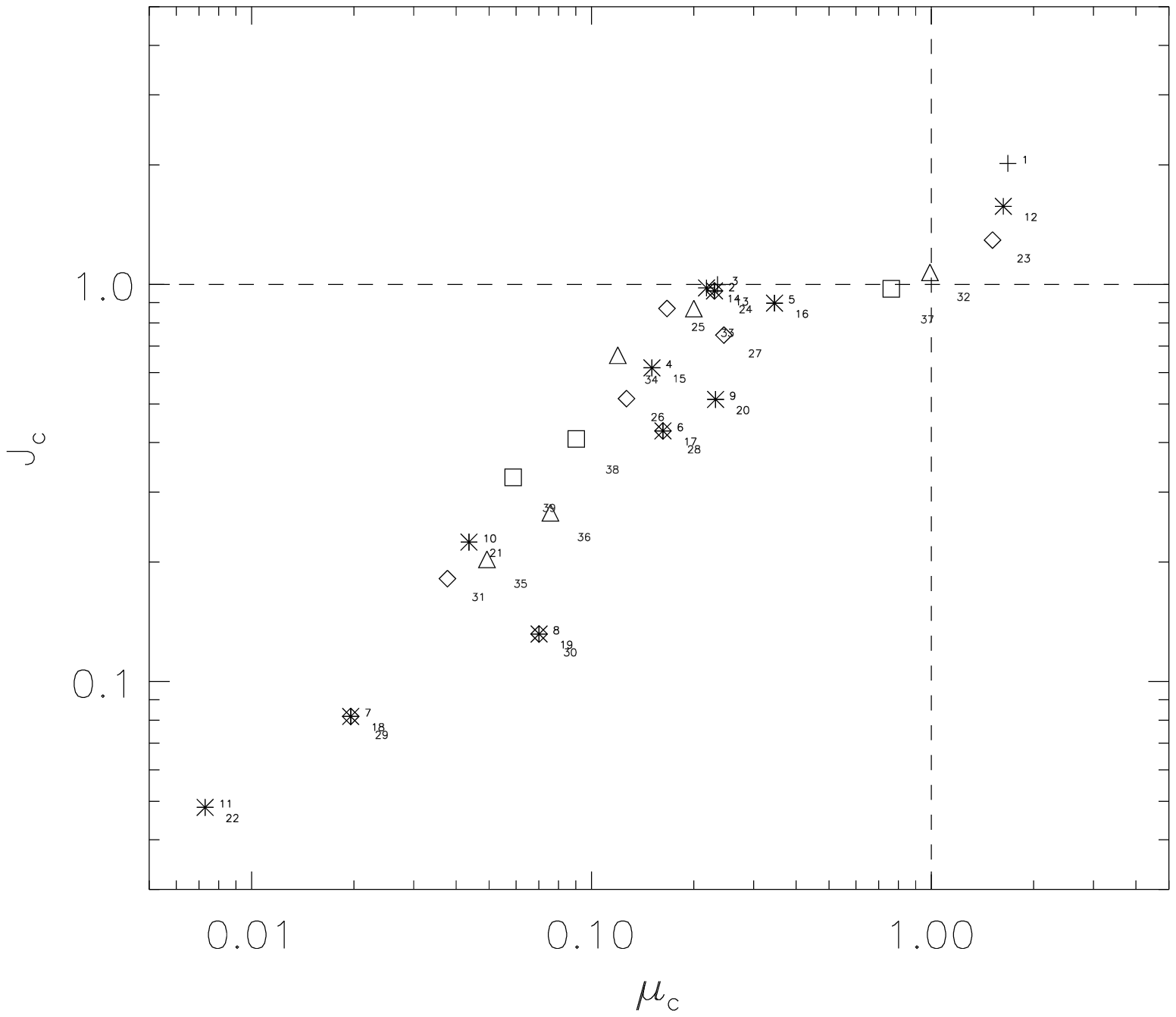}
\caption{Same as Fig.~\ref{fig13} but for the strongly supercritical run M10J4$\beta$1 at $t=30=1.2$ Myr.}
\label{fig23}
\end{figure}

\begin{figure}
\plotone{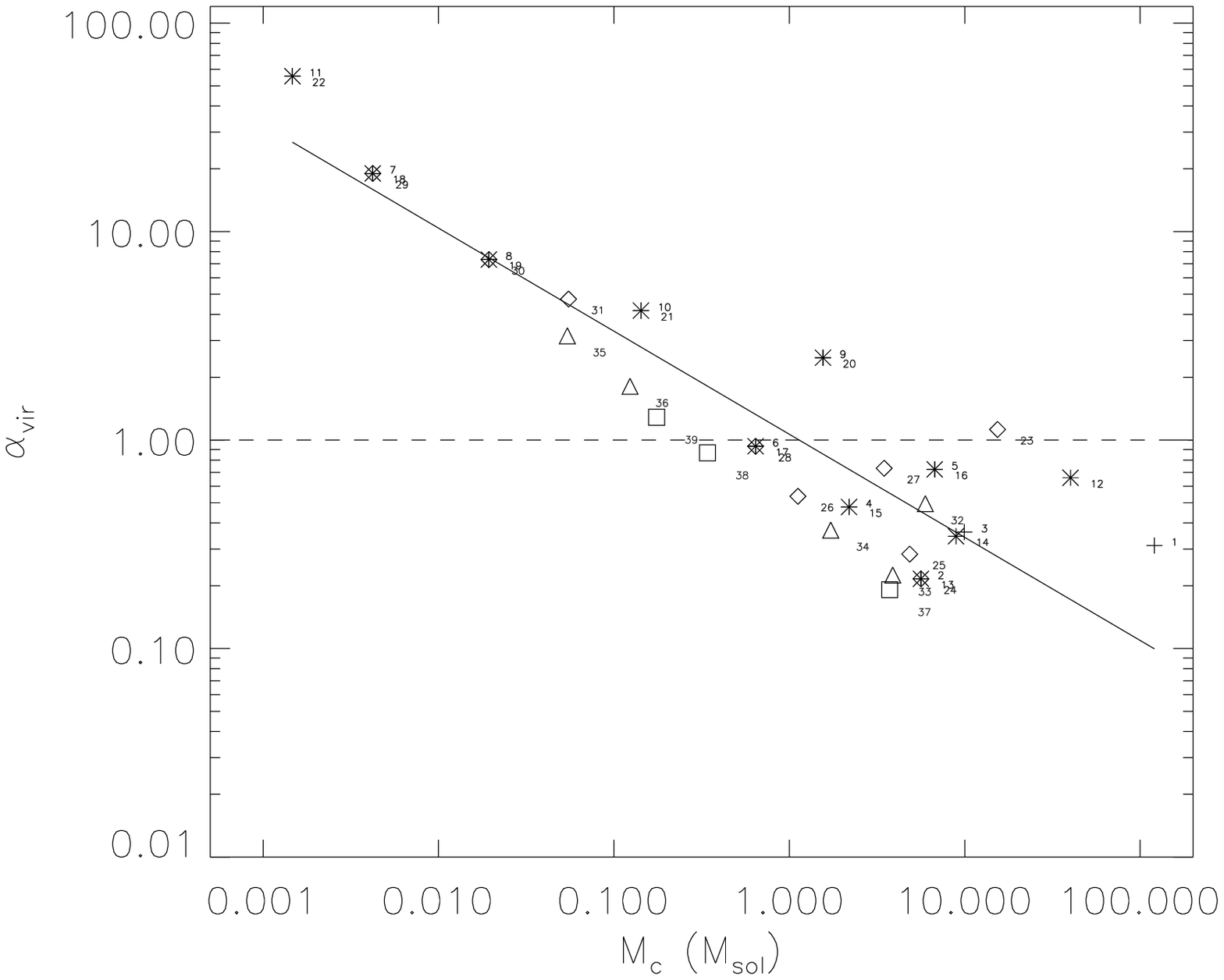}
\caption{Same as Fig.~\ref{fig14} but for the strongly supercritical run M10J4$\beta$1 at $t=30=1.2$ Myr.}
\label{fig24}
\end{figure}

Fig.~\ref{fig22} shows the virial balance for the ensemble of CCs at the selected timestep. The different condensations show a diverse behavior. The central regions of the condensation that contains the highest density peak (i.e., object 1,12,23,32,37) are observed to be gravitationally bound (i.e., cores 32 and 37) whereas when considered with its larger envelope, it is gravitationally unbound (i.e., objects 1,12, and 23). Similarly to the cases discussed in the previous sections, the  values of the classical indicators, $J_{c}, \mu_{c}$, and $\alpha_{vir}$ for this condensation suggest that it is bound at all threshold levels, and magnetically more supercritical when considered with its extended envelope (i.e., Figs.~\ref{fig23} and \ref{fig24}). The second condensation (i.e., clumps 2,13,24,33,38) displays a rather complex behavior. The classical indicators suggest that it is gravitationally bound (i.e., $\alpha_{vir}$ estimates in Fig.~\ref{fig24}) or marginally bound (i.e., $J_{c}$ estimates in Fig.~\ref{fig23}), whereas its virial balance tends to indicate that it is confined by forces other than gravity (thermal pressure, dynamical compression and the magnetic force). At the second highest threshold level (i.e., $60~\bar{n}$), it is observed to be marginally bound (i.e., core 33, $|W|/|\Theta_{VT}| \lesssim 1.5$). This is one of the condensation which are observed to be long lived. The third condensation (i.e., clumps 3,14,25,34, and 39) is found to be gravitationally bound at all threshold levels in the virial analysis and is one of the objects that later proceeds to gravitational collapse. Another condensation exhibits the same trend as the first two condensations of having a gravitationally bound or marginally bound central region (i.e., cores 26 and 35) and an envelope that is unbound (i.e., clumps 4 and 15).

\begin{figure}
\plotone{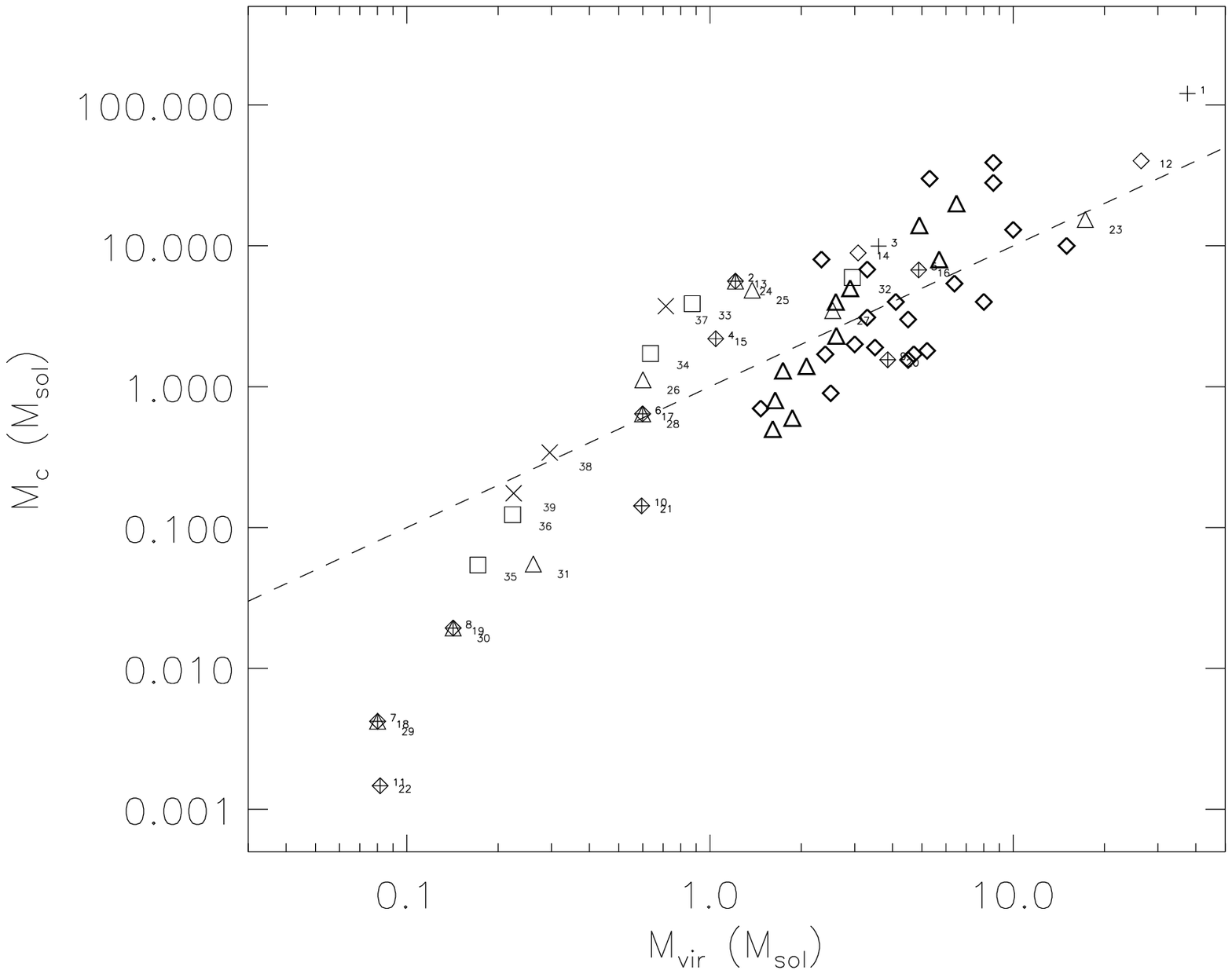}
\caption{Same as Fig.~\ref{fig15} but for the strongly supercritical run M10J4$\beta$1 at $t=30=1.2$ Myr.}
\label{fig25}
\end{figure}

 We note that in this simulation as well, the $\alpha_{vir}$ estimates (Fig.~\ref{fig24}) tend to suggest a larger number of gravitationally bound clumps than what is indicated by the virial analysis. The exponent $\epsilon$ in the $\alpha_{vir}-M_{c}$ relation shows in this case a shallower value of $\epsilon \sim -0.49 \pm 0.03$ which is smaller than any of the exponents deduced from the observations (see Tab.~\ref{tab3}). The $M_{vir}-M_{c}$ relation shown in Fig.~\ref{fig25} exhibits a larger number of objects that depart from the unity line and that are gravitationally bound (i.e., $M_{vir} < M_{c}$) down to masses of $\sim 0.6$ M$_{\odot}$, which is not observed in the sample of Caselli et al. (2002). The region where there is a good reproduction of the observational data is for masses $M_{vir} \sim M_{c} \gtrsim 2.5$ M$_{\odot}$. Even in that mass regime, the fact that $M_{vir}$ is comparable to the real mass is merely an indication of equipartition between the volume kinetic+thermal energies and the gravitational term. The assumption that these objects are in virial equilibrium is in disagreement with their virial analysis. We also note that the $M_{vir}$ estimates of the objects at the lowest thresholds are larger than those at higher threshold levels. A fact reported by Tachihara et al. (2000) from their $^{13}$CO and C$^{18}$O survey of cores in Ophiuchus. 

\subsection{Nonmagnetic cloud}\label{nobcloud}  

\begin{figure}
\plotone{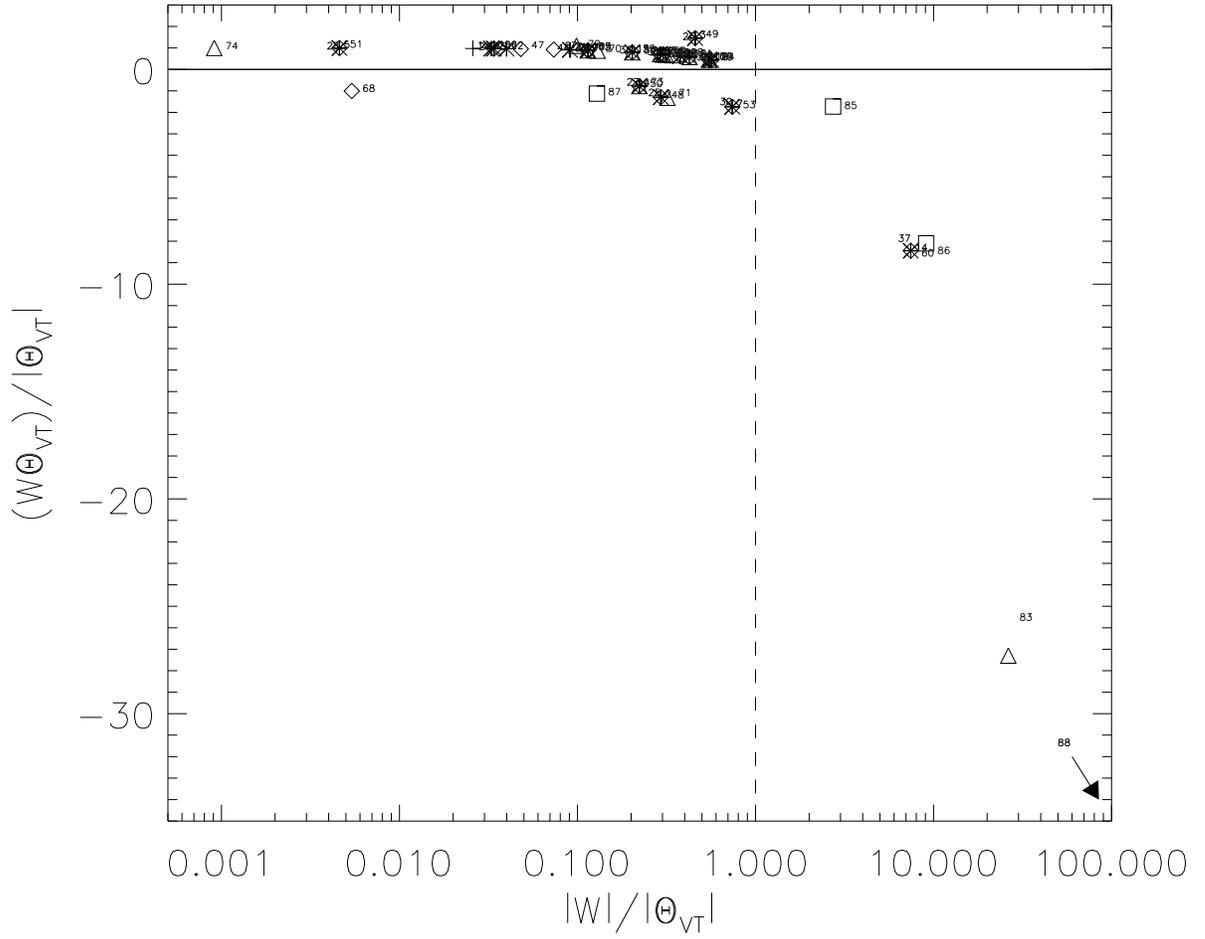}
\caption{Same as Fig.~\ref{fig11} but for the non-magnetic run M10J4$\beta \inf$ at $t=20=0.8$ Myr.}
\label{fig26}
\end{figure}

In the non-magnetic cloud model (i.e., run M10J4$\beta \inf$), the clumps are observed to evolve quickly towards gravitational collapse (see Fig. 2 in V\'{a}zquez-Semadeni et al. 2005a and the animation therein). At $t=30=1.2$ Myr after the start of the simulation, there are already 5 independent cores whose average density is higher than $500~\bar{n}$ (i.e., Tab.~\ref{tab1}). We should remind that in our simulations, whenever the density in a cell violates the Truelove criterion (Truelove 1997) and becomes higher than $\sim 256~\bar{n}$, the dynamics around this cell becomes numerically unreliable. Therefore, we show the results of the virial analysis at an earlier epoch (i.e., time-step $t=20=0.8$ Myr), at which only one object has collapsed. Fig.~\ref{fig26} displays the virial balance of CCs at this epoch. The most massive object corresponds to clumps 1,24,47,70,85, and 88 at the threshold levels of 10,15,30,60,100, and 500 $\bar{n}$, respectively. The inner regions of this object are observed to be undergoing gravitational collapse (i.e., cores 85 and 88 since e.g., $|W_{88}| >> |2(E_{k,88}+E_{th,88}-\tau_{k,88}-\tau_{th,88})|$) while the object as a whole is found to be unbound (clumps, 1,24,47 and 70). Two other condensations corresponding to clumps 2,25,48,71, and 86 (at threshold levels of 10,15,30,60, and 100, respectively) and 14,37,60, and 83 (at threshold levels of 10,15,30, and 60 $\bar{n}$, respectively) are also observed to have a gravitationally bound core (i.e., cores 86 and 83, respectively) and a more dilute, unbound envelope. 

\begin{figure}
\plotone{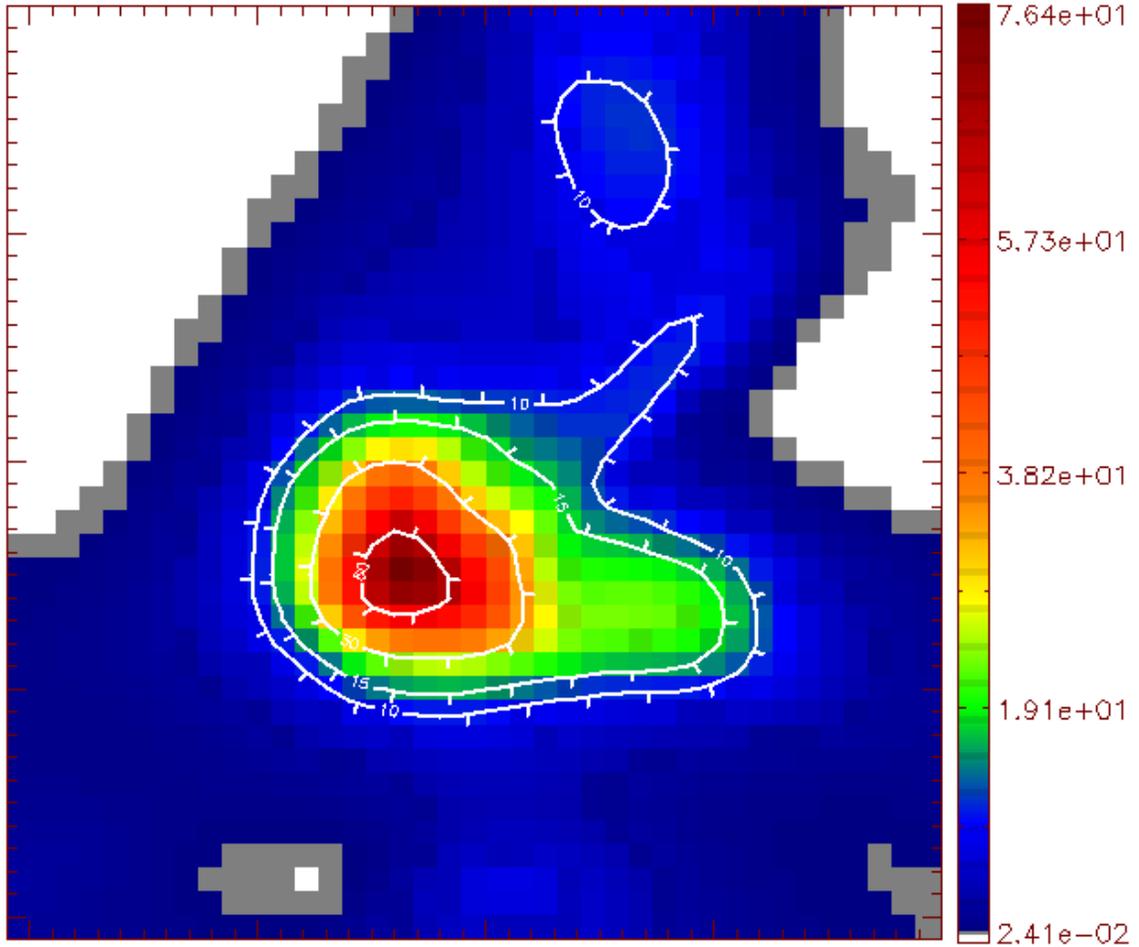}
\caption{Cut at the position of the density maximum in the x-direction of the box for the object corresponding to clumps (14,37,60, and 83 at the thresholds levels of 10,15,30 and 60 $\bar{n}$, respectively) in the non-magnetic run M10J4$\beta \inf$ at $t=20=0.8$ Myr. The dynamical and thermodynamical properties as well as the bean-like morphology of this object resemble those of the starless core Barnard 68 (Alves et al. 2001a,b). In order to better highlight the structure of the cloud, the resolution of the map has been artificially multiplied by a factor 8 and the map smoothed.}
\label{fig27}
\end{figure}

\begin{figure}
\plotone{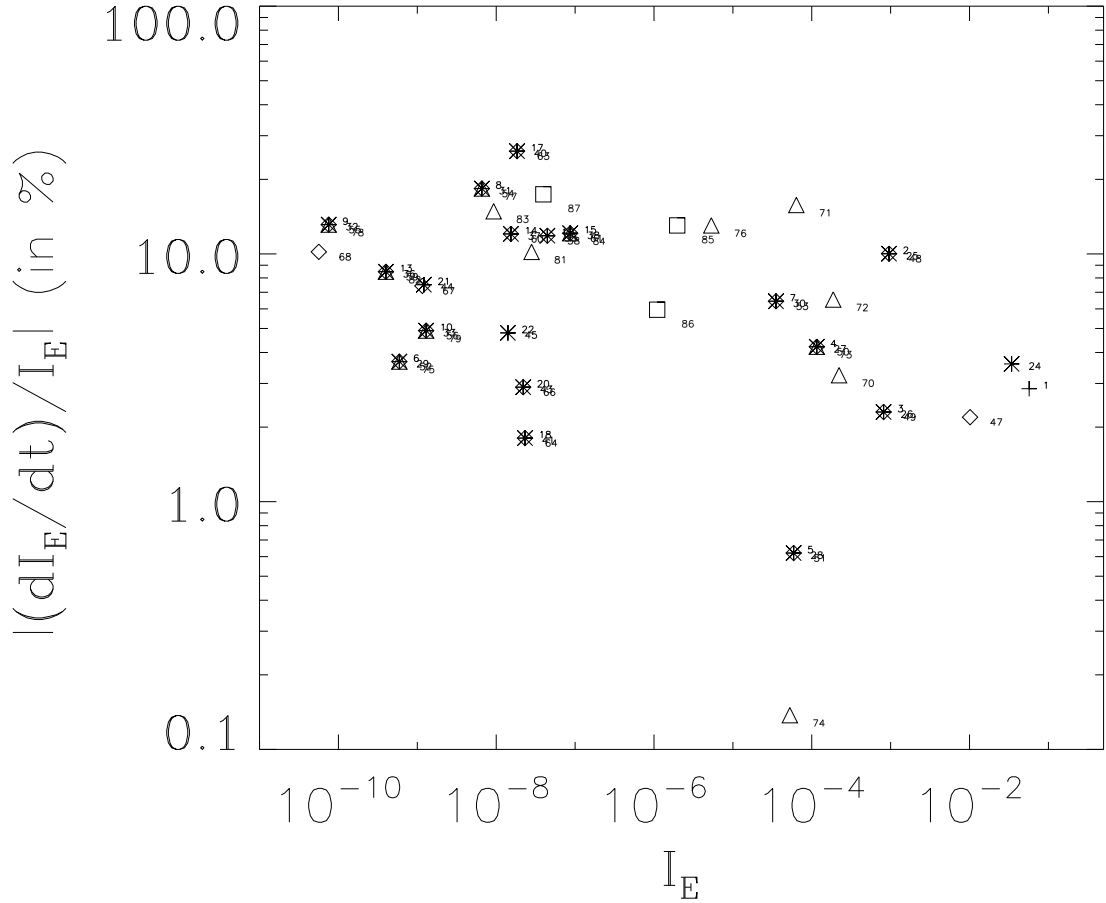}
\caption{Same as Fig.~\ref{fig7} but for the non-magnetic run M10J4$\beta \inf$ at $t=20=0.8$ Myr. The Barnard 68-like object corresponding to clumps (14,37,60, and 83 at the threshold levels of 10,15,30 and 60 $\bar{n}$), shows a temporal rate of change of it's moment of inertia of the order of $\sim 10 \%$.}
\label{fig28}
\end{figure}

A peculiar condensation is the one corresponding to clumps (14,37,60, and 83, at threshold levels 10,15,30, and 60 $\bar{n}$, respectively), which is found to be gravitationally bound at all threshold levels (Fig.~\ref{fig26}). This object (i.e., Fig.~\ref{fig27}) is small i.e., 64 cells in total, has 3-5 cells in each direction which correspond to a physical size of 0.046-0.078 pc, has sharp boundaries, an average number density of $\sim 65~\bar{n} \sim 3.2 \times 10^{4}$ cm$^{-3}$ which is nearly-independent of the selected density threshold level, a peak number density of $76.4~\bar{n} \sim 4 \times 10^{4}$ cm$^{-3}$, a mass of $\sim 1.5$ M$_{\sun}$ and a non-thermal velocity dispersion of 0.036 km s$^{-1}$. This implies a one-dimensional velocity dispersion of $\sim 0.036/3^{1/2} \sim 0.02$ km s$^{-1}$, which is roughly one tenth of the thermal sound speed. Thus, and despite the fact that this core is highly dynamical (see the temporal rate of change of its moment of inertia in Fig.~\ref{fig28}), very interestingly, all of its structural and dynamical properties resemble very closely those of Bok globules and particularly those of the Bok globule Barnard 68 (e.g., Hotzel et al. 2002a,b; Lada et al. 2003). Unfortunately, due do the low number of cells present in each direction, it is not possible to draw, for this core, the velocity profile in order to check if they present any of the observational evidence of spatial asymmetry in the blue-shifted and red-shifted components of molecular lines such as in the case of B68 (Redman et al. 2006).  

\begin{figure}
\plotone{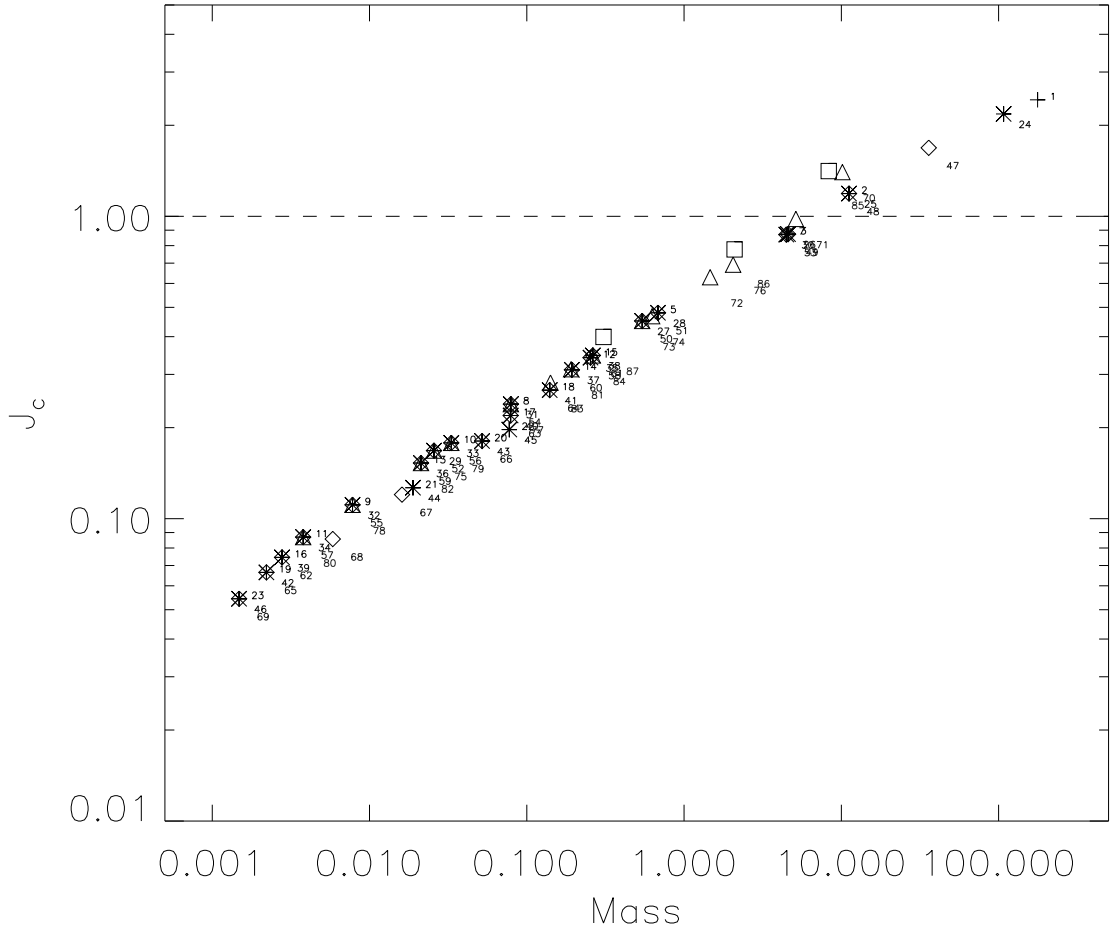}
\caption{Jeans number-Mass relations for the ensemble of clumps and cores in run M10J4$\beta \inf$ at $t=20=0.8$ Myr.}
\label{fig29}
\end{figure}

\begin{figure}
\plotone{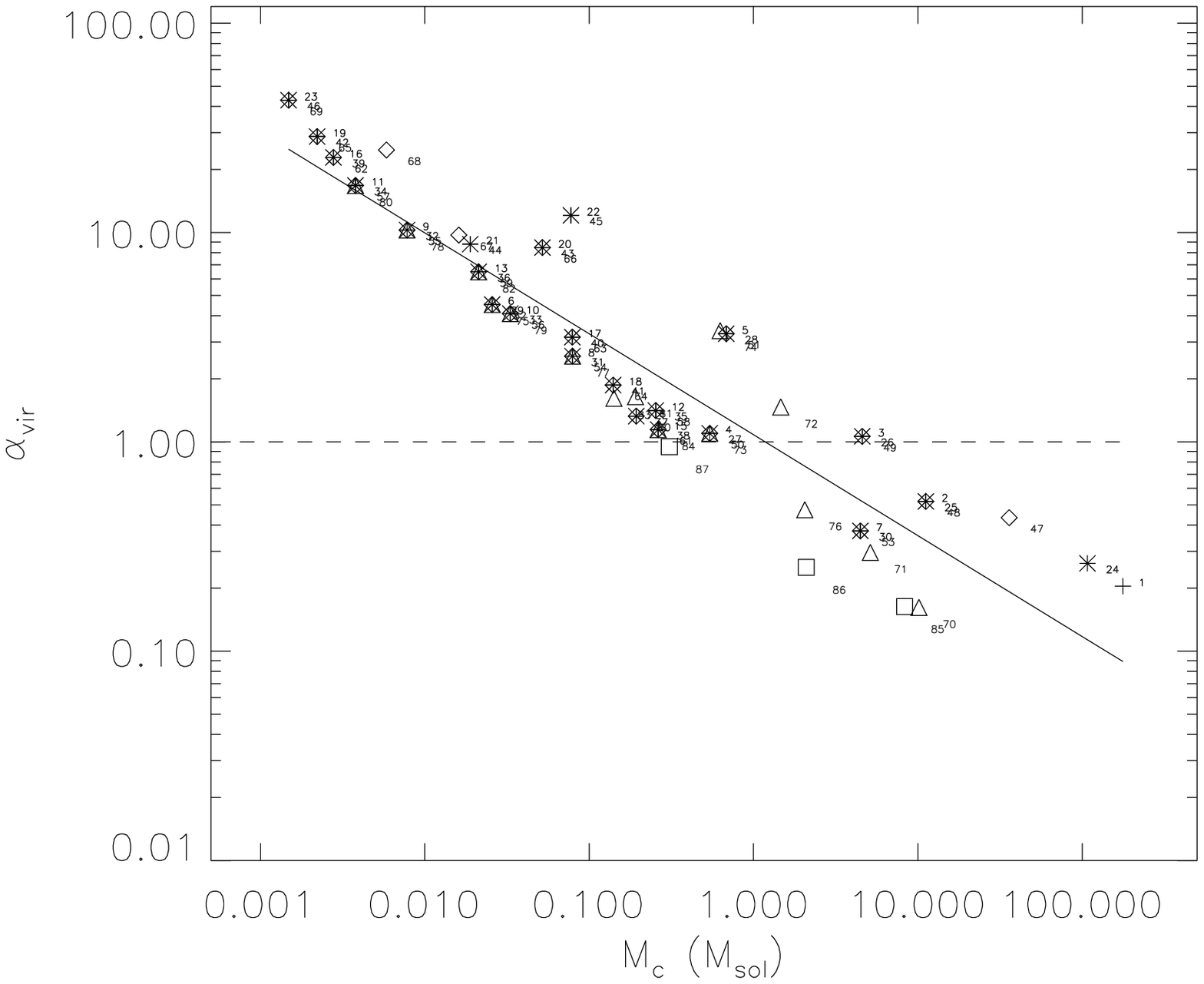}
\caption{Same as Fig.~\ref{fig14} but for run M10J4$\beta \inf$ at $t=20=0.8$ Myr.}
\label{fig30}
\end{figure}

One noticeable fact about this non-magnetic cloud simulation is that the number of condensations that are found to be gravitationally bound or readily collapsing is not particularly larger than the number of gravitationally bound objects found in the magnetized runs M10J4$\beta$.1 and M10J4$\beta$1. However, though the mass range of CCs in all simulations is roughly the same, the total number of CCs and particularly the number of small CCs is much larger in the non-magnetized cloud than in the magnetized ones (see numbers in Tab.~\ref{tab1}). Figs.~\ref{fig29} and ~\ref{fig30}, which display the $J_{c}$ and $\alpha_{vir}$ estimates in this run, respectively, illustrate this effect; i.e., same mass range of the clumps as in the magnetized runs but with larger numbers of small size CCs. Therefore, the presence of a magnetic field in the cloud seems to influence the star formation process by essentially reducing the number of formed cores within a certain volume rather than by substantially preventing or delaying the collapse process in individual clouds once gravity has overcome all other forces. 

\begin{figure}
\plotone{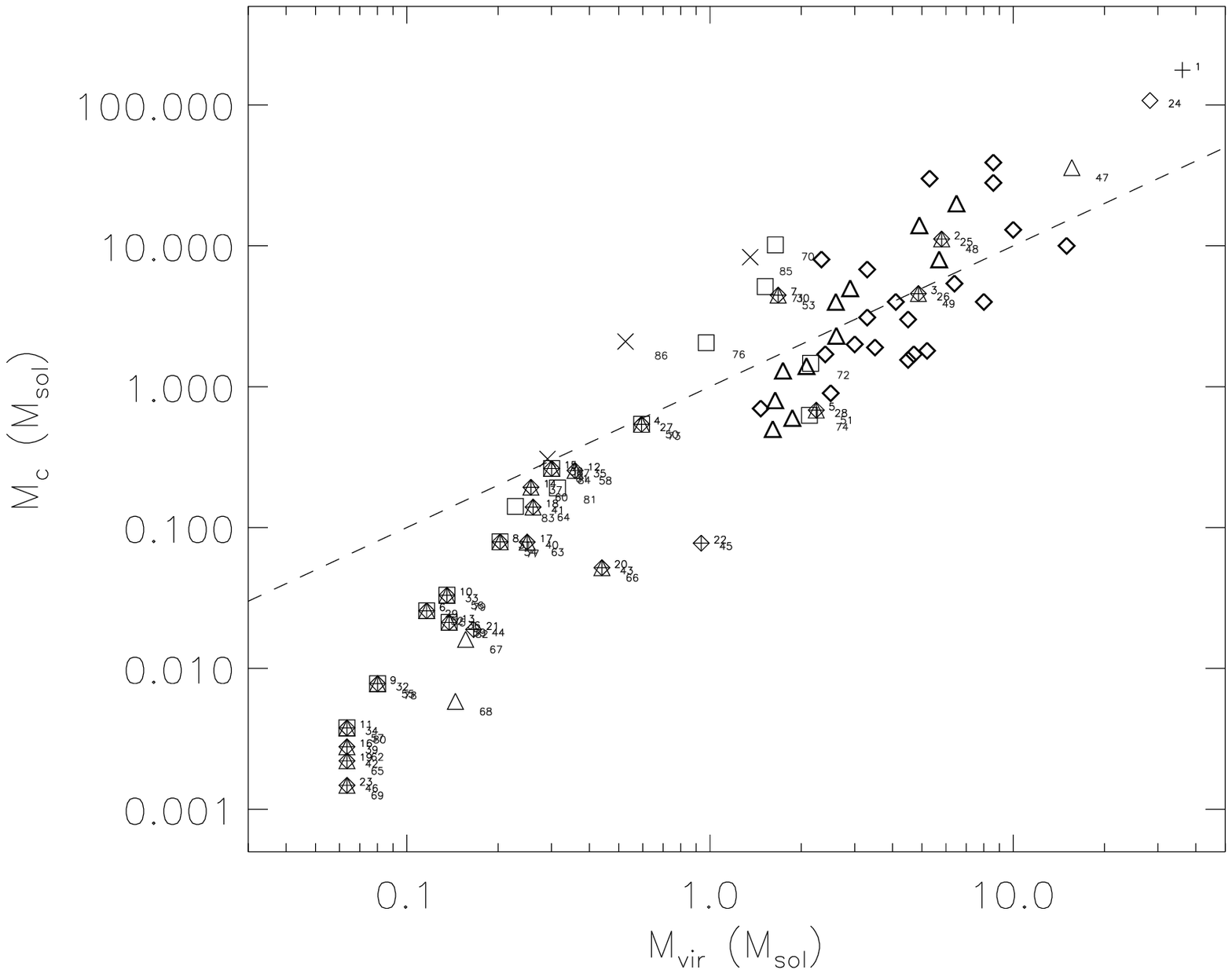}
\caption{Same as Fig.~\ref{fig15} but for run M10J4$\beta \inf$ at $t=20=0.8$ Myr.}
\label{fig31}
\end{figure}

Similarly to the cases of the magnetized clouds, the $J_{c}$ and $\alpha_{vir}$ estimates for the ensemble of CCs in this simulation (i.e., Figs.~\ref{fig29} and \ref{fig30}) suggest that the most massive condensations (condensation 1 corresponding to clumps 1,24,47,70,85 and 88 and condensation 2 corresponding to clumps 2,25,48,71,86) are gravitationally bound at all threshold levels, being also more bound for the cloud as a whole (i.e., when defined at a lower threshold level). This is in disagreement with the virial analysis which suggests that only the inner regions of these condensations are gravitationally bound. Also, in this non-magnetic case, the $\alpha_{vir}$ estimates would catalog some CCs as being gravitationally bound (e.g., clump 76) or marginally bound (e.g., clump 87), in disagreement with the detailed virial analysis, which suggests they are not. We find that the $\alpha_{vir}-M_{c}$ relation for this model (over-plotted to the data in Fig.~\ref{fig30}) is $\alpha_{vir} \propto M_{c}^{-0.48 \pm 0.02}$. The exponent, $\epsilon$, is in this case close to the value obtained for the strongly supercritical model and is generally in less good agreement with the observations (see Tab.~\ref{tab3}). The $M_{vir}-M_{c}$ relation for this non-magnetic cloud model is shown in Fig.~\ref{fig31} and displays similar characteristics than the supercritical cloud model. The different values of the exponent $\epsilon$ between the weak magnetic field (strongly supercritical) and non magnetic models on the one hand and the stronger magnetic field cases (subcritical and near critical) suggests that the presence of the magnetic field in the cores modifies both the density profile and the velocity dispersion-radius relations in the cores upon which the value of $\alpha_{vir}$ is dependent. Understanding the exponent of the $\alpha_{vir}-M_{c}$ relation requires a careful study of the radial density profiles of the cores drawn along their principal axis and is left to a future detailed study. 

After discussing the virial balance in the sample of magnetized and non-magnetized simulations, a few additional points are worth mentioning: a) The maximum density that can be reached in our simulations inside a core before the dynamics in these cores violates the Truelove criterion is $256~\bar{n} \sim 1.2 \times 10^{5}$ cm$^{-3}$ (we recall $\bar{n}=500$ cm$^{-3}$) which is close to the critical densities for the NH$_{3}$ and N$_{2}$H$^{+}$ molecules of $\sim 1-2 \times 10^{5}$ cm$^{-3}$. All of the cores that have peak densities of the order of the critical density for exciting the NH$_{3}$ and N$_{2}$H$^{+}$ lines are gravitationally bound and proceed later to gravitational collapse. Examples are core 38 in model M10J4$\beta$.1 and core 37 in model M10J4$\beta$1 and the collapsing cores in the non magnetic model. The former two cores have peak number densities of 336.7 and 390.5 $\bar{n}$ and average number densities of 142.7 and 164.2 $\bar{n}$, respectively. They possess small $\alpha_{vir}$ values in the range of $0.15-0.20$ (see Figs.~\ref{fig20} and \ref{fig24}) and, thus, are bound according to their $\alpha_{vir}$ values as well as the virial analysis. Hence our results tend to show that NH$_{3}$ or N$_{2}$H$^{+}$ cores are gravitationally bound. However the number of cells in those cores which have densities larger than $256~\bar{n}$ is small and the estimates of the virial parameter in such cases are likely to be dominated by numerical noise; b) We do not observe, in any of the analyzed simulation, CCs that are positioned in the upper right quadrant in the $(W+\Theta_{VT})-(|W|/|\Theta_{VT}|)$ diagrams. This simply means that the extended distribution of mass around the CCs has very little influence on the CCs gravitational boundedness and that the gravitational term $W$ is always close to the CCs gravitational energy; c) We observe that some CCs, which are dynamical in essence, have $W+\Theta_{VT} \sim 0$ and $|W|/|\Theta_{VT}| \sim 1$ (i.e., Figs.~\ref{fig17} and \ref{fig22}). This could lead them to be cataloged as being in a state of {\it magneto-static equilibrium} or 'virialized', which is not supported by their virial analysis as they have non-zero temporal terms. In reality they are just merely 'crossing the line' between the instantaneously unbound and bound regimes.   
        
\section{CONCLUSIONS AND DISCUSSION}\label{conc}

Understanding the gravitational boundedness of clumps and cores (CCs) in molecular clouds (MCs) is a key element in better describing the process of fragmentation of a cloud which, itself, is the necessary step on the way of better understanding the process of star formation in the dense regions of a MC, the origin of stellar multiplicity, and the initial stellar mass function (see, e.g., Padoan \& Nordlund 2002,2004; Mac Low \& Klessen 2004; Ballesteros-Paredes et al. 2007; Klein et al. 2006). 
 
In this paper, we have analyzed the instantaneous virial balance of CCs that form in three-dimensional, isothermal, magnetohydrodynamical, driven, and self-gravitating MC models. The analyzed simulations vary by the strength of the magnetic field initially present in the box ranging from subcritical to non-magnetic regimes. For each simulation, the selection of the CCs and the virial analysis is performed at different timesteps. However, in order to follow more accurately the internal dynamics inside the analyzed CCs, results are preferentially presented at timesteps that precede the appearance of collapsed objects in the simulation box. CCs have been identified by a friend-of-friend algorithm as being an ensemble of connected cells that have a density higher than a defined density threshold. We have used a range of selected thresholds (i.e., from 7.5 to 100 times the average density) that covers a wide dynamical range that mimics the usage of various density tracers sensitive to different average gas densities in the observations. Once an object has been defined, we calculate for this object all the quantities that appear in the virial theorem in its Eulerian form (Eq.~\ref{eq1}), along with other quantities such as its mass, volume, average density, internal velocity dispersion and three quantities commonly used in observational studies as gravitational binding indicators, namely the Jeans number $J_{c}$, the mass-to magnetic flux ratio (normalized to the critical value for collapse) $\mu_{c}$, and the virial parameter $\alpha_{vir}$. In the virial theorem formalism, we define a CC as being gravitationally bound when the gravitational term is larger than the amount for the system to be virialized, which is more stringent than the condition that it be large enough to make the total volume energy negative (see \S~2). Our results show that :

a) CCs are not in virial equilibrium, and are dynamical out-of-equilibrium structures as indicated by their complex changing geometry over time and the importance of the flux of moment of inertia term in the virial equation, which is a quantitative confirmation of the highly dynamical, morphing nature of CCs in a MC. 

b) The surface energy terms are of the same order than the volume energy term (by up to a factor $\sim 5$ smaller or larger) and, thus, are very important in determining the virial balance in CCs. Particularly important are the anisotropic kinetic and magnetic surface energy terms, which are related to the complex topology of the velocity and magnetic fields at the boundary of the CC. We observe that the difference between the volume and surface term is larger whenever the velocity or magnetic field does not cross smoothly the surface of the object, due for example to the presence of a high density peak.  

c) We observe that CCs can be either in the process of being compressed by the velocity field and in that case they have a net kinetic energy $E_{k}-\tau_{k} < 0$ or of being dispersed with $E_{k}-\tau_{k} > 0$. Yet, not all CCs that have $E_{k} - \tau_{k} < 0$ are necessarily gravitationally bound, but are simply in the process of assembling from less dense gas by turbulent ram pressure.  

d) Despite their dynamical nature, some CCs are observed to have $W+\Theta_{VT} \sim 0$ ($\Theta_{VT}=2~(E_{th}+E_{k}-\tau_{th}-\tau_{k})+E_{mag}+\tau_{mag}$) and $|W|/|\Theta_{VT}| \sim 1$ which can lead them, erroneously, to be cataloged as being in a state of {\it virial or magnetostatic equilibrium} or 'virialized', despite their dynamical nature. Additionally, we find, in all simulations, that the gravitational term must always be close to the gravitational energy and that the mass distribution outside the CCs does not have a significant contribution in distorting the CCs structure by external gravitational torques. 

e) There is no one-to-one correspondence between the state of gravitational boundedness of a CC as described by the virial balance analysis (i.e., gravity versus other virial energy terms) and as implied by the classical gravitational binding indicators $J_{c}$, $\mu_{c}$, and $\alpha_{vir}$. In general, from the virial analysis we observe that only the inner regions of the clumps (i.e., the dense cores selected at high density thresholds) are gravitationally bound, whereas Jeans number estimates of the same clumps tend to show that the objects are gravitationally bound at all threshold levels. On the other hand, the calculated $\alpha_{vir}$ values not only shows that the clumps are more gravitationally bound at the lower threshold levels as the Jeans numbers, but also indicate a number of gravitationally bound objects always in excess of what is yielded by the virial analysis. Preliminary results by Dib \& Kim (2006) show that the $J_{c}$ estimates and the corresponding energy ratio, when including the thermal surface energy ($E_{th}-\tau_{th}/|W|$) are well correlated, and that the virial parameter $\alpha_{vir}$ and the corresponding energy ratio $(E_{k}+E_{th}-\tau_{k}-\tau_{th})/|W|)$ or ($(E_{k}-\tau_{k})/|W|$) are poorly correlated. A future study (Dib et al., in preparation) will investigate these relations in more detail for different models as well as the correlation between $\mu_{c}$ and the magnetic energy to gravitational term ratio. 

f) The exponent, $\epsilon$, of the virial parameter-mass relation, $\alpha_{vir} \propto M_{c}^{\epsilon}$, shows a better agreement with observationally derived values for the case of the moderately supercritical cloud in which the most massive CCs are near critical ($\mu_{c} \sim 1$) , and a lesser good agreement for the subcritical (in which most CCs are subcritical) and strongly supercritical (in which the most massive object is supercritical) and non-magnetic cases. This result is not extremely conclusive yet because of the large uncertainties in the observational values of $\epsilon$ and the limited statistics in our simulations. Yet, it might reflect an agreement with the recent predictions that most observed CCs have a mass-to magnetic flux ratio that is supercritical or subcritical by factors of 2 with the median value being around $\mu_{c} \sim 1$ (Crutcher \& Troland 2006).    

g) In the non-magnetic simulation, we observe the formation of a core which possesses all the structural and dynamical properties of the Bok globule Barnard 68 (B68). This core is gravitationally bound. Such bound cores may survive the ionizing front of a \ion{H}{2} region formed elsewhere in the cloud and which expands in the cloud's clumpy distribution (e.g., Mellema et al. 2006; Will Henney, private communication). The ionizing front can evacuate the gas around the core, and leave it confined by a surrounding warm gas such as in the case of B68.        

 In the set of simulations presented in this work, we have used an isothermal equation of state to describe the gas physics. However, several authors have argued recently that deviations from isothermality in molecular clouds might lead them to have distinct structural and/or dynamical properties than isothermal clouds (e.g., Larson 2005). Li et al. (2003) and Jappsen et al. (2004) showed that the fragmentation process is dependent on the polytropic exponent which describes the equation of state. Dib et al. (2004) showed that the average density-size relation for non-isothermal clouds follows the observed $\bar{n} \propto R_{c}^{-1}$ Larson relations (Larson, 1981) unlike the isothermal clouds and their substructure is which this relation is not found (e.g., Ballesteros-Paredes \& Mac Low 2002; Li et al. 2004). Hennebelle \& Inutsuka (2006) discuss the possibility that warm gas can survive inside MCs. However, Pavlovski et al. (2006), showed that deviations from isothermality are only visible in terms of the high temperature zones behind shock waves. The detailed effects of a non-isothermal equation of state on the properties of individual cores has not been investigated in detail so far, and is left to a future work. 

\acknowledgements

 We have benefited from interesting discussions with Paolo Padoan, George Helou, Caroline Bot, Sean Carey, Alberto Noriega-Crespo, Alexei Kritsuk, Frank Shu, J\"{u}rgen Steinacker, Richard Crutcher, Patrick Hennebelle, Mark Krumholz, Paola Caselli, and Eve Ostriker. We are grateful to the referee, Christopher McKee, for insightful comments which helped improve this paper in many aspects. S. D. is supported by a UNAM postdoctoral fellowship and is grateful to the hospitality of the Infrared Processing and Analysis Center (IPAC) at Caltech and to the Center for Astrophysics and Space Sciences (CASS) at the University of California, San Diego, during which parts of this article have been written. J. K. is supported by the Astrophysical Research Center for the Structure and Evolution of the COSMOS (ARCSEC) of the Korea Science and Engineering Foundation through the Science Research Center (SRC) program, E.V.-S by CONACyT grant 47366-F, and M. S. by the Programme for Research in Third Level Institutions (PRTLI) administered by the Irish Higher Education Authority under the National Development Plan and with partial support from the European Regional Development Fund. The numerical simulations analyzed in this paper were performed on the Linux cluster at KASI, with funding from KASI and ARCSEC. This research has made use of NASA's Astrophysics Data System Bibliographic Services. S.D. dedicates this work to the memory of his grandmother Soubhi\'{e} Hijazi-Dib.    
 
{}

\clearpage 
\begin{deluxetable}{cccccccc}
\tabletypesize{\footnotesize}
\tablecaption{Number of clumps and cores found in the simulations for the various density thresholds at selected time-steps. \label{tab1}}
\tablewidth{0pt}
\tablehead{
\colhead{Model} & \colhead{time-step} & \colhead{ $n_{thr}=7.5~\bar{n}$} & \colhead{$n_{thr}=15~\bar{n}$} & \colhead{$n_{thr}=30~\bar{n}$} & \colhead{$n_{thr}=60~\bar{n}$} &  \colhead{$n_{thr}=100~\bar{n}$} & \colhead{$n_{thr}=500~\bar{n}$} }
\startdata

M10J4$\beta$.01     & 223 &  11 & 8  & 2  & 1 & 1  & 0  \\
                    &     &     &    &    &   &    &    \\

M10J4$\beta$.1      & 30  &  19  & 18 & 8  & 5 & 4 & 0      \\
                    & 40  &  16  & 11 & 7  & 3 & 2 & 0      \\
                    & 50  &  14  & 14 & 9  & 2 & 1 & 1      \\
                    & 210 &  12  &  9 & 9  & 4 & 3 & 2      \\
                    &     &      &    &    &   &   &        \\  

M10J4$\beta$1       & 20  &  16  & 14 & 7 & 2  & 2  & 0     \\
                    & 30  &  11  & 11 & 9 & 5  & 3  & 0     \\
                    & 40  &  15  & 13 & 8 & 6  & 1  & 0     \\ 
                    & 50  &  14  & 10 & 5 & 3  & 3  & 1     \\
                    & 60  &  16  & 15 & 5 & 4  & 3  & 1     \\
                    & 70  &  15  & 12 & 7 & 5  & 3  & 1     \\
                    &     &      &    &    &   &    &       \\

M10J4$\beta \inf$   & 20  &  23$^a$  & 23  & 23  & 15 & 3  & 1  \\
                    & 30  &  19$^a$  & 19  & 19  & 12 & 12 & 5  \\ 
                    & 40  &  22$^a$  & 22  & 22  & 18 & 12 & 6  \\

\enddata
\tablecomments{
               (a) due to the large number of cells found in dense structures in the non-magnetic run M10J4$\beta \inf$, the lowest density threshold used is 10 $\bar {n}$.}

\end{deluxetable}

\clearpage
\pagestyle{empty} 

\setlength{\voffset}{2cm}

\begin{deluxetable}{llllllllllllllllllllllll}
\rotate
\tabletypesize{\tiny}
\tablecaption{Numbering of the condensations found in run M10J4$\beta$.01 at t=223=8.92 Myr, run M10J4$\beta$.1 at t=40=1.6 Myr, and run M10J4$\beta$1 at t=30=1.2 Myr. Each column corresponds to a separate object which is attributed a distinct number at the different density thresholds. In the case the inner parts of an object are fragmented into 2 clumps, each core is assigned a different number at the corresponding higher density threshold. The numbering of the non-magnetic run is not shown due to the large number of columns needed (i.e., 23 different objects, see Tab.~\ref{tab1}), the numbering of CCs in this model follows the same rules as for the magnetized runs. Between parentheses are shown the cubic roots of the total number of cells present in each object. This only represents the characteristic number, not the number of cells along each of the object's principal axis since many objects are elongated. In our scaling, one cell represents a physical scale of 4/256=0.015625 pc.\label{tab2}}
\tablewidth{0pt}
\tablehead{
\colhead{n$_{thr}$} & \colhead{o.1} & \colhead{o.2} & \colhead{o.3} & \colhead{o.4} & \colhead{o.5} & \colhead{o.6} & \colhead{o.7} & \colhead{o.8} & \colhead{o.9} & \colhead{o.10} & \colhead{o.11} & \colhead{o.12} & \colhead{o.13} & \colhead{o.14} & \colhead{o.15} & \colhead{o.16} }
\startdata

M10J4$\beta$.01 at t=223=8.92 Myr\cr
                &    &    &    &    &    &    &    &    &    &    &    &    &    &    &    &   \cr             
7.5~$\bar{n}$   & 1~(53.0) &  2~(14.9) & 3~(16.1)  &  4~(15.2) &  5~(1.7) &  6~(11.51) &  7~(3.8) & 8~(2.9) & 9~(7.8) & 10~(1.2) & 11~(1) &    &    &    &    &   \cr
                &    &    &    &    &    &    &    &    &    &    &    &    &    &    &    &   \cr
15~$\bar{n}$    & 12~(26.1) & 13~(12.0) & 14~(14.1) & 15~(15.4) & 16~(1.7) & 17~(11.0) & 18~(3.8) & 19~(2.9) &    &    &    &    &    &    &    &   \cr
                &    &    &    &    &    &    &    &    &    &    &    &    &    &    &    &   \cr 
30~$\bar{n}$    & 20~(19.0) & 21~(4.8) &    &    &    &    &    &    &    &    &    &    &    &    &    &   \cr
                &    &    &    &    &    &    &    &    &    &    &    &    &    &    &    &   \cr  
60~$\bar{n}$    & 22~(12.4) &    &    &    &    &    &    &    &    &    &    &    &    &    &    &   \cr
                &    &    &    &    &    &    &    &    &    &    &    &    &    &    &    &   \cr
100~$\bar{n}$   & 23~(5.2) &    &    &    &    &    &    &    &    &    &    &    &    &    &    &   \cr 
                &    &    &    &    &    &    &    &    &    &    &    &    &    &    &    &   \cr
                &    &    &    &    &    &    &    &    &    &    &    &    &    &    &    &   \cr
M10J4$\beta$.1 at t=40=1.6 Myr   \cr
                &    &    &    &    &    &    &    &    &    &    &    &    &    &    &    &   \cr
7.5~$\bar{n}$   & 1~(58.9) &  2~(3.5) &  3~(14.0) & 4~(16.0)  & 5~(6.4)  & 6~(18.5)  & 7~(7.8)  & 8~(10.2)  & 9~(11.5)  & 10~(1) & 11~(2.1) & 12~(1) & 13~(3.8) & 14~(2.3) & 15~(1.2) & 16~(1.4) \cr
                &    &    &    &    &    &    &    &    &    &    &    &    &    &    &    &   \cr
15~$\bar{n}$    & 17~(33.2) & 18~(3.5) & 19~(14.0) & 20~(16.0) & 21~(6.4) & 22~(18.5) & 23~(7.8) & 24~(10.2) & 25~(7.0) & 26~(1) & 27~(2.1) &    &    &    &    &   \cr
                &    &    &    &    &    &    &    &    &    &    &    &    &    &    &    &   \cr 
30~$\bar{n}$    & 28~(19.1) & 29~(3.5) & 30~(10.4) & 31~(9.5) & 32~(6.4) & 33~(9.6) & 34~(7.6) &    &    &    &    &    &    &    &    &   \cr
                &    &    &    &    &    &    &    &    &    &    &    &    &    &    &    &   \cr  
60~$\bar{n}$    & 35~(13.8) & 36~(3.5) & 37~(5.1) &    &    &    &    &    &    &    &    &    &    &    &    &   \cr
                &    &    &    &    &    &    &    &    &    &    &    &    &    &    &    &   \cr
100~$\bar{n}$   & 38,39~(9.8,3.5) &    &    &    &    &    &    &    &    &    &    &    &    &    &   \cr 
                &    &    &    &    &    &    &    &    &    &    &    &    &    &    &    &   \cr
                &    &    &    &    &    &    &    &    &    &    &    &    &    &    &    &   \cr
M10J4$\beta$1 at t=30=1.2 Myr \cr 
                &    &    &    &    &    &    &    &    &    &    &    &    &    &    &    &   \cr
7.5~$\bar{n}$   & 1~(59.1) & 2~(12.1) & 3~(19.9) & 4~(11.5) & 5~(16.7) & 6~(7.0) & 7~(1.2) & 8~(2.2) & 9~(11.7) & 10~(5.6) & 11~(1.2) &    &    &    &    &   \cr
                &    &    &    &    &    &    &    &    &    &    &    &    &    &    &    &   \cr
15~$\bar{n}$    & 12~(32.3) & 13~(12.1) & 14~(18.5) & 15~(11.5) & 16~(16.7) & 17~(7.0) & 18~(1.2) & 19~(2.2) & 20~(11.7) & 21~(5.6) & 22~(1.2) &    &    &    &    &   \cr
                &    &    &    &    &    &    &    &    &    &    &    &    &    &    &    &   \cr 
30~$\bar{n}$    & 23~(18.3) & 24~(12.1) & 25~(2.7) & 26~(8.4) & 27~(12.4) & 28~(7.0) & 29~(1.2) & 30~(2.2) & 31~(3.3) &    &    &    &    &    &    &   \cr
                &    &    &    &    &    &    &    &    &    &    &    &    &    &    &    &   \cr  
60~$\bar{n}$    & 32~(10.3) & 33~(10.2) & 34~(7.8) & 35~(2.6) & 36~(3.5) &    &    &    &    &    &    &    &    &    &    &   \cr
                &    &    &    &    &    &    &    &    &    &    &    &    &    &    &    &   \cr
100~$\bar{n}$   & 37~(7.8) & 38~(4.1) & 39~(3.2) &    &    &    &    &    &    &    &    &    &    &    &    &   \cr 
             
\enddata
\end{deluxetable}
\clearpage
\pagestyle{plaintop}

\setlength{\voffset}{0cm}

\begin{deluxetable}{lc}
\tabletypesize{\footnotesize}
\tablecaption{Exponent, $\epsilon$, of the Virial parameter-Mass relation, $\alpha_{vir} \propto M_{c}^{\epsilon}$ obtained from the observations of a number of molecular clouds (Bertoldi \& McKee 1992$^a$; Williams et al. 2000) and from our models. \label{tab3}}
\tablewidth{0pt}
\tablehead{
\colhead{Cloud/Model} & \colhead { $\epsilon$ }}
\startdata

G216-2.5    (Williams et al.)    &   $\sim -0.72$     \\    
                                 &                    \\
Rosette     (Williams et al.)    &   $\sim -0.61$     \\    
                                 &                    \\
Rosette     (Bertoldi \& McKee)  &   $-0.50 \pm 0.10$ \\
                                 &                    \\
Ophiucus    (Bertoldi \& McKee)  &   $-0.68 \pm 0.03$ \\
                                 &                    \\
Orion B     (Bertoldi \& McKee)  &   $-0.67 \pm 0.24$ \\
                                 &                    \\
Cepheus OB3 (Bertoldi \& McKee)  &   $-0.54 \pm 0.08$ \\
                                 &                    \\         
model M10J4$\beta$.01            &   $-0.57 \pm 0.03$ \\
                                 &                    \\
model M10J4$\beta$.1             &   $-0.60 \pm 0.03$ \\
                                 &                    \\  
model M10J4$\beta$1              &   $-0.49 \pm 0.03$ \\
                                 &                    \\
model M10J4$\beta \inf$          &   $-0.48 \pm 0.02$ \\
                    
\enddata
\tablecomments{
               (a) Bertoldi \& McKee (1992) did not include the massive objects in their fits to the $\alpha_{vir}-M_{c}$ relations which would have the effect of giving, for some clouds they have analyzed, a slightly shallower value of $\epsilon$}
\end{deluxetable}

\clearpage 

\clearpage
\thispagestyle{empty}
\clearpage

\clearpage

\clearpage
\thispagestyle{empty}
\clearpage

\clearpage

\end{document}